\documentclass[12pt]{article}
\pdfoutput=1

\usepackage[dvipdfmx]{graphicx}
\usepackage{graphicx,bm,color}
\usepackage{amsmath}
\usepackage{amssymb}
\usepackage{amsfonts}
\usepackage{cases}
\usepackage{cancel}
\usepackage{hyperref}
\usepackage{ulem}
\usepackage{cite}

\newcommand{\dd}[2]{\frac{\mathrm{d} #1}{\mathrm{d}#2}}

\newcommand{\gsim}{\lower.7ex\hbox{$\;\stackrel{\textstyle>}{\sim}\;$}}
\newcommand{\lsim}{\lower.7ex\hbox{$\;\stackrel{\textstyle<}{\sim}\;$}}

\definecolor{RFcolor}{rgb}{0.48,0,0.76}

\begin{document}

\vspace*{8mm}

\begin{center}

{\Large\bf  Non-supersymmetric SO(10) models}

\vspace*{2mm}

{\Large\bf with Gauge and Yukawa coupling unification}

\vspace*{9mm}

\mbox{\sc Abdelhak Djouadi$^{1,2}$, Renato Fonseca$^1$,  Ruiwen Ouyang$^2$, Martti Raidal$^2$ }\vspace*{3mm}

{\small

$^1$ Centro Andaluz de Fisica de Particulas Elementales (CAFPE) and \\ Depto.~de F\'isica Te\'orica y del Cosmos, Universidad de Granada, E--18071 Granada, Spain\\ \vspace{0.3cm}

$^2$ Laboratory of High Energy and Computational Physics, NICPB, R\"avala pst. 10,\\ 10143 Tallinn, Estonia\\ \vspace*{0.3cm}

}

\end{center}

\vspace{20pt}

\begin{abstract}

\noindent We study a non-supersymmetric SO(10) Grand Unification Theory with a very high energy intermediate symmetry breaking scale in which not only gauge but also Yukawa coupling unification are enforced via suitable threshold corrections and matching conditions. For gauge unification, we focus on a few symmetry breaking patterns with the intermediate gauge groups ${\rm SU(4)_C \times SU(2)_L \times SU(2)_R}$ (Pati-Salam) and ${\rm SU(3)_C \times SU(2)_L \times SU(2)_R\times U(1)_{B-L}}$ (minimal left-right symmetry) assuming an additional global U(1) Peccei--Quinn symmetry, and having the Standard Model supplemented by a second Higgs doublet field at the electroweak scale. We derive the conditions as well as the approximate analytical solutions for the unification of the gauge coupling constants at the two-loop level and discuss the constraints from proton decay on the resulting high scale.  Specializing to the case of the Pati-Salam intermediate breaking pattern, we then impose also the unification of the  Yukawa couplings of third generation fermions at the high scale, again at the two-loop level. In the considered  context, Yukawa unification implies a relation between the fermion couplings to the 10- and 126-dimensional scalar representations of the SO(10) group. We consider one such possible relation which is obtainable in an ${\rm E_6}$ model where the previous two scalar fields are part of a single multiplet. Taking into account some phenomenological features such as the absence of flavor changing neutral currents at tree-level, we derive constraints on the parameters of the low energy model, in particular on the ratio of the two Higgs doublets vacuum expectation values $\tan\beta$.  

\end{abstract}

\clearpage

\section{Introduction}

A Grand Unified Theory (GUT) which describes the four fundamental forces that are present in Nature has always been the Holy Grail of particle physics. Leaving aside the gravitational force which has a rather special status, it has been shown already in the 1970's \cite{Georgi,PS} that the concept of gauge symmetries makes it possible to combine in a very elegant manner the electromagnetic, weak and strong interactions of the Standard Model (SM) into a single force at a very high energy scale \cite{Weinberg}. This would have been the case of SU(5), the simplest and most economical gauge symmetry group that contains the ${\rm SU(3)_C \! \times \!  SU(2)_L \!  \times \! U(1)_Y}$ group of the SM as a subgroup. Alas, when the three SM gauge couplings are evolved with the energy scale, starting from their experimentally measured values and including the SM particle content, they shortly fail to meet at a single point, the presumed GUT scale $M_U$ \cite{SUSY-Unif}. 

One solution to this problem was to invoke Supersymmetry (SUSY) \cite{SUSY}, a theory that predicts the existence of a partner to each SM particle and has an extended Higgs sector consisting of two complex scalar fields to break the electroweak symmetry down to the electromagnetic U(1) group \cite{HMSSM}. The new particle content modifies the slopes of the renormalisation group evolution (RGE) of the gauge couplings such that they meet at a GUT scale that is high enough, $M_U\! \approx \!2 \! \times \! 10^{16}$ GeV, to prevent a too fast decay of the proton \cite{SUSY-Unif}. Another virtue of SUSY, which made it extremely popular in the past four decades, is that it solves the problem of the large hierarchy between the weak and Planck scales that induces quadratic ``divergences" to the observed Higgs boson mass. However, in order to  resolve both the unification and hierarchy problems, SUSY needs to be broken at an energy not too far from the electroweak scale and, hence, should involve superpartners with masses of a few hundred GeV to order a TeV at most. Unfortunately, such a low SUSY-breaking scale has been excluded for most superparticles (in particular the strongly interacting ones that are copiously produced) by dedicated and non-conclusive searches at the CERN LHC \cite{SUSY-LHC}. Thus, the theory lost some of its appeal as it appears now to be less ``natural".

In principle, the existence of extra particles with the appropriate masses and quantum numbers to give the necessary contributions to the RGEs is all what is needed to achieve unification of the three gauge couplings. However, postulating the existence of extra fields for this reason alone might be considered a somewhat contrived solution to the problem. A more appealing possibility is to consider symmetry groups larger than SU(5)  which break down to the SM gauge group via a chain that involves intermediate symmetries. In this case, the new scalar multiplets that break these intermediate symmetries (and some of the associated new gauge bosons) will generate additional contributions at the intermediate scale $M_I$, which will modify the RG evolution of the gauge couplings. Taking into account these threshold corrections, it is then possible to unify the three couplings at a scale $M_U$ \cite{Threshold}. 

Such a unification with an intermediate step can be realized in the context of SO(10) \cite{SO10}; see Refs.~\cite{Unif,Meloni} This group is particularly interesting as it has a representation of dimension 16 which can accommodate the 15 SM chiral fermions of each generation, as well as an additional Majorana neutrino. If the mass of this neutrino is very large, of the order of $10^{12\!-\!14}$ GeV, other very pressing problems in particle physics can also be addressed. This is, for instance, the case of  the complicated pattern of the SM neutrino masses and mixings which can be explained by the see-saw mechanism. This is also the case of the baryon asymmetry in the Universe which could be achieved through a leptogenesis triggered by the additional Majorana neutrino. Hence, SO(10) with an intermediate scale of ${\cal O}(10^{12\!-\!14})$ GeV, could explain  the most acute problems of the SM that call for new physics beyond it, leaving aside the hierarchy problem  and introducing a suitable axion that could account for the particle that forms the dark matter in the Universe; see Refs.~\cite{Babu,Bajc,Altarelli} for reviews. 

Another issue for which low-energy SUSY theories gained popularity, is the unification of the Yukawa couplings of third generation fermions \cite{SUSY-YUC}. This additional step in the unification paradigm is accomplished in the minimal supersymmetric extension of the SM (MSSM), thanks to the presence of the two-Higgs doublets fields that are required by the extended symmetry. In constrained scenarios, such as the minimal supergravity model with universal ``soft" SUSY--breaking parameters \cite{mSUGRA}, the top, bottom and tau Yukawa couplings can be unified at the same scale $M_U$ that allows for gauge coupling unification. Indeed, for large values of the ratio of the vacuum expectation values of the two Higgs fields, $\tan\beta$, one  can generate the required hierarchy for the top and bottom quark masses, $\tan\beta \! \approx \! m_t/m_b  \! \approx \! 60$, and the RG evolution that allows the couplings to also meet at $M_U$.

In a recent letter, we have contemplated the possibility of Yukawa coupling unification in the context of a non-supersymmetric SO(10) model as well \cite{Letter}. Focusing on the most widely studied scenarios with intermediate symmetry breaking, namely the Pati-Salam scenario with the intermediate group ${\rm SU(4)_C \! \times \!  SU(2)_L \!  \times \! SU(2)_R}$ \cite{PS} and the minimal left-right symmetry group ${\rm SU(3)_C \!  \times SU(2)_L \! \times \!  SU(2)_R } $ $ {\rm \times U(1)_{B-L}}$ \cite{LR}, we have shown that in a two-doublet Higgs model (2HDM) extension present at the electroweak scale, exactly like in the MSSM, one can first obtain the correct hierarchy for the  masses $m_t$ and $m_b$  by again taking a ratio $\tan\beta$ that is sufficiently high. In both schemes, it is then possible to arrange such that the RG running of third generation Yukawa couplings, with suitable matching conditions at the same intermediate scale $M_I$ for which gauge coupling unification occurs, leads to Yukawa coupling unification at the same GUT scale $M_U$. This can be achieved while preserving important phenomenological features such as reproducing third family fermion and weak gauge boson masses, ensuring the stability of the electroweak vacuum up to the high scales and keeping the Yukawa couplings perturbative.  

In this paper, we perform a more exhaustive analysis of the possibility of simultaneous gauge and Yukawa coupling unification, extending the earlier analysis \cite{Letter} in several directions. Firstly, the present discussion is more thorough and general, as our results are valid for any breaking chain of non-SUSY SO(10) models with only one intermediate scale and we consider the interplay between gauge coupling unification, proton decay, the perturbativity of the Yukawa couplings and, more importantly, the absence of flavor changing neutral currents at tree-level. Secondly, for gauge coupling unification, we present some approximations which highly simplify the analytical discussions of the RGEs and we discuss unification in models in which one adds a global U(1) Peccei-Quinn symmetry \cite{Peccei-Quinn} that would allow the resulting axion to address the dark matter problem; this will have important repercussions on the breaking pattern, the RGE running of the couplings as well as on the fermionic mass pattern. {A third difference when compared to Ref.~\cite{Letter} is that in the present work, we study the case in which the condition for Yukawa couplings unification at the high scale is inspired by the existence of an even larger ${\rm E_6}$ gauge symmetry.}

The rest of the paper is organized as follows. In the next section, we introduce the non-SUSY SO(10) model, discuss its various intermediate breaking schemes and the weak scale 2HDM structure.  In section 3, we enforce gauge coupling unification using threshold effects and discuss some approximations. In section 4, we analyze the issue of simultaneously unifying the gauge and third generation fermion Yukawa couplings. A short conclusion is made in Section 5 and some analytical complementary material is given in an Appendix.

\section{Non-SUSY SO(10) with an intermediate scale}\label{sec:2}

In this section, we will summarize how unification of the three gauge interactions of the SM can be achieved in a non-supersymmetric  SO(10) GUT with a spontaneous symmetry breaking pattern  that involves an intermediate gauge group at a very high scale which breaks down to the SM gauge group. A very interesting aspect of the SO(10) model is that all fermions can be embedded into a single representation of the symmetry group.

Indeed, the SO(10) group possesses a fundamental 16-dimensional representation ${\bf 16_F}$ in which, for each generation, the 15 SM chiral fermions\footnote{For a single generation of the SM fermions, one has two chiralities time six colored quarks and one charged lepton, plus a left-handed neutrino; this makes 15 degrees of freedom in total.} as well as one right-handed neutrino can be embedded. In this case, the allowed Yukawa couplings of the scalar bosons to pairs of these fermions belong to the direct product representation ${\bf 16_F} \times {\bf 16_F}$, which can be decomposed into
\begin{eqnarray}
{\bf 16_F} \times {\bf 16_F} = {\bf 10} + {\overline{\bf 126}} + {\bf 120} \, .
\end{eqnarray}
Thus, the most general Yukawa interactions are given by the expression
\begin{eqnarray}
-{\cal L}_{\rm Yukawa} = {\bf 16_F} (Y_{10} {\bf 10_H} + Y_{126} {\bf \overline{126}_H} + Y_{120} {\bf 120_H} ) {\bf 16_F} \, ,
\label{eq:LY-general}
\end{eqnarray}
where ${\bf 10_H}$, ${\bf \overline{126}_H}$, and ${\bf 120_H}$ denote the scalar representations of SO(10) group. However, among the large number of scalar field components in these representations, we will assume all those that do not participate in the symmetry breaking mechanism by acquiring vacuum expectations values (vevs) will have masses of the order of the SO(10) symmetry-breaking scale. This is known as the extended survivial hypothesis~\cite{Survival}, by which one can safely decouple most of the redundant ingredients in the SO(10) scalar representations at the GUT scale and be left only with the light Higgs boson spectrum of the low-energy effective theory which is present at the electroweak scale. The hypothesis helps to drastically reduce the number of scalar fields that couple to fermions and, hence,  to simplify the structure of the Yukawa sector of the model.  

As was discussed in many instances, see for instance Ref.~\cite{Bajc}, the Yukawa sector of the SO(10) model must consist of a ${\bf \overline{126}_H}$ representation, to trigger the see-saw mechanism via the breaking of the left-right symmetry at an intermediate scale $M_I$. One additional scalar representation, either the ${\bf 10_H}$ or the ${\bf 120_H}$, is needed to break the SM gauge symmetry. Because the main difference between these two representations is that the ${\bf 120_H}$ decomposes into four scalar doublets under the SM group, while the ${\bf 10_H}$ representation decomposes only in two scalar doublets, based on minimality one should consider the Yukawa sector of SO(10) with only the ${\bf 10_H}$ and the ${\bf \overline{126}_H}$ representations, leading to the so-called minimal SO(10) models. Note that more scalar representations, which do not affect fermion masses, are needed to achieve the correct symmetry breaking pattern down to the Standard Model group, as will be discussed in the breaking patterns below.

Given that the 10-dimensional representation of SO(10) is real, the field ${\bf 10_H}$ could in principle be real. However, it was shown~\cite{Bajc} that a scalar sector composed of a real ${\bf 10_H}$ and a complex ${\bf \overline{126}_H}$ leads to an unrealistic mass spectrum for the second and third generation fermions. \footnote{If the ${\bf 10_H}$ field is real and there is more than one generation of fermions, the mass ratio of isospin up and down--type quarks is predicted to be of order 1 \cite{Bajc}, in contradiction with the observed quark masses: $m_t/m_b\gg1$. This indicates that a more complicated scenario should be considered.}

The simplest possible extension is to complexify the original real ${\bf 10_H}$, which leads to the following minimal SO(10) model with complex scalar fields 
\begin{eqnarray}
-{\cal L}_Y = {\bf 16_F} (Y_{10} {\bf 10_H}  + Y_{10^*} {\bf 10_H^*} + Y_{126} {\bf \overline{126}_H} ) {\bf 16_F} \, .\label{eq:LY-complex-10}
\end{eqnarray}
The price to pay is that we need to introduce a new Yukawa coupling $Y_{10^*}$ which makes the theory less predictive.  To avoid the extra independent Yukawa coupling associated with the ${\bf 10^*_H}$, we will assume in this paper an additional global U(1) Peccei-Quinn symmetry~\cite{Peccei-Quinn} with the following charge assignment for some real parameter $\alpha$ 
\begin{eqnarray}
{\bf 16_F} \to e^{i \alpha} {\bf 16_F} \, , \quad {\bf 10_H} \to e^{-2i \alpha} {\bf 10_H} \, , \quad {\bf \overline{126}_H} \to e^{-2i \alpha} {\bf \overline{126}_H}\,. \label{eq:U1-charges}
\end{eqnarray}
It reduces eq.~(\ref{eq:LY-complex-10}) to
\begin{eqnarray}
-{\cal L}_Y = {\bf 16_F} (Y_{10} {\bf 10_H}  + Y_{126} {\bf \overline{126}_H} ) {\bf 16_F} \, .
\label{eq:LY-minimal}
\end{eqnarray}
There are two additional motivations for adding such a global symmetry to our model. A first one is that, when the ${\rm U(1)_{PQ} }$ symmetry is broken by assigning a vev to a SO(10) scalar, 
it influences the symmetry breaking pattern  and the renormalization group running of the gauge couplings, as we will see shortly. Another and more phenomenological motivation is that it implies the existence of an axion which can solve  the strong CP problem \cite{Peccei-Quinn} and, at the same time, provide a good candidate for the dark matter in the Universe. 

The breaking of SO(10) to the SM gauge group, which we will denote for shortness by
\begin{equation}
{\rm G_{SM} \equiv G_{321} = SU(3)_C \times SU(2)_L \times U(1)_Y} \, , 
\end{equation} 
can be triggered in several ways. As mentioned in the introduction, in the case of non-SUSY SO(10) models, the existence of intermediate scales $M_I$ play an important role in the unification of the gauge couplings at some scale $M_U$. More precisely,  as the evolution of the U(1) coupling needs to be strongly modified to meet with the other two couplings at a unique scale that should be high enough, large contributions from additional gauge bosons are required. These contributions can be provided by the particles that are present at the intermediate breaking step, i.e.  at the scale $M_I$. We will stick to the breaking chains with only one intermediate-step involving a left-right (LR) symmetry --- with a ${\rm SU(2)_R}$ group --- to invoke the see-saw mechanism for neutrinos. The embedding of the ${\rm SU(2)_R}$ symmetry above the intermediate scale $M_I$ strongly affects the gauge coupling evolution.
 
 In our analysis, we will interested in the following breaking patterns:
\begin{eqnarray}
 \text{422}: && \quad \text{SO(10)}|_{M_{U} } \xrightarrow{\langle \mathbf{210_H} \rangle} \, \, \, {\cal G}_{422}|_{M_{I} } \, \, \, \xrightarrow{\langle \mathbf{\overline{126}_H}\rangle
} {\cal G}_{321}|_{M_Z}\xrightarrow{\langle \mathbf{10_H} \rangle} {\cal G}_{31}  \, ; \label{breakingchain1} \\
 \text{422D}: && \quad \text{SO(10)}|_{M_{U} } \xrightarrow{\langle \mathbf{54_H} \rangle} {\cal G}_{422} \times {\cal D}|_{M_{I} } \xrightarrow{\langle \mathbf{\overline{126}_H}\rangle
} {\cal G}_{321}|_{M_Z}\xrightarrow{\langle \mathbf{10_H} \rangle} {\cal G}_{31}  \, ; \label{breakingchain2} \\
 \text{3221}: && \quad \text{SO(10)}|_{M_{U}} \xrightarrow{\langle \mathbf{45_H} \rangle} \, \, \, {\cal G}_{3221}|_{M_{I}} \, \, \, \xrightarrow{\langle \mathbf{\overline{126}_H} \rangle} {\cal G}_{321}|_{M_Z} \xrightarrow{\langle \mathbf{10_H} \rangle} {\cal G}_{31} \, ; \label{breakingchain3}
\\
 \text{3221D}: && \quad \text{SO(10)}|_{M_{U}} \xrightarrow{\langle \mathbf{210_H} \rangle} {\cal G}_{3221} \times {\cal D}|_{M_{I}} \xrightarrow{\langle \mathbf{\overline{126}_H} \rangle} {\cal G}_{321}|_{M_Z} \xrightarrow{\langle \mathbf{10_H} \rangle} {\cal G}_{31} \, ,
\label{breakingchain4}
\end{eqnarray}
where ${\cal D}$ refers to a left-right discrete symmetry, called ${\cal D}$ parity, transforming spinors of opposite chirality~\cite{Kibble-et-al,Aulakh:2002zr}.  We have used the following abbreviations for the gauge groups
\begin{eqnarray} 
{\cal G}_{422} & \equiv & {\rm SU(4)_C \times SU(2)_L \times  SU(2)_R} \, , \nonumber \\
{\cal G}_{3221}  & \equiv &
{\rm SU(3)_C \times SU(2)_L \times SU(2)_R\times U(1)_{B-L}} \, , 
\end{eqnarray}
with the former being the Pati-Salam (PS) group and the latter being the minimal left-right (LR) gauge group.  

To achieve the desired symmetry breaking in these scenarios, one would necessarily need to introduce scalar multiplets that acquire vevs at the corresponding high scales. In the breaking chains above, the scalar content that acquires vevs at the intermediate scale $M_I$ or at the electroweak scale $M_Z$ consists of, respectively, the $\mathbf{\overline{126}_H}$ and $\mathbf{10_H}$ representations; while at the GUT scale $M_U$, the relevant representations that break the SO(10) symmetry are $\mathbf{\overline{210}_H, \overline{54}_H}$ and $\mathbf{\overline{45}_H}$; the latter will not enter our discussion here.

Despite the large number of scalars, under the extended survival hypothesis, most of them have a mass of the order of $M_U$, and only certain scalar components from the $\mathbf{10_H}$ and $\mathbf{\overline{126}_H}$ representations acquire masses below the GUT scale; they are the only ones to contribute to the running of the various couplings between the two scales $M_I$ and $M_U$. In the different scenarios, these are:
${\bf (1, 2, 2)}$ or ${\bf (1, 2, 2, 0)}$ ($\Phi_{10}$) from $\mathbf{10_H}$ and ${\bf (15, 2, 2)} \oplus {\bf (10, 1, 3)}$ or ${\bf (1, 2, 2, 0)} \oplus {\bf (1, 1, 3, 2)}$ ($\Sigma_{126}\oplus \Delta_R $) from $\mathbf{ \overline{126}_H}$ for the gauge group ${\cal G}_{422}$ or ${\cal G}_{3221}$, correspondingly.

On the other hand, the global ${\rm U(1)_{PQ} }$ symmetry in these chains can be simultaneously broken at a distinct scale by assigning a PQ charge to an SO(10) scalar~\cite{PQ-charge,fermionsrunningSO10}. 
For the Pati-Salam model, one of the options could be that the ${\bf 45_H}$ scalar representation from SO(10) acquires a vev, in addition to the vev of $\mathbf{\overline{126}_H}$ that allows to break the linear combination of ${\rm PQ}$, $B-L$ and $T_{3R}$ at the intermediate scale~\cite{Bajc,Mohapatra-Holman}. This allows for the breaking of the Peccei-Quinn symmetry and the Pati-Salam symmetry with the minimal ingredients from SO(10) and, at the same time,  avoids the unnecessary fine-tuning of introducing an SO(10) singlet~\cite{Meloni}. For the latter, the only price we need to pay is to have an extra scalar field ${\bf (1, 1, 3)}$ or ${\bf (1, 1, 3,0)}$ ($\Delta_{45R}$) from the ${\bf 45_H}$ that contributes to the running of gauge couplings between the intermediate and the GUT scales. 

For the breaking chain involving an intermediate ${\cal D}$ parity, namely 422D or 3221D in our case, similar arguments can be invoked, except that we have to also add the representation  ${\bf (10, 3, 1)}$ or ${\bf (1, 3, 1, -2)}$ ($\Delta_L $) from $\mathbf{ \overline{126}_H}$ representation as well as the ${\bf (1, 3, 1)}$ or ${\bf (1, 3, 1, 0)}$ ($\Delta_{45L}$) from ${\bf 45_H}$ representation in order to preserve the ${\cal D}$ parity. Lastly, for the special case of the 3221 chain in eq.~(\ref{breakingchain3}) and because the ${\bf 45_H}$ is assigned a vev directly at the GUT scale to break down both the SO(10) and the Pati-Salam symmetry to their  ${\cal G}_{3221}$ subgroup, the PQ symmetry is expected to be also broken at the GUT scale together with the SO(10) symmetry, and thus no scalars from the ${\bf 45_H}$ scalar representation survive at the intermediate scale.  Table \ref{tab:survive} gives, for each breaking chain, the surviving scalars at the intermediate scale.

\begin{table}[h!]
\renewcommand{\arraystretch}{1.25}
    \centering
\begin{tabular}{|c|c|}
\hline
   Intermediate symmetry & Scalar Multiplets \\
   \hline
   422  &  $\Phi_{10} \oplus \Sigma_{126} \oplus \Delta_R  \oplus \Delta_{45R}$\\
   \hline
   422D  &  $\Phi_{10} \oplus \Sigma_{126} \oplus \Delta_L \oplus \Delta_R \oplus \Delta_{45L} \oplus \Delta_{45R}$\\
   \hline
   3221  & $\Phi_{10} \oplus \Sigma_{126} \oplus \Delta_R$ \\
   \hline
   3221D  & $\Phi_{10} \oplus \Sigma_{126} \oplus \Delta_L \oplus \Delta_R \oplus \Delta_{45L} \oplus \Delta_{45R}$ \\
   \hline
\end{tabular}
\caption{List of scalar multiplets containing light fields, for each intermediate symmetry. They are the only ones which are not integrated out below the SO(10) symmetry breaking scale mass $M_U$.}
\label{tab:survive}
\end{table}

As a result, in the different breaking patterns, the number of Higgs multiplets that contribute to the renormalization group running of the gauge couplings above the intermediate scale is different. Broadly speaking, the larger the number of Electroweak scalars that contribute to the RGEs is, the faster the couplings will evolve, and the lower the intermediate or the GUT scale will be. As we will shortly see, this will have very important consequences in the scenarios that we are adopting in our analysis, which has an extended Higgs sector at the electroweak scale.

Above the intermediate scale $M_I$, only the two Higgs bi-doublet fields, decomposing from the $\mathbf{10_H}$ and $\mathbf{ \overline{126}_H}$ representations, will couple to the fermions. Starting from eq.~(\ref{eq:LY-minimal}), the Yukawa Lagrangian for fermions at the intermediate scale $M_I$ can be written in each of considered schemes as 
\begin{eqnarray}
-{\cal L}_Y^{422} &=& \bar{F}_L(Y_{10} \Phi_{10}+Y_{126} \Sigma_{126})F_R +  Y_R F_R^T C \overline{\Delta_R} F_R  +{\rm h.c.} \, , \nonumber \\
-{\cal L}_Y^{422D} &=& \bar{F}_L(Y_{10} \Phi_{10}+Y_{126} \Sigma_{126})F_R + Y_L  F_L^T  C \overline{\Delta_L} F_L + Y_R  F_R^T  C \overline{\Delta_R} F_R  +{\rm h.c.} \, , \nonumber \\
-{\cal L}_Y^{3221} &=& \bar{Q}_L(Y_{10,q}\Phi_{10}+Y_{126,q}\Sigma_{126})Q_R + \bar{L}_L(Y_{10,l}\Phi_{10}+Y_{126,l}\Sigma_{126})L_R \nonumber\\
&+&  Y_R L_R ^T  i \sigma_2 \Delta_R L_R  +{\rm h.c.} \, ,  \nonumber \\
-{\cal L}_Y^{3221D} &=& \bar{Q}_L(Y_{10,q}\Phi_{10}+Y_{126,q}\Sigma_{126})Q_R + \bar{L}_L(Y_{10,l}\Phi_{10}+Y_{126,l}\Sigma_{126})L_R \nonumber\\
&+&    Y_L L_L^T i \sigma_2 \Delta_L L_L +  Y_R L_R ^T i \sigma_2 \Delta_R L_R    +{\rm h.c.} \, , 
\label{eq:Yuk-int}
\end{eqnarray}
where $F_{L,R}$ are generic left or right-handed SU(4) fermion fields, $Q,L$ are quark/lepton fields, and $\sigma_2$ one of the Pauli matrices. In both cases, we have assumed that terms like $\bar{F}_L^T \Tilde{\phi} F_R$ with $\phi=\Phi$ or $\Sigma$ and $\Tilde{\phi}=  \sigma_2^T \phi^* \sigma_2$ are forbidden by suitably chosen ${\rm U(1)_Y}$ charges~\cite{Fukuyama:2002vv}.

At the intermediate scale, the corresponding left-right symmetry is broken by the right-handed triplet. As we assume that both the ${\bf 10_H}$ and the $\mathbf{\overline{126}_H}$ representations are complex, their vevs should be aligned in the following way to break the intermediate ${\rm SU(2)_L \times SU(2)_R \times U(1)_{B-L}}$ symmetry down to the electromagnetic ${\rm U(1)_{EM}}$ symmetry
\begin{eqnarray}
\nonumber
    &&\langle \Phi_{10} \rangle = \frac{1}{\sqrt{2}}
        \begin{pmatrix} 
        \kappa_{10}^u e^{i \theta_{10}^u}& 0 \\
        0 & \kappa_{10}^d e^{i \theta_{10}^d} \\
        \end{pmatrix} \, , \quad 
    \langle \Sigma_{126} \rangle = \frac{1}{\sqrt{2}}
        \begin{pmatrix} 
        \kappa_{126}^u e^{i \theta_{126}^u}& 0 \\
        0 & \kappa_{126}^d e^{i \theta_{126}^d} \\
        \end{pmatrix} \, , \quad 
        \\ 
    && \langle \Delta_L \rangle = \frac{1}{\sqrt{2}}
        \begin{pmatrix} 
        0  & 0 \\
        \kappa_L e^{i \theta_L}  & 0\\
        \end{pmatrix} \, , \quad \quad \quad \quad
    \langle \Delta_R \rangle = \frac{1}{\sqrt{2}}
        \begin{pmatrix} 
        0  & 0 \\
        \kappa_R e^{i \theta_R}  & 0\\
        \end{pmatrix} \, ,
\label{eq:vevalignment}
\end{eqnarray}
where for later convenience we use the following notation to denote the ${\bf 10_H}$ and $\mathbf{\overline{126}_H}$  vevs
\begin{eqnarray}
v_a^b = \kappa_a e^{i\theta_a^b} \, (a=10,126; b=u,d;).
\end{eqnarray}
In our vevs assignment, the left-handed triplet should also acquire a tiny but nonzero vev $v_L \sim 0$, while the right handed triplet should have an intermediate-scale vev $v_R \sim M_I$.

Below the intermediate scale, the low-energy models include, besides the triplet field $\Delta_R$ that gives masses to the heavy right-handed neutrino species, four Higgs doublet fields $\phi_{1, \cdots 4}$: the two doublets $\phi_1$ and $\phi_3$ from the bi-doublet $\Phi_{10}$ and which have opposite hypercharge $Y_\phi = \pm 1$ and the doublets $\phi_2$ and $\phi_4$ again with opposite hypercharge from the bi-doublet $\Sigma_{126}$. The fields $\phi_1$ and $\phi_2$ will couple to up-type quarks and the heavy right-handed neutrinos, while the fields $\phi_3$ and $\phi_4$ will couple to down-type quarks and the light leptons. While the triplet fields acquire a very large vev, $ \langle \Delta_R \rangle = v_R \sim {\cal O}(M_I)$,  the bi-doublet fields acquire vevs of the order of the electroweak scale. This should imply the relation $\sum_{i=1}^4 v_i^2 = v_{\rm SM}^2 \simeq (246\;{\rm GeV})^2$ between vevs, when their running is neglected. This ensures that the right-handed gauge bosons are very heavy,  $M_{W_R}, M_{Z_R} \approx g v_{R}$, while the ${\rm SU(2)_L}$ $W$ and $Z$ bosons have weak scale  masses, $M_{W}, M_{Z} \approx g v_{\rm SM}$. 

In fact, one should arrange such that only two linear combinations of the four scalar doublet fields $\phi_1, \cdots \phi_4$ acquire masses of of the order of the electroweak scale, while the masses of the two other field combinations should be close to the very high scale $M_I$. The two fields with weak scale masses will be ultimately identified with the doublets $H_u$ and $H_d$ of the low energy 2HDM that we adopt here.  At the intermediate scale $M_I$, these fields should match the $\Phi_{10}$ and $\Sigma_{126}$ fields, the interactions of which have been given in eqs.~(\ref{eq:Yuk-int}), and will be discussed in details in section 4.

To achieve this peculiar configuration, one has to tune the parameters of the scalar potential of the model and a discussion of this issue, together with the constraints to which these parameters should obey, has been made in e.g. Refs.~\cite{Deshpande:1990ip,Dev:2018foq} and we refer to them for the relevant details. 

Hence, for each decomposing bi-doublet, only one Higgs doublet remains light  and the rest of the scalar multiplets acquire intermediate scale masses. 
In this respect, at low energies, we will have in fact a model with two Higgs doublet fields $H_u$ and $H_d$ that couple separately to isospin $+\frac12$ and $- \frac12$ fermions and acquire vevs $v_u$ and $v_d$
\begin{eqnarray}
\left \langle H_u \right \rangle = \frac{1}{\sqrt{2}} \binom{0}{v_u} \quad , \quad \left \langle H_d \right \rangle = \frac{1}{\sqrt{2}} \binom{0}{v_d} \, ,
\label{vev1}
\end{eqnarray}
to generate masses to the $W$ and $Z$ bosons, thus implying the relation $\sqrt{v_u^2 + v_d^2 }= v_{SM} \simeq 246\;$GeV. We further define the ratio of these two vevs to be $\tan\beta = v_u/v_d$. The most general renormalizable scalar potential of this two Higgs doublet model may be written~\cite{2HDM}
\begin{eqnarray}
V_H &=& m_{11}^2 H_d^\dagger H_d + m_{22}^2 H_u^\dagger H_u -\left(m_{12}^2 H_d^\dagger H_u+ {\rm h.c.}\right) \, \nonumber \\
&& + \lambda_1 \left(H_d^\dagger H_d \right)^2 + \lambda_2 \left(H_u^\dagger H_u \right)^2 + \lambda_3 \left(H_d^\dagger H_d \right) \left(H_u^\dagger H_u \right) +  \lambda_4 \left(H_d^\dagger H_u \right) \left(H_u^\dagger H_d \right) \, \nonumber \\
&& + \left[  \lambda_5 \left(H_d^\dagger H_u \right)^2 + \lambda_6 \left(H_d^\dagger H_d \right) \left(H_d^\dagger H_u \right) + \lambda_7 \left(H_u^\dagger H_u \right) \left(H_d^\dagger H_u \right) +{\rm h.c.} \right] \, .
\end{eqnarray}
We will later discuss in details the above scalar sector, in particular when it comes to the perturbativity of the various couplings and  the stability of the corresponding vacua. These impose severe constraints on the model as we will see. 

The Yukawa interactions of the fermions are those of a Type-II 2HDM \cite{2HDM} with a Lagrangian given by
\begin{eqnarray}
-{\cal L}_{Y}^{\text{2HDM}}= Y_{u}\bar{Q}_L H_u \; u_R +Y_{d}\bar{Q}_L H_d \; d_R + Y_{e}\bar{L}_L {H_d}\; e_R + {\rm h.c.} \, ,
\label{eq:Yuk-2HDM}
\end{eqnarray}
with $Q_L/L_L$ the quark/lepton left-handed doublets and $f_R$ the right-handed singlets. In our discussion, only the third generation fermions will be considered and the small Yukawa couplings of the first two generations will be neglected. The relations between the masses and Yukawa couplings are then simply  given by
\begin{equation}
m_t =\frac{1}{\sqrt{2}}Y_{t} v_{u} \, , \  \ \
m_b =\frac{1}{\sqrt{2}}Y_{b} v_{d} \, , \ \  \
m_\tau =\frac{1}{\sqrt{2}}Y_{\tau} v_{d} \, .
\label{eq:fmasses}
\end{equation}
 
Having introduced these essential elements, we can now discuss the unification of the gauge and Yukawa couplings. 


\section{Gauge couplings unification with thresholds}

\subsection{Approximate solutions of the RGEs}

In this section, we present some analytical expressions for the renormalization group evolution of the three SM gauge couplings, which can be used to derive the unification scale $M_U$ and the universal coupling constant $\alpha_U$ at this scale for any breaking pattern of the non-SUSY SO(10) GUTs with an intermediate scale $M_I$. The RGEs with an energy scale $\mu$ of the couplings $\alpha_i= g_i^2/4\pi$, where $g_i$ are the coupling constants of the SU(3), SU(2) and U(1) groups for respectively $i=3,2,1$, are given by the following differential equations
\begin{eqnarray}
\dd{\alpha_i^{-1}(\mu)}{\ln \mu} = -\frac{a_i}{2\pi} - \sum_{j}\frac{b_{ij}}{8\pi^2\alpha_j^{-1}(\mu)} \, .
\end{eqnarray}
Including the Yukawa interactions, the solutions take the following approximate form in terms of a reference scale $\mu_0$
\begin{eqnarray}
\alpha_{i}^{-1}(\mu) = \alpha_{i}^{-1}(\mu_0) -\frac{a_i}{2\pi}\ln \frac{\mu}{\mu_0} - \frac{1}{4\pi}\sum_{j}\frac{b_{ij}}{a_j}\ln \frac{\alpha_j(\mu)}{\alpha_j(\mu_0)} + \Delta_{Y}^i\, .
\label{eq:RGE2l}
\end{eqnarray}
The one- and two-loop $\beta$ coefficients (as they are usually called; not to be confused with the ratio of vevs $\tan \beta$), $a_i$ and $b_{ij}$, are given explicitly in Appendix \hyperref[subsecA1]{A1} for the symmetry groups and representations that we are considering. $\Delta_Y^i$ stands for the Yukawa couplings contributions that enter at two-loops but, as they only have a very small impact on the running of the gauge couplings compared to the other two-loop contributions, we will neglect them in our computation. The detailed calculation including the two-loop Yukawa contributions to the gauge couplings can be found in Ref~\cite{Langacker:1992rq} for instance.

In addition, at the intermediate symmetry breaking scales, threshold effects \cite{Threshold} due to all the particles that have masses in the vicinity of these scales and,  in particular, all the scalar fields that develop vevs at these scales, will be active. These higher order corrections will modify the matching conditions of the gauge couplings at the symmetry breaking scale, depending on the particle content. For a general symmetry breaking from a group ${\cal G}$ to a subgroup ${\cal H}$ at the scale $\mu$, the matching conditions with the threshold corrections included take the form
\begin{eqnarray}
\alpha_{i,{\cal G}} ^{-1} (\mu )= \alpha_{i,{\cal H}} ^{-1} (\mu ) + \frac{\lambda_{i, {\cal H}} ^{\cal G}}{12 \pi} \, ,
\label{eq:threshold}
\end{eqnarray}
where $\lambda_{i, {\cal H}} ^{\cal G}$ are weighted by the parameters $\eta_i=\ln ({M_{i}}/{\mu})$ with $M_{i}$ being the masses of the heavy particles integrated out at the low energy scale. The complete expressions for the one-loop threshold corrections $\lambda_{i, {\cal H}} ^{\cal G}$ at the relevant scale are given in Ref.~\cite{Meloni} for the models that we are considering here. 

At this stage, combining eq.~(\ref{eq:RGE2l}) and eq.~(\ref{eq:threshold}), the gauge couplings at a low scale $\alpha^{-1}_{i,{\cal H}} (\mu_0)$ can be evolved to an arbitrary high scale $\mu$ where the gauge couplings are embedded into a higher symmetric group ${\cal G}$ as 
\begin{eqnarray}
\alpha_{i,{\cal G}}^{-1}(\mu) = \alpha_{i,{\cal H}}^{-1}(\mu_0) -\frac{a_i^{\cal H}}{2\pi}\ln \frac{\mu}{\mu_0} - \frac{1}{4\pi}\sum_{j}\frac{b_{ij}^{\cal H}}{a_j^{\cal H}}\ln \frac{\alpha_{j,{\cal H}}(\mu)}{\alpha_{j,{\cal H}}(\mu_0)} +  \frac{\lambda_{i, {\cal H}} ^{\cal G}}{12 \pi}\, ,
\label{eq:RGEfull}
\end{eqnarray}
where the two-loop corrections can be approximated with the following relation 
\begin{eqnarray}
-\frac{1}{4\pi}\sum_{j}\frac{b_{ij}^{\cal G}}{a_j^{\cal G}}\ln \frac{\alpha_{j,{\cal G}}(\mu)}{\alpha_{j,{\cal G}}(\mu_0)} \approx -\frac{\alpha_{U}}{8\pi^2}
 \theta_{i}^{\cal G}
\ln \frac{\mu}{\mu_0}  \, ,
\label{eq:approximation}
\end{eqnarray}
where $\alpha_U$ is the universal gauge coupling at the GUT scale and the coefficient
\begin{eqnarray}
\theta_{i}^{\cal G} \equiv \sum_j b_{ij}^{\cal G}\frac{\ln (1+a_{j}^{\cal G}\alpha_U t)}{a_{j}^{\cal G}\alpha_U t}~~~~ {\rm and}~~~ t=\frac{1}{2\pi}\ln \frac{\mu}{\mu_0} 
\end{eqnarray}
are defined with the same way as in Ref.~\cite{Langacker:1992rq}. The exact forms of the coefficients $\theta_{i}^{\cal G}$ are given in Appendix A2 for all the considered  symmetry groups.  

The unification of the gauge couplings at the scale $M_U$ sets the boundary conditions for the RGEs, which are valid for any breaking pattern of SO(10) with an intermediate gauge group ${\cal G}_I$
\begin{eqnarray}
\alpha_U^{-1} = \alpha_{i,{\cal G}_I}^{-1} (M_U) +\frac{\lambda_{i, {\cal G}_I} ^{\rm SO(10)}}{12 \pi} \, ,
\label{eq:boundary}
\end{eqnarray}

At the intermediate scale $M_I$, depending on the symmetry breaking chain, the gauge couplings are related with the ones at low-energy by proper normalization of the generators. As an example, in the particular symmetry breaking chains that we consider, one has
\begin{eqnarray}
422/422D: & &
\alpha_{4,{\cal G}_{422}}^{-1} (M_I) = \alpha_{3,{\cal G}_{321}}^{-1} (M_I) \, , \ 
\alpha_{2_L,{\cal G}_{422}}^{-1} (M_I) = \alpha_{2,{\cal G}_{321}}^{-1} (M_I)  \, , \nonumber \\ 
 && \alpha_{2_R,{\cal G}_{422}}^{-1} (M_I) =\frac53 \alpha_{1,{\cal G}_{321}}^{-1} (M_I)-\frac23 \alpha_{3,{\cal G}_{321}}^{-1} (M_I)  \, ,  
 \label{eq:match-gauge1}
\end{eqnarray} 
in the 422/422D cases and, in the case of the 3221 and 3221D breaking chains, 
\begin{eqnarray}
3221: & &
\alpha_{3,{\cal G}_{3221}}^{-1} (M_I) = \alpha_{3,{\cal G}_{321}}^{-1} (M_I) \, , \ 
\alpha_{2_L,{\cal G}_{3221}}^{-1} (M_I) = \alpha_{2,{\cal G}_{321}}^{-1} (M_I) \, , \nonumber \\ 
&&  \alpha_{B\!-\!L,{\cal G}_{3221}}^{-1} (M_I) = \kappa \alpha_{2_R,{\cal G}_{3221}}^{-1} (M_I) =
 \big( \frac{2\kappa+ 3}{5 \kappa} \big)^{-1} \alpha_{1,{\cal G}_{321}}^{-1} (M_I)  \, ,   ~~\nonumber \\ 
3221D: & &
\alpha_{3,{\cal G}_{3221}}^{-1} (M_I) = \alpha_3^{-1} (M_I) \, , \ 
\alpha_{2_L,{\cal G}_{3221}}^{-1} (M_I) =\alpha_{2_R,{\cal G}_{3221}}^{-1} (M_I) = \alpha_{2,{\cal G}_{321}}^{-1} (M_I) \, , \  \nonumber \\
&& \alpha_{B\!-\!L,{\cal G}_{3221}}^{-1} (M_I) = \frac{5}{2}\alpha_{1,{\cal G}_{321}} ^{-1} (M_I)-\frac{3}{2}\alpha_{2,{\cal G}_{321}}^{-1} (M_I)  \, . 
\label{eq:match-gauge2}
\end{eqnarray} 
In the 422D chain we also require $\alpha_{2_L,{\cal G}_{422}}^{-1}(M_I)=\alpha_{2_R , {\cal G}_{422}}^{-1}(M_I)$ to preserve the ${\cal D}$ parity, and in the 3221 chain, we assume $\alpha_{B\!-\!L,{\cal G}_{3221}}^{-1} (M_I) = \kappa \alpha_{2_R,{\cal G}_{3221}}^{-1} (M_I)$ as we are matching three couplings to four. This normalization factor $\kappa$ of ${\cal O}(1)$ is to be solved for together with the scales $M_I$ and $M_U$. 

For the purposes of achieving unification, it is enough to consider the differences between the various gauge couplings, $\alpha_{i,{\cal G}}^{-1}-\alpha_{j,{\cal G}}^{-1}$, whose running depends only on the parameters
\begin{eqnarray}
\Delta_{ij}^{\cal G}=\frac{a_i^{\cal G}-a_j^{\cal G}}{2\pi}+\frac{\theta_i^{\cal G}-\theta_j^{\cal G}}{8\pi^2} \alpha_U\,.
\label{eq:def-Delta}
\end{eqnarray}
In fact, it turns out that for each intermediate symmetry ${\cal G}_I$, it is enough to consider only one combination of the various $\Delta_{ij}^{\cal G}$, which we will call $C_{{\cal G}_I}$. For the cases ${\cal G}_{I}={\cal G}_{422}$ and ${\cal G}_{I}={\cal G}_{3221}$ they read
\begin{eqnarray}
C_{{\cal G}_{\rm 422}}=\frac{3}{5}\frac{\Delta_{42_{R}}^{{\cal G}_{{\rm 422}}}}{\Delta_{42_{L}}^{{\cal G}_{{\rm 422}}}} \; \textrm{ and} \ \
C_{{\cal G}_{{\rm 3221}}}=\frac{3\Delta_{32_{R}}^{{\cal G}_{{\rm 3221}}}+2\Delta_{3B\!-\!L}^{{\cal G}_{{\rm 3221}}}}{5\Delta_{32_{L}}^{{\cal G}_{{\rm 3221}}}} \, .
\label{eq:def-C}
\end{eqnarray}

With the boundary conditions defined in eq.~(\ref{eq:boundary}) and the matching conditions at $M_I$ for different breaking chains (e.g. eqs.~(\ref{eq:match-gauge1})-(\ref{eq:match-gauge2})), the RGEs of SO(10) in eq.~(\ref{eq:RGEfull}) can be transformed into the following general equations, where the intermediate scale $M_I$, the unification scale $M_U$, and the universal SO(10) coupling $\alpha_U$, are related to the initial conditions of the gauge couplings in the SM
\begin{eqnarray}
\ln \left( \frac{M_I}{M_Z} \right) &=&
 \frac{(\alpha_{1_{\rm EW}}^{-1}-\alpha_{3_{\rm EW}}^{-1}) - C_{{\cal G}_{I}} (\alpha_{2_{\rm EW}}^{-1}-\alpha_{3_{\rm EW}}^{-1})+D_{{\cal G}_I}}{C_{{\cal G}_{I}}\Delta_{32}^{{\cal G}_{321}}-\Delta_{31}^{{\cal G}_{321}}}  \, ,
 \label{eq:mi}
\end{eqnarray}
\begin{eqnarray}
\ln \left( \frac{M_U}{M_I} \right) &=&
-\frac{\alpha_{2_{\rm EW}}^{-1}-\alpha_{3_{\rm EW}}^{-1}}{\Delta_{3_{I}2L_{ I}}^{{\cal G}_{I}}} -\frac{\Delta_{32}^{{\cal G}_{321}}}{\Delta_{3_{I}2L_{ I}}^{{\cal G}_{I}}} 
\ln \left( \frac{M_I}{M_Z} \right)-\frac{D_{{\cal G}_I}'}{\Delta_{ 3_I 2L_I}^{{\cal G}_I}}  \, ,
\label{eq:mu}
\end{eqnarray}
\begin{eqnarray}
\alpha_U^{-1} && \simeq \alpha_{3_{\rm EW}}^{-1} - \frac{1}{C_{{\cal G}_{I}}\Delta_{32}^{{\cal G}_{321}}-\Delta_{31}^{{\cal G}_{321}}} 
\Bigg[ \nonumber \\
&& \!\!\!\!\!\! \left( \frac{a_3^{{\cal G}_{321}}}{2\pi}- \frac{\Delta_{32}^{{\cal G}_{321}}}{\Delta_{3_{I}2L_I}^{{\cal G}_{I}}} \frac{a_{3_I}^{{\cal G}_I}}{2\pi} +{\cal O}\left( \frac{\alpha_U \theta_i^{{\cal G}}}{8\pi^2}\right) \right) (\alpha_{1_{\rm EW}}^{-1}-\alpha_{3_{\rm EW}}^{-1})  \nonumber \\
-&& \!\!\!\!\!\! \left( C_{{\cal G}_I}\frac{a_3^{{\cal G}_{321}}}{2\pi}- \frac{\Delta_{31}^{{\cal G}_{321}}}{\Delta_{3_{I}2L_I}^{{\cal G}_{I}}} \frac{a_{3_I}^{{\cal G}_I}}{2\pi} +{\cal O}\left( \frac{\alpha_U \theta_i^{{\cal G}}}{8\pi^2}\right) \right) (\alpha_{2_{\rm EW}}^{-1}-\alpha_{3_{\rm EW}}^{-1}) \Bigg] \, .
\label{eq:au}
\end{eqnarray}
The four constant terms $C_{{\cal G}_{I}}$, $\Delta_{31}^{{\cal G}_{321}}$, $\Delta_{32}^{{\cal G}_{321}}$ and $\Delta_{3_{I}2L_{I}}^{{\cal G}_{I}}$ (in this last term, $3_I$ and $2L_I$ refer to the corresponding gauge couplings in the intermediate gauge group ${\cal G}_I$,  containing the SM ${\rm SU(3)}_C$ and ${\rm SU(2)}_L$ components;  for instance, if ${\cal G}_I={\cal G}_{422}$, the factor refers to $\Delta_{42_L}^{{\cal G}_{422}}$) are all determined by the $\beta$ coefficients of the low-energy models ${\cal G}_{321}$ and the intermediate-scale model ${\cal G}_I$ from eqs.~(\ref{eq:def-C})-(\ref{eq:def-Delta}). 

We have calculated these factors using the work of Ref.~\cite{two-loop-RGEs} for each breaking chains we consider and we list them in the Appendix \hyperref[subsecA3]{A3}; they can easily be calculated from the quantum number of the light fields in the different breaking chains of SO(10). The shorthand notation ($\alpha^{-1}_{1_{\rm EW}}, \alpha^{-1}_{2_{\rm EW}},\alpha^{-1}_{3_{\rm EW}}$) was used to denote the gauge couplings at the electroweak scale ($\alpha^{-1}_{1,{\cal G}_{321}}(M_Z),\alpha^{-1}_{2,{\cal G}_{321}}(M_Z),\alpha^{-1}_{3,{\cal G}_{321}}(M_Z)$). The factors $D_{{\cal G}_I}$ and $D_{{\cal G}_I}'$ include all  threshold corrections for each breaking chain and are given by
\begin{eqnarray}
 D_{{\cal G}_{422}} &=&
D_{13,{\cal G}_{321}}^{{\cal G}_{422}} +\frac35 D_{2_R 4,{\cal G}_{422}}^{\rm SO(10)}  -C_{{\cal G}_{422}} D_{{\cal G}_{422}}' \, , \nonumber \\  
D_{{\cal G}_{422}}' &=& D_{23,{\cal G}_{321}}^{{\cal G}_{422}}+D_{2_L4,{\cal G}_{422}}^{\rm SO(10)} \, , \nonumber \\
 D_{{\cal G}_{3221}} &=&
D_{13,{\cal G}_{321}}^{{\cal G}_{3221}} +\frac35 D_{2_R3,{\cal G}_{3221}}^{\rm SO(10)} + \frac25 D_{B\!-\!L3,{\cal G}_{3221}}^{\rm SO(10)} -C_{{\cal G}_{3221}} D_{{\cal G}_{3221}}' \,  , \nonumber \\
D_{{\cal G}_{3221}}' &=& D_{23,{\cal G}_{321}}^{{\cal G}_{3221}}+D_{2_L3,{\cal G}_{3221}}^{\rm SO(10)} \, ,
\label{eq:thres}
\end{eqnarray}
where the parameter $D_{ij,{\cal H}}^{\cal G}$ depicts the difference between the threshold corrections of the gauge couplings $\alpha_i^{\cal H}$ and $\alpha_j^{\cal H}$ defined as
\begin{eqnarray}
D_{ij,{\cal H}}^{\cal G} = \frac{1}{12\pi} \left( \lambda_{i,{\cal H}} ^{\cal G} -\lambda_{j,{\cal H}}^{\cal G} \right) \, .
\end{eqnarray}

\subsection{Uncertainties of the calculation at the two-loop order}

The initial conditions on the SM gauge couplings
($\alpha^{-1}_{1_{\rm }}, \alpha^{-1}_{ 2_{\rm }},\alpha^{-1}_{3_{\rm }}$), evaluated in the $\overline{\text{MS}}$ renormalization scheme with two-loop accuracy, are the coupling values at the electroweak scale that we take to be the $Z$ boson mass $M_Z=91.2$ GeV, namely~\cite{PDG},
\begin{eqnarray}
\Big( \alpha^{-1}_{1_{\rm EW}}, \  \alpha^{-1}_{2_{\rm EW}}, \  \alpha^{-1}_{3_{\rm EW}}\Big) = \Big( 59.0272 , \ 29.5879, \ 8.4678 \Big) \, ,
\label{eq:gis}
\end{eqnarray}
where the hypercharge coupling $\alpha_Y$ has been normalized with the usual GUT condition leading to $\alpha_1 / \alpha_Y=5/3 $.  In the equation above, we have neglected for convenience the experimental errors on the inverse couplings constants (as well as the estimated theoretical uncertainties) and kept only the central values. These errors, in particular the one that affects the strong coupling $\alpha_3$ will lead to an uncertainty on the obtained scales $M_U$ and $M_I$ of the order of a few percent at most and will therefore not affect our discussion in a significant way. 

With the above initial conditions, the solutions to eqs.~(\ref{eq:mi})-(\ref{eq:mu}) can be derived order by order. At one-loop order, the two-loop coefficients can be ignored, which is equivalent to setting $\alpha_U$ to zero in eqs.~(\ref{eq:def-C})-(\ref{eq:def-Delta}). Neglecting also the one-loop threshold corrections $D_{{\cal G}_I}$ and $D_{{\cal G}_I}'$, the solutions of eqs.~(\ref{eq:mi})-(\ref{eq:mu}) in this case, denoted as  $\ln\left({M_I}/{M_Z}\right)_1$ and $\ln\left({M_U}/{M_I} \right)_1$, are determined by the one-loop values of the four constants ($C_{{\cal G}_{I}}$, $\Delta_{31}^{{\cal G}_{321}}$, $\Delta_{32}^{{\cal G}_{321}}$ and   $\Delta_{3_{I}2L_{I}}^{{\cal G}_{I}}$) (see Appendix A3 for details). The universal coupling at one-loop order $\alpha_U^{\rm 1-loop}$ can also be obtained in a similar way by substituting back the one-loop values of these four constants in the right-handed side of eq.~(\ref{eq:au}). 

We summarize the results for the three one-loop quantities  $\ln\left({M_I}/{M_Z}\right)_1$, $\ln\left({M_U}/{M_I}\right)_1$ and $\alpha_U^{\rm 1-loop}$ in the first panel of Table \ref{tab:sol} for some considered  breaking chains when the threshold corrections (as well as the Yukawa couplings) are neglected.

At two-loop order, eqs~(\ref{eq:mi})-(\ref{eq:mu}) can be seen as implicit functions of the independent variables $\alpha_U$, $\ln\left({M_I}/{M_Z} \right)$ and $\ln\left({M_U}/{M_I} \right)$. Denoting the right-handed sides of these  equations as $F\left(\alpha_U,\ln\left({M_I}/{M_Z} \right),\ln\left({M_U}/{M_I} \right)\right)$ and $G\left(\alpha_U,\ln\left({M_I}/{M_Z} \right),\ln\left({M_U}/{M_I} \right)\right)$ correspondingly, eqs~(\ref{eq:mi})-(\ref{eq:mu}) can be rewritten as
\begin{eqnarray}
F\left(\alpha_U,\ln\left(\frac{M_I}{M_Z} \right),\ln\left(\frac{M_U}{M_I} \right)\right)-\ln\left(\frac{M_I}{M_Z} \right)&=&0 \, ,  \label{eq:F}
\\  G\left(\alpha_U,\ln\left(\frac{M_I}{M_Z} \right),\ln\left(\frac{M_U}{M_I} \right)\right)-\ln\left(\frac{M_U}{M_I} \right)&=&0 \, . \label{eq:G}
\end{eqnarray}
Because the one-loop solutions $\ln\left({M_I}/{M_Z}\right)_1$ and $\ln\left({M_U}/{M_I}\right)_1$ when $\alpha_U=0$ are exact solutions to the above eqs.~(\ref{eq:F})-(\ref{eq:G}), the small required corrections can be found by performing the following variations to the one-loop solutions
\begin{eqnarray}
\left.\frac{\partial F}{\partial \alpha_U}\right|_{\alpha_U=0}
\hspace*{-5mm} \delta \alpha_U + \left[\left.\frac{\partial F}{\partial \ln \left( \frac{M_I}{M_Z} \right) }\right|_{\alpha_U=0 } \hspace*{-5mm} -1\right]\delta \ln \left( \frac{M_I}{M_Z} \right) + \left.\frac{\partial F}{\partial \ln \left( \frac{M_U}{M_I} \right) }\right|_{\alpha_U=0 } \hspace*{-5mm} \delta \ln \left( \frac{M_U}{M_I} \right) &=& 0 \, , \\
\left.\frac{\partial G}{\partial \alpha_U}\right|_{\alpha_U=0} \hspace*{-5mm} \delta \alpha_U + \left.\frac{\partial G}{\partial \ln \left( \frac{M_I}{M_Z} \right) }\right|_{\alpha_U=0} \hspace*{-5mm} \delta \ln \left( \frac{M_I}{M_Z} \right) + \left[ \left.\frac{\partial G}{\partial \ln \left( \frac{M_U}{M_I} \right) }\right|_{\alpha_U=0 } \hspace*{-5mm} -1\right]\delta \ln \left( \frac{M_U}{M_I} \right) &=& 0 \, .
\end{eqnarray}

A careful investigation of the above differential forms reveal that indeed all the other derivatives vanish when $\alpha_U=0$ except for ${\partial F}/{\partial \alpha_U}$ and ${\partial G}/{\partial \alpha_U}$ due to the fact that the derivatives of the two-loop factor $\Delta_{ij}^{\cal G}$ satisfies the relation ${\partial \Delta_{ij}^{\cal G}}/{\partial t}|_{\alpha_U =0} =0$. Therefore, the two-loop solutions $\ln\left({M_I}/{M_Z}\right)_2$ and $\ln\left({M_U}/{M_I}\right)_2$ can be approximated by
\begin{eqnarray}
\ln\left(\frac{M_I}{M_Z}\right)_2 & = &  \ln\left(\frac{M_I}{M_Z}\right)_1 + \delta \ln\left(\frac{M_I}{M_Z}\right) 
\approx \ln\left(\frac{M_I}{M_Z}\right)_1 + \left.\frac{\partial F}{\partial \alpha_U}\right|_{\alpha_U=0}\delta \alpha_U \, ,
\label{eq:approx2.1} \\
\ln\left(\frac{M_U}{M_I}\right)_2 & = &  \ln\left(\frac{M_U}{M_I}\right)_1 + \delta \ln\left(\frac{M_U}{M_I}\right) 
\approx \ln\left(\frac{M_U}{M_I}\right)_1 + \left.\frac{\partial G}{\partial \alpha_U}\right|_{\alpha_U=0}\delta \alpha_U \, ,
\label{eq:approx2.2}
\end{eqnarray}
where $ \delta \alpha_U= \alpha_U^{\rm 1-loop}$ should be substituted in the above equation. The universal grand unified coupling at the two-loop level $\alpha_U^{\rm 2-loop}$ can also be solved numerically from eq.~(\ref{eq:au}) by substituting the one-loop value of $\ln\left({M_I}/{M_Z}\right)_1$ and $\ln\left({M_U}/{M_I}\right)_1$ into the parameters $C_{{\cal G}_I}$, $\Delta_{31}^{{\cal G}_{321}}$, $\Delta_{32}^{{\cal G}_{321}}$, and $\Delta_{3_{I}2L_{I}}^{{\cal G}_{I}}$\footnote{At the two-loop level, the most important contributions to $\alpha_U$ are due to ($C_{{\cal G}_I}$, $\Delta_{31}^{{\cal G}_{321}}$, $\Delta_{32}^{{\cal G}_{321}}$, and $\Delta_{3_{I}2L_{I}}^{{\cal G}_{I}}$) corrected by the two-loop $\beta$ coefficients (see explicitly Appendix A3), so that one can safely neglect the terms proportional to $({\alpha_U \theta_i ^{\cal G}}/{8\pi^2})$ in eq.~(\ref{eq:au}), which can be considered as higher-order corrections~\cite{Langacker:1992rq}.}. 

\begin{table}[t!]
\renewcommand{\arraystretch}{2}
    \centering
    \begin{tabular}{|c|c||c|c|c||c|c|c|}
    \hline 
    ${\cal G}_{321}$ & $\, {\cal G}_I \, $ & $\log \left( \frac{M_{I1}}{\rm GeV} \right)$ & $\log \left( \frac{M_{U1}}{\rm GeV} \right)$ & $\, \alpha_U^{\rm 1-loop} \, $  & $\log \left( \frac{ M_{I2}}{\rm GeV} \right)$ & $\log \left( \frac{ M_{U2}}{\rm GeV} \right)$ & $\, \alpha_U^{\rm 2-loop}$  \, \\ 
    \hline 
    SM & $\, {\cal G}_{422} \, $ &  ${11.102}$  & ${16.314}$  & 0.0275 & ${9.627}$ & ${16.718}$ & 0.0313 \\ 
    \hline
    SM & $\, {\cal G}_{3221} \, $ &  ${9.807}$  & ${16.165}$  & 0.0223 & ${9.942}$ & ${15.929}$ & 0.0262 \\ 
    \hline
    2HDM & $\, {\cal G}_{422} \, $ &  ${11.429}$  & ${15.988}$  & 0.0273 & ${10.133}$ & ${16.346}$ & 0.0304 \\ 
    \hline
    2HDM & $\, {\cal G}_{3221} \, $ &  ${10.234}$  & ${15.896}$  & 0.0226 & ${10.398}$ & ${15.652}$ & 0.0230 \\ 
    \hline
    \end{tabular}
\vspace*{2mm}
    \caption{A summary table of our approximate analytical estimates of the intermediate scales $M_I$, the unification scales $M_U$, and the values of the universal SO(10) coupling constant $\alpha_U$ for different intermediate breaking groups ${\cal G}_I$ and low-energy models ${\cal G}_{\rm SM}$, where the Yukawa contributions and the threshold corrections are neglected.}
    \label{tab:sol}
\vspace*{-.1mm}
\end{table}

 In summary, neglecting all the threshold corrections, as the coupling constant $\alpha_U$ is rather small, it is a good approximation to expand the coefficients $\Delta_{ij}^{{\cal G}_I}$ in terms of this coupling to find, first the one-loop solutions. The two-loop solutions are then obtained by inserting the one-loop solutions into eqs.~(\ref{eq:approx2.1}) and (\ref{eq:approx2.2}). 
 
 We summarize our results for the one-loop and two-loop predictions for $\ln\left({M_I}/{M_Z}\right)$, $\ln\left({M_U}/{M_I}\right)$, and $\alpha_U$ separately in Table~\ref{tab:sol}. This approximation is in a good agreement with the numerical results to be discussed in subsection 3.4.  One can observe from Table~\ref{tab:sol} that the one-loop solutions agree with the numerical results given in Ref.~\cite{Meloni} at the 4$\sigma$ confidence level, while the two-loop solutions agree with the numerical results in Table.~\ref{tab:scales} from subsection 3.4 at the 2$\sigma$ confidence level. 

In practice, one can solve these equations iteratively, as is done for instance in Ref.~\cite{Langacker:1992rq}, to obtain more accurate predictions of the scales $M_I$ and $M_U$. However, as we are assuming the approximation in eq.~(\ref{eq:approximation}) for a unification of gauge couplings, without integrating out higher derivatives, the approximation by the first derivatives in eqs.~(\ref{eq:approx2.1})-(\ref{eq:approx2.2}) already includes uncertainties of the order of one-percent, which is also comparable with the contributions from the Yukawa couplings that we neglect in our computation. Besides the uncertainties from our approximations and leaving aside the Yukawa contributions, the largest uncertainty actually comes from the threshold corrections $D_{{\cal G}_I}$ and $D'_{{\cal G}_I}$, which are shown in several analyses to be able to modify the predictions of the unification scales by more than an order of magnitude; see for instance Refs.~\cite{Chakrabortty:2019fov,Meloni,Letter}. 

Finally, we should note that in principle, analytical expressions cannot be derived when a  multi-step symmetry breaking with more than one intermediate scale is present, unless additional constraints on the intermediate scale are imposed. Our analytical results generalize the formulae derived in Ref.~\cite{Langacker:1992rq} for SUSY-SO(10) GUTs to the non-SUSY case and to the case with one intermediate symmetry breaking\footnote{This formalism can be generalized to the supersymmetric case discussed for example in Ref.~\cite{Pokorski:2019ete} where a general analytical method is applied for a SUSY SU(5) GUT. For SUSY SO(10) GUTs like the ones discussed in Ref.~\cite{Ellis:2018khn}, one can identify the intermediate scale to be the SUSY-breaking scale and use the formalism presented here to derive the unification scale $M_U$.}.

\subsection{Impact of proton decay}

Before moving to the numerical results,  let us first have a brief discussion\footnote{For a detailed account, see the recent and more  general discussion given in Ref.~\cite{Chakrabortty:2019fov}.}  on the constraints  that come from  proton decay on our SO(10) GUTs with intermediate breaking, and more precisely on the values of the unification scale $M_U$ and unification coupling $\alpha_U$.  The most-constraining decay channel on the proton lifetime is the one in which one has a pion and a positron in the final state \cite{Nath-Babu}. In this particular mode,  the proton lifetime in years can be roughly estimated to be~\cite{Meloni}: 
\begin{eqnarray}
\tau (p \to e^+ \pi ^0 ) \simeq (7.47 \times 10^{35} {\rm yr}) \left( \frac{M_{U}}{10^{16}\; {\rm GeV}} \right)^4 \left( \frac{0.03}{\alpha_{ U}} \right)^2 \, .
\end{eqnarray}  

The strongest current experimental constraint, including other decay channels, for proton decay come from the Super-Kamiokande experiment~\cite{proton-decay-experiment}
which sets the bounds on the proton lifetime
\begin{equation}
\tau (p \to e^+ \pi ^0 ) > 1.67 \times 10^{34} {\rm yr}
\end{equation}
at the 90\% confidence level, which yields the following bound 
\begin{eqnarray}
\ln \left( \frac{M_U}{M_Z}\right) + \frac12 \ln \left( \alpha_U^{-1} \right) > 33.1 \, ,
\label{eq:bounds-protondecay}
\end{eqnarray}
where the unification scale $\ln \left( \frac{M_U}{M_Z}\right)$ and coupling $\alpha_U^{-1}$ can be obtained from eqs.~(\ref{eq:mi})-(\ref{eq:au}) with values that are summarized in Table~\ref{tab:sol} given in the previous subsection. 

In the  general case, the analytical expressions for the $\beta$ coefficients can be found in Ref.~\cite{two-loop-RGEs}, where the dependence on the number of fermion families and Higgs doublets is explicitly given. We can thus express all $\beta$ coefficients as a function of the number of scalars running from the electroweak scale to the intermediate scale. One can generally state that the more colorless scalars contribute to the gauge couplings, the lower the unification scale would be and, thus, the shorter the proton lifetime would be. 

For the low-energy model ${\cal G}_{321}$ studied in our paper, namely the 2HDM, and without including the threshold corrections as is shown for example in Table~\ref{tab:sol}, the only two breaking chains that survive the constraint from proton decay are the 422 and the 3221 breaking chains, with the latter one sitting right on the edge of the proton decay bounds that could be spoiled easily by slightly going beyond our approximation. Including the threshold corrections could raise the unification scale by an order of magnitude to avoid a too fast proton decay. The shift of scales $M_I$ and $M_U$ when including the threshold corrections numerically is discussed in the next subsection.  

\subsection{Numerical results}

In this subsection, we will give more precise results that we obtain numerically by deriving and solving the RGEs for each considered breaking chain up to two-loop order, using the Mathematica package SARAH~\cite{SARAH}. In our present case, from the electroweak scale $M_Z$ to the intermediate scale $M_I$, the low-energy model ${\cal G}_{321}$ is not the SM but is assumed to be the 2HDM  whose two-loop RGEs are also given in Appendix B.  Note also that in our numerical treatment, the contributions of the Yukawa couplings, determined from the fermion masses at the electroweak scale and the parameter $\tan \beta$ of the 2HDM, have been also included.

We also include the one-loop threshold corrections numerically at the scales $M_I$ and $M_U$, by randomly sampling the parameters $\eta_i = \ln(M_i/\mu)$ of eq.~(\ref{eq:threshold}) within the range of values  $\eta_i \in [-1,1]$. The systems of two-loop RGEs would then be solved together with the given one-loop threshold corrections to determine the values of the two scales $M_I$ and $M_U$ for each sampling parameter set, by requiring all the gauge couplings to match at the grand unified scale $M_U$ including the threshold corrections when appropriately adjusting the intermediate scale $M_I$. We took at least 10,000 points for the parameters $\eta_i$ within the selected range of $\eta_i$ values and determined the sets of all scales $(M_I,M_U)$ that allow for gauge coupling unification for each breaking chain. 

\begin{figure}[!t]
\begin{center}
\mbox{
\includegraphics[width=7.75cm,height=8cm,clip]{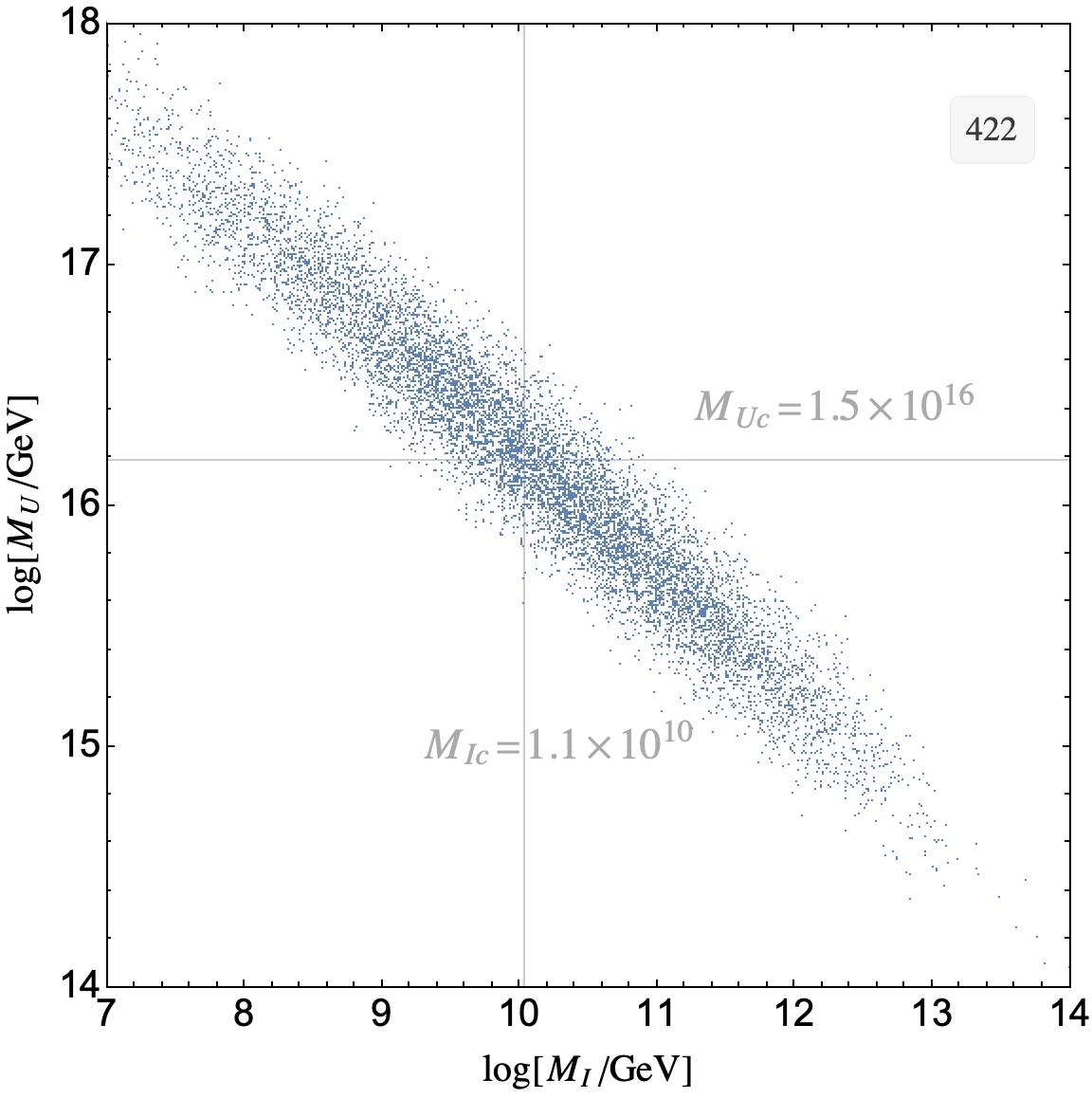}~~ 
\includegraphics[width=7.75cm,height=8cm,clip]{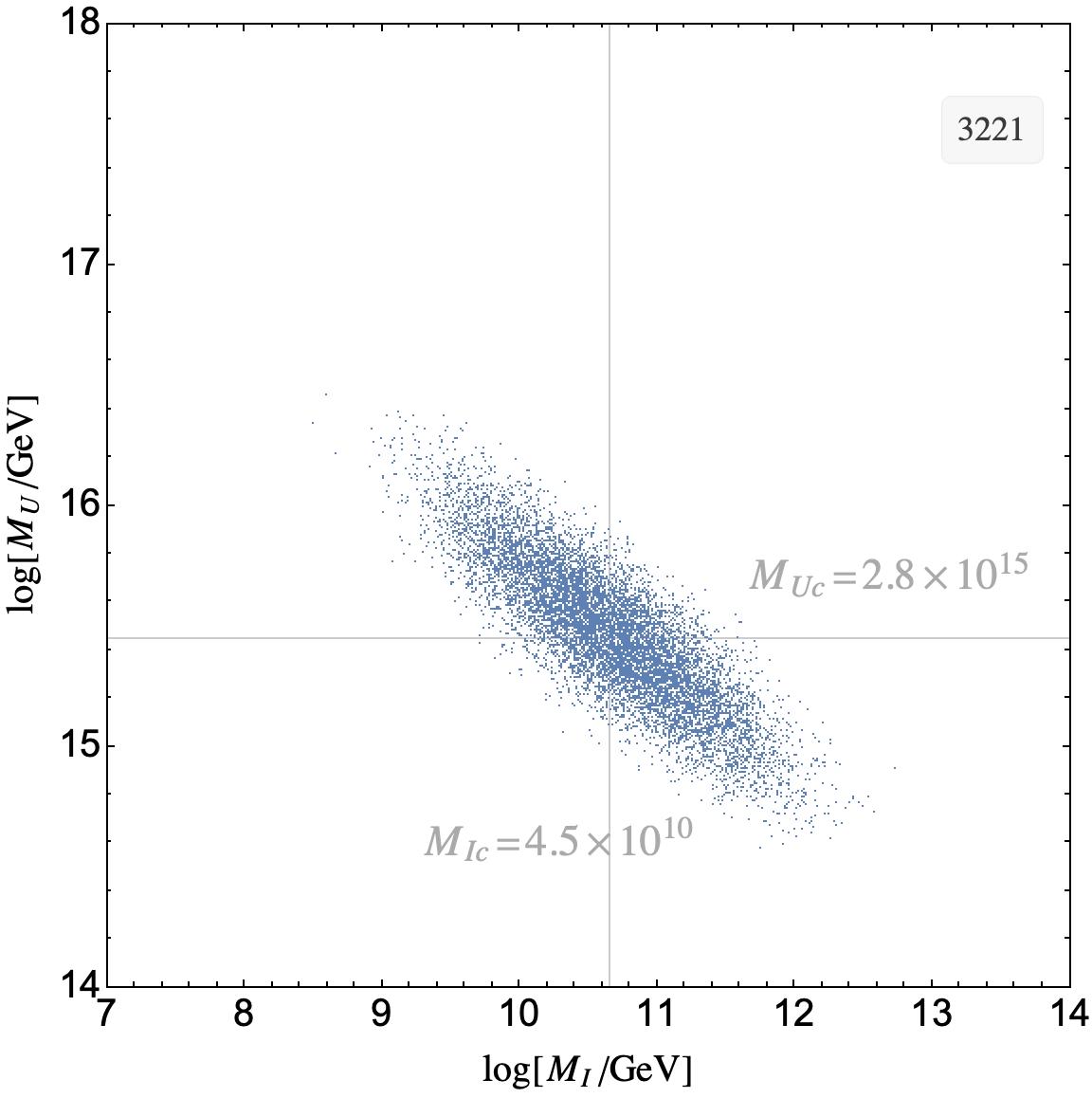} }
\\[5mm]
\mbox{\includegraphics[width=7.75cm,height=8cm,clip]{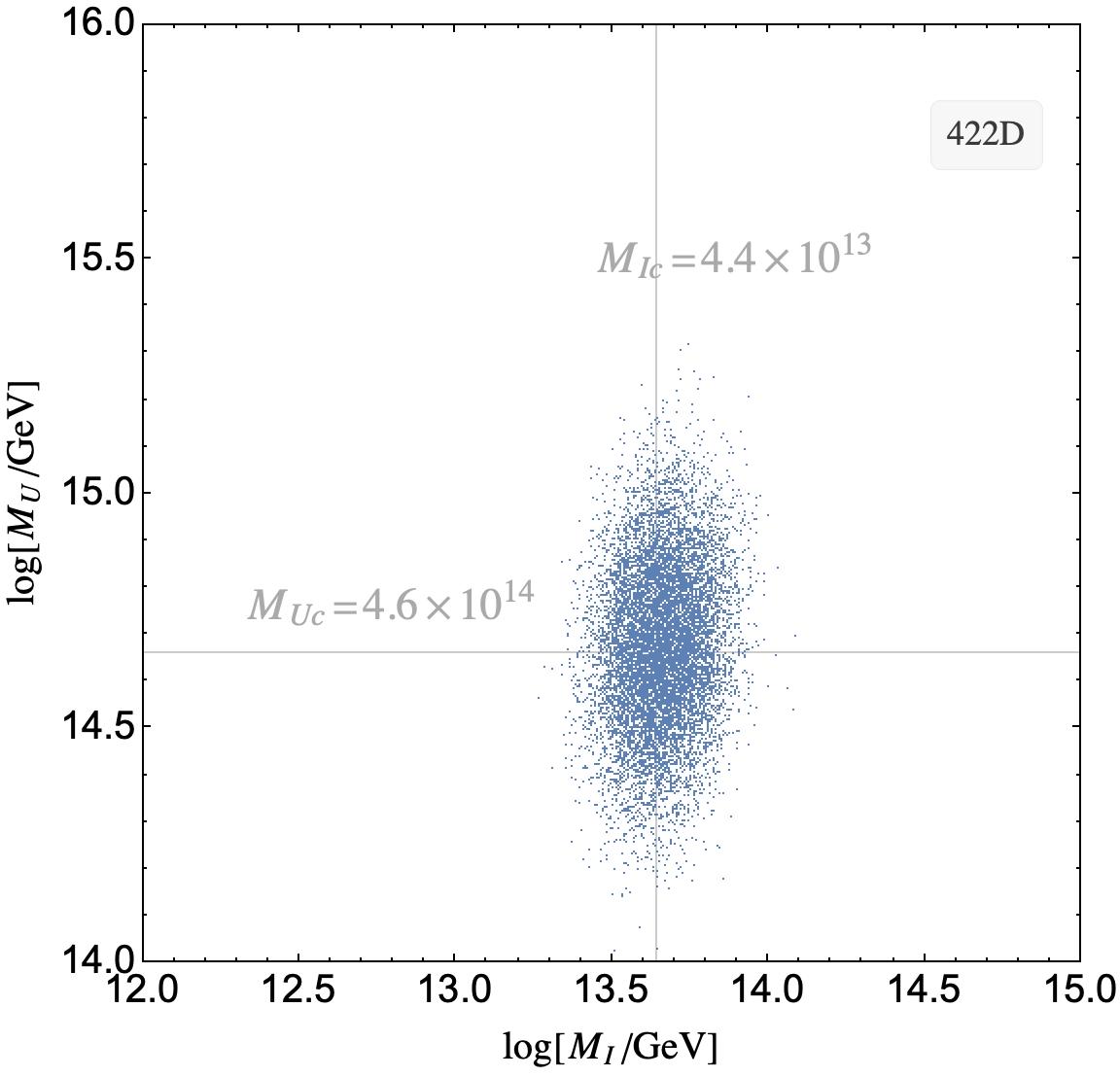}~~ 
\includegraphics[width=7.75cm,height=8cm, clip]{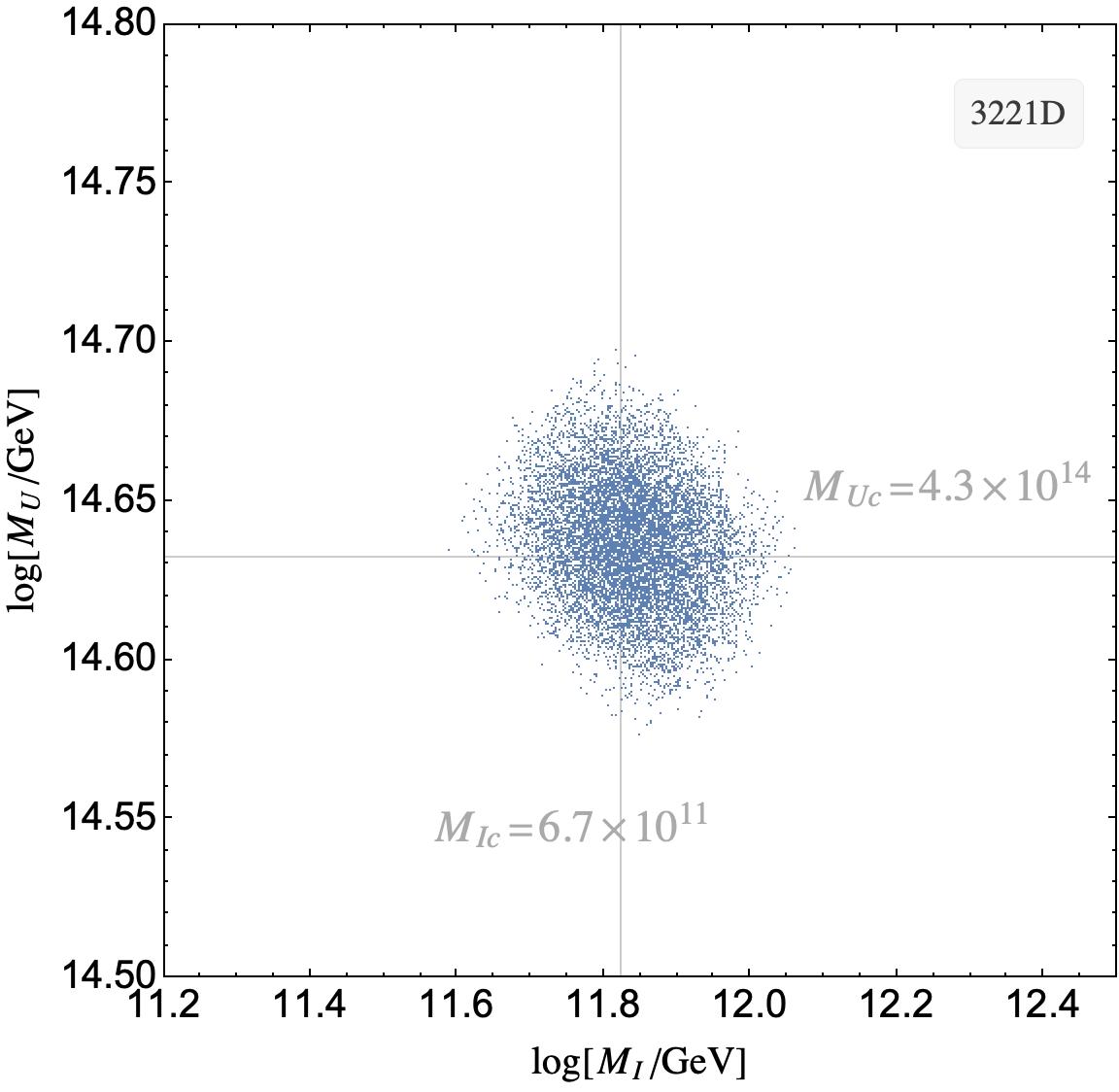}
}
\end{center}
\caption{The scatter plots for the set of (logarithms of the) scales $(M_I,M_U)$ of the four breaking patterns considered, with randomly sampled threshold corrections for $\eta_i \in [-1,1]$ when the ratio of vevs of the two Higgs fields is chosen to be $\tan \beta=65$. Note that these results are not sensitive to $\tan \beta$. The central values $(M_{Ic},M_{Uc})$ that we indicate represent the intermediate and the GUT scales with all threshold corrections taken to be zero, $\eta_i=0$.}
\label{fig:scatter}
\vspace*{-1mm}
\end{figure}

The results are given by the four panels of  Fig.~\ref{fig:scatter} which shows for the four considered breaking patterns, the scatter plots for the set of scales $(M_I,M_U)$ with the randomly sampled threshold corrections, when the ratio of the 2HDM vevs is chosen to be $\tan \beta=65$. The intermediate and the GUT scales when all the threshold corrections are taken to be zero ($\eta_i = 0$) are defined as the central values $(M_{Ic},M_{Uc})$ that are  specified in each plot. As we have already noticed in Ref.~\cite{Letter}, both the two-loop corrections and the threshold corrections have a significant impact. 

In particular, our results in Fig.~\ref{fig:scatter} show the effect of the additional Higgs doublet contributions to the gauge coupling running for our considered breaking chains (again the $\beta$ coefficients are given in Appendix \hyperref[subsecA1]{A1}), when comparing for instance to the work of Ref.~\cite{Meloni} (especially to their Figure 3), where only the SM particle content is used in the running at low energy. The extra  contributions in the 2HDM to the running of the gauge couplings, even though not very large, results in a unification scale $M_U$ that is significantly smaller.  In fact, for some of the breaking scenarios that we consider, in particular the 422D and 3221D chains, the resulting $M_U$ values could easily fall into the values excluded by proton-decay bounds even when large threshold corrections are included.

\begin{figure}[!t]
\begin{center}
\includegraphics[width=7.75cm,height=8cm,clip]{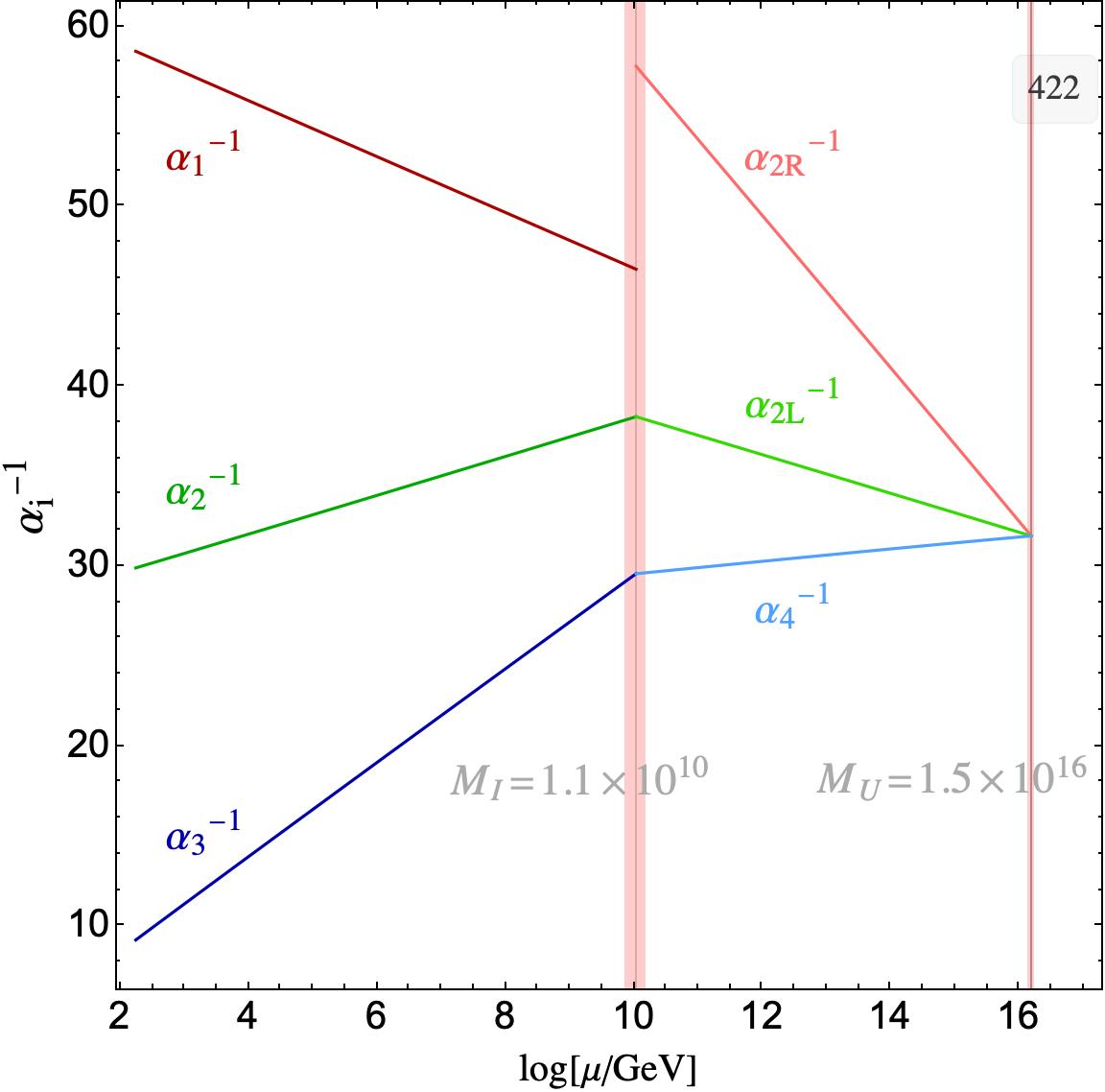}~~
\includegraphics[width=7.75cm,height=8cm,clip]{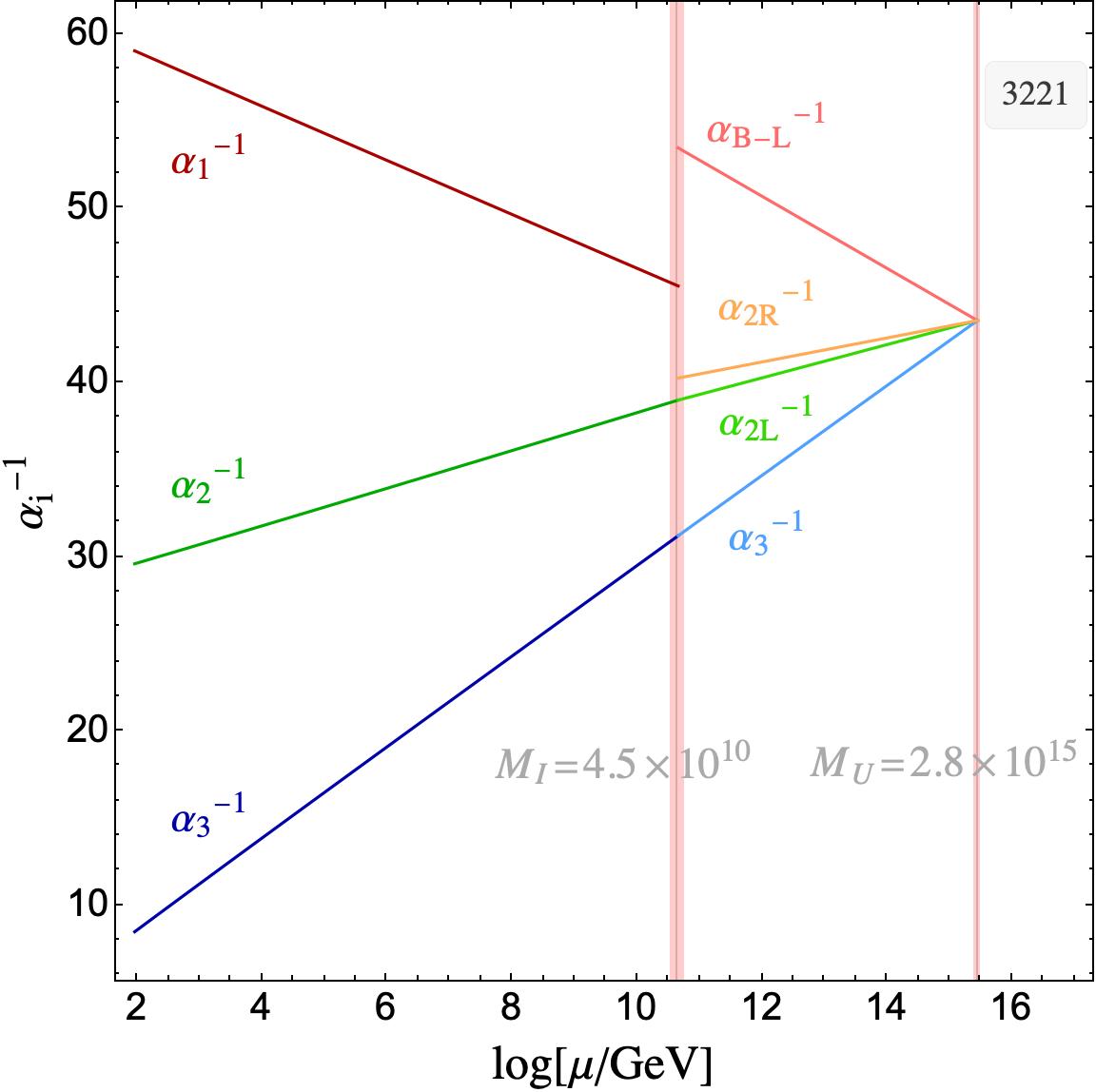} \\[5mm]
\includegraphics[width=7.75cm,height=8cm,clip]{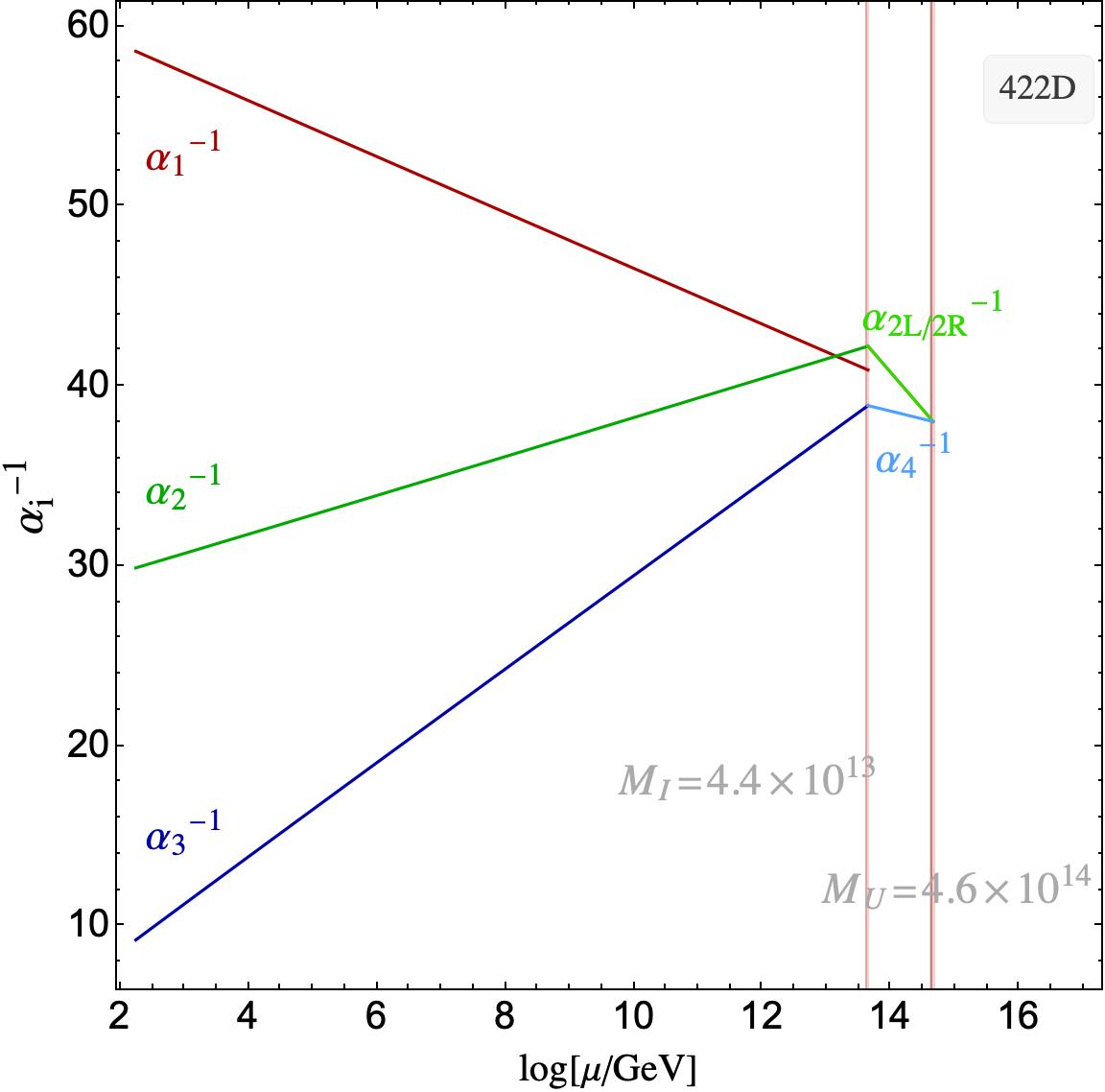}~~
\includegraphics[width=7.75cm,height=8cm,,clip]{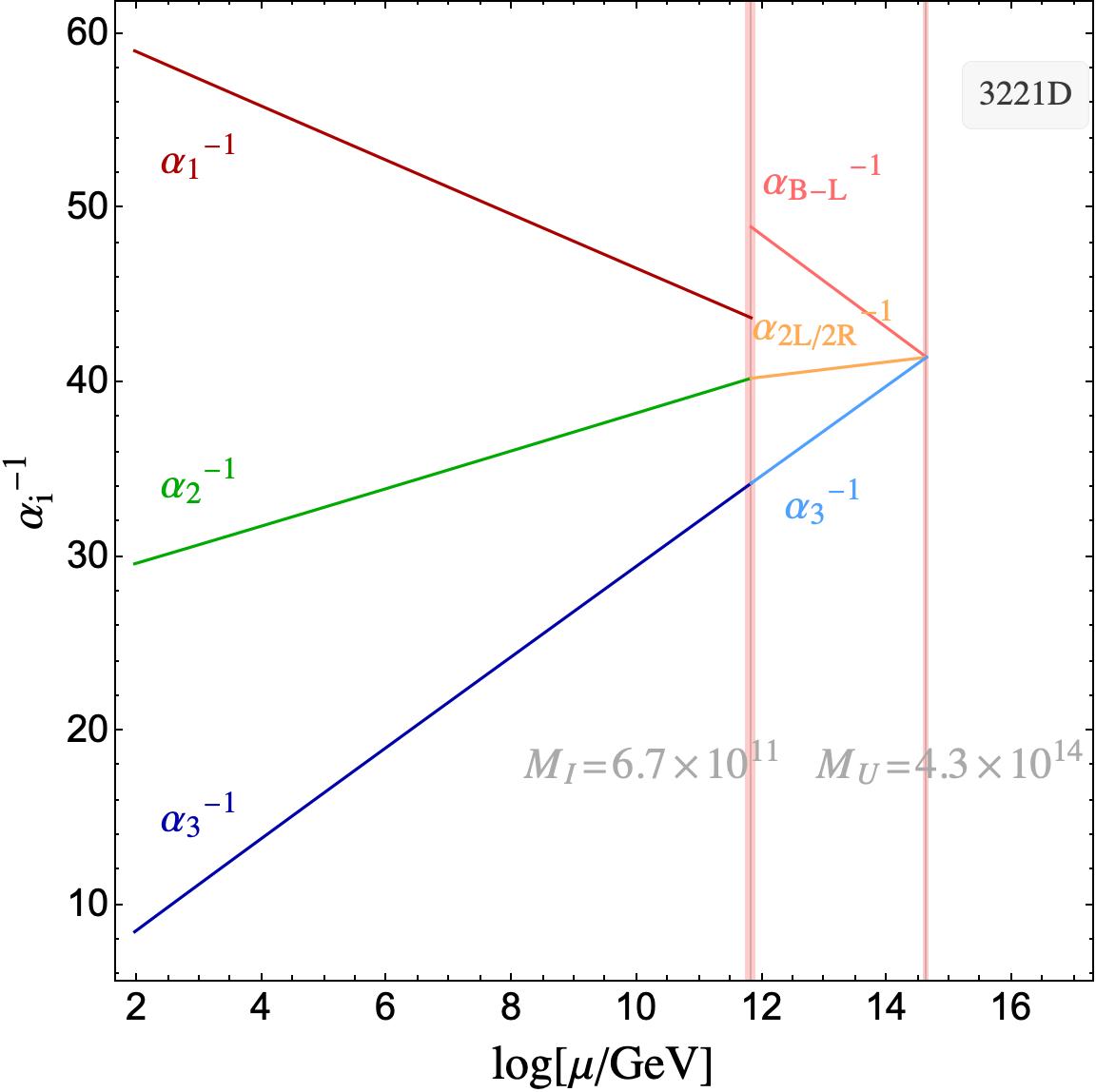}
\end{center}
\vspace*{-2mm}
\caption{The evolution of the inverse of the gauge coupling constants squared $\alpha_i^{-1}$ as a function of the (logarithm of the) energy scale $\mu$ for the value $\tan \beta=65$, when all threshold corrections $\eta_i$ are taken to be zero from the electroweak scale to the GUT scale in the 2HDM+422 (upper left), 2HDM+3221 (upper right), 2HDM+422D (bottom left) and 2HDM+3221D (bottom right) models. The red vertical bands reveal the uncertainty on the measurement of gauge couplings at the electroweak scale.}
\label{fig:gauge-unif}
\vspace*{-2mm}
\end{figure}

As an example, the evolution of the inverse of the gauge coupling constants squared $\alpha_i^{-1}$ for the selected 2HDM ratio of vevs $\tan \beta=65$ when all threshold corrections $\eta_i$ are taken to be zero from the scale $M_U$ down to the scale $M_I$ and then down to the weak scale $M_Z$ is shown in Fig.~\ref{fig:gauge-unif} as a function of the (logarithm of the) energy scale $\mu$. We have used the program SARAH in which we have implemented the full two-loop RGEs for the considered breaking patterns 422 (upper left), 422D (bottom left), 3221 (upper right) and 3221D (bottom right).  While the three couplings are clearly different at the scale $M_I$, of the order of a few times $10^{10-13}$ GeV, the slope are significantly modified at this energy by the additional contributions so that the couplings meet at a scale $M_U$ of the order $10^{14-16}$ GeV.  The small impact of the experimental errors on the couplings is illustrated by the narrow vertical red bands that are drawn at the scales $M_I$ and $M_U$.

Finally, we also summarize in Table~\ref{tab:scales} our numerical results for our four considered breaking patterns, when the threshold corrections are not included. The relevant intermediate and unification scales at the two-loop level $M_{Ic}$ and $M_{Uc}$ as well as the unification coupling $\alpha_{U}$, are to be compared with those given in Tab.~\ref{tab:sol}; in addition, we display the estimated  proton lifetime in each scenario.

\begin{table}[h!]
\renewcommand{\arraystretch}{1.8}
    \centering
    \begin{tabular}{|c|c|c|c|c|}
    \hline
    Breaking chain  & $\log \left( \frac{M_{Ic} } {\rm GeV} \right)^{\rm 2-loop}$ & $\log \left( \frac{ M_{Uc} } {\rm GeV} \right)^{\rm 2-loop} $ & $\alpha_{U}^{\rm 2-loop}$ & $\tau (p \to e^+ \pi ^0 )/{\rm yr}$ \\ \hline
    422  & 10.03 & 16.19 & 0.032 & $3.82\times 10^{36}$  \\ \hline 
    3221  & 10.66 & 15.45  & 0.023  & $7.84\times 10^{33}$  \\ \hline 
    422D  & 13.65 & 14.66  & 0.026 & $4.22\times 10^{30}$ \\ \hline
    3221D  & 11.82 & 14.63 & 0.024  & $3.89\times 10^{30} $ \\ \hline 
    \end{tabular}
    \caption{A summary table of the numerical results of the intermediate scale, the unification scale, and the universal gauge coupling at the two-loop level, neglecting all the threshold corrections as well as the estimated proton lifetimes obtained for each considered breaking chain  with two Higgs doublets at the electroweak scale. The ratio of vevs is fixed to $\tan\beta=65$ as the results do not change significantly for lower values of $\tan\beta$.}
    \label{tab:scales}
\vspace*{-2mm}
\end{table}

From this table, one can see that when the threshold corrections are switched off, only the breaking chain 422 with the Pati-Salam symmetry as an intermediate step and a 2HDM at the low energy scale, survives the proton decay bound from Kamiokande, namely $\tau (p \to e^+ \pi ^0 ) > 1.67 \times 10^{34} {\rm yr}$. In addition, even though the 3221 chain seems to lie at the edge of the dangerous region excluded by proton decay, any small amount of threshold corrections at a given  symmetry breaking scale could easily rescue it, by raising the unification scale by an order of magnitude, as can be seen from Fig.~\ref{fig:scatter}. The same situation occurs in the 422D breaking chain, but large threshold corrections ($\eta_i \simeq 1$) would be needed to prevent fast decay of proton in this case. Finally, for the 3221D breaking chain resulting to a 2HDM at the low energy scale, we find that the bound from proton decay is violated unless extremely large (an potentially unrealistic) threshold corrections ($\eta_i \gg 1$) are taken into account.

Before we close this section, let us make a brief comment on the fact that gauge coupling unification in non-SUSY SO(10) models with only one intermediate scale suffers from the severe constraints from proton decay, if no large threshold corrections are imposed, except for the 422 breaking chain. As we have seen above, the two-loop RGEs of these SO(10) models have approximate analytical solutions which are completely determined by the $\beta$ coefficients for any breaking chain. In other words, there are no free parameters in determining the symmetry breaking scales except for the threshold corrections (e.g. in the $D$ factors given in  eqs.~(\ref{eq:thres})) and the value of the parameter $\tan \beta$ of the low energy 2HDM. The latter parameter generally affects only marginally gauge coupling  unification, but it will be strongly constrained when the Yukawa interactions of the fermions are included as we will see in the next section. This will be particularly the case when one invokes the requirement of the perturbativity of the Yukawa couplings (the absence of Landau poles) and by the consistency of the values for the fermion masses that can be obtained at the intermediate scale. 

Thus, the surviving parameter space for non-SUSY minimal SO(10) models with an intermediate scale is rather small, thus rendering the model quite predictive. We move now to the unification of third generation Yukawa couplings. In this case, we will ignore the models 422D and 3221D with a PQ symmetry as they lead to a low unification scale and, hence, are in conflict with the limits from proton decay.

\section{Yukawa coupling unification}

\subsection{Yukawa unification in non-SUSY SO(10)}

Following the paradigm of the unification of the gauge coupling constants, one is tempted to push the idea further and to consider also the possibility of unifying the fermionic Yukawa couplings in the framework of the same GUT symmetry group. In this context, one is forced to ignore the rather small Yukawa couplings of the first- and second-generation fermions as the masses of these particles are below the few GeV scale which allows them to be realistically described without being affected by the strong interaction uncertainties that are encountered at the corresponding mass scale. In our work, we will thus consider only the Yukawa couplings of third-generation fermions, the top quark, the bottom quark and the tau lepton, with the additional simplification of neglecting all possible mixings. The three fermions will be assumed to have a common Yukawa coupling at the GUT scale $M_U$ within the natural context of SO(10) unification where the fermions are embedded into a single irreducible representation $16_F$ of the symmetry group.

Below the intermediate scale $M_I$ and down to the electroweak scale, the Yukawa interactions of these fermions are those of a Type-II 2HDM with a Lagrangian given by eq.~(\ref{eq:Yuk-2HDM}), which leads to the masses given in eq.~(\ref{eq:fmasses}) in terms of the two vevs $v_u$ and $v_d$ defined at the electroweak scale. The choice of the 2HDM as the low-energy scale directly follows from the requirement that the top and bottom Yukawa couplings should be comparable and this cannot be achieved in the context of the SM with its single Higgs doublet field. In turn, in extended Higgs sectors, the large ratio  between the top and bottom quark masses could be due to a large ratio of the vevs of the  Higgs multiplets that give rise to the masses of the up- and down-type fermions. The simplest of such an extension\footnote{The idea of Yukawa coupling unification emerged and was developed in the late 1980s  in the context of supersymmetric theories, to predict the mass of the not yet discovered top quark and to understand the origin of the top-bottom mass hierarchy; see e.g. Ref.~\cite{Rattazzi:1994bm}. For consistency reasons,  the minimal supersymmetric standard model or MSSM required two Higgs doublets fields of Type-II.} is a 2HDM of Type II. More specifically, one would have for the parameter $\tan\beta$ which is defined as the ratio of the two vevs $v_u$ and $v_d$ of the fields $H_u$ and $H_d$ that break the electroweak symmetry
\begin{eqnarray}
\tan \beta = {v_u}/{v_d} \sim {m_t}/{m_b} \approx {\cal O}(60) \, .
\end{eqnarray}

In the context of the SO(10) unification group that we are considering here, with either a ${\bf 10_{H}}$ or a ${\bf \overline{126} _{H} }$ scalar representation coupling to fermions,  the third generation masses must depend on a single parameter for consistency reasons. However, instead of just one SO(10) scalar representation, we consider the possibility that both a complex ${\bf 10_{H}}$ and a ${\bf \overline{126}_{H}}$ scalar interact with fermions; see eq.~(\ref{eq:LY-minimal}). Fermion masses can therefore receive non-negligible contributions from two Yukawa couplings. Thus, a discussion on Yukawa unification implies that $Y_{10}$ and $Y_{126}$ are somehow related, which in turn implies some connection between the two scalars in our model. One tantalizing possibility is that both the ${\bf 10_{H}}$ and the ${\bf \overline{126}_{H}}$ are part of a single irreducible representation of an even larger gauge group. A natural candidate is the exceptional group ${\rm E_6}$ for the following reasons:

\begin{itemize}
\item The smallest non-trivial representation of ${\rm E_6}$, the one of dimension $\boldsymbol{27}$, decomposes as $\boldsymbol{16}+\boldsymbol{10}+\boldsymbol{1}$ and, therefore, contains the SM fermions plus vector-like ones.

\item A scalar representation ${\bf 351^{\prime}_{H}}$ can couple to the bilinear product of fermions in the representation $\boldsymbol{27}\times\boldsymbol{27}$ and, furthermore, it decomposes as ${\bf 10_{H}}+{\bf \overline{126}_{H}}+\cdots$ under the SO(10) group. Note that $\boldsymbol{351^{\prime}}$ is a complex representation and, therefore, ${\bf 10_{H}}$ must be associated to a complex field.

\item The ${\rm E_6}$-symmetric Yukawa interaction $Y\times{\bf 27_{F}}\cdot\boldsymbol{{\bf 27_{F}}}\cdot\boldsymbol{351_{H}^{\prime}}$ can be written as a sum of terms which,  individually, are symmetric only under SO(10): $c_{10}Y\times{\bf 16_{F}}\cdot{\bf 16_{F}} \cdot{\bf 10_{H}}+c_{126}Y\times{\bf 16_{F}}\cdot{\bf 16_{F}}\cdot{\bf \overline{126}_{H}} +\cdots$ for specific Clebsch-Gordon factors $c_{10},c_{126},\cdots$. These last numbers are therefore a prediction of an ${\rm E_6}$-symmetric theory, hence the enlarged symmetry enforces a particular ratio $Y_{10}/Y_{126}$ since
\begin{equation}
\frac{Y_{10}}{Y_{126}}=\frac{c_{10}Y}{c_{126}Y}=\frac{c_{10}}{c_{126}}\,.
\label{eq:yukawa-ratio}
\end{equation}

\item An exactly ${\rm E_6}$-symmetric theory does not involve the coupling ${\bf 16_{F}}\cdot {\bf 16_{F}}\cdot\boldsymbol{10_{H}^{*}}$ and, hence, there is not such an interaction  at leading order. Its absence can be understood by the fact that ${\rm E_6}$ contains an extra U(1) subgroup which commutes with SO(10), under which the fields are changed precisely in the manner described in eq.~\eqref{eq:U1-charges}.
\end{itemize}

The crucial ratio of Yukawa couplings discussed above turns out to be
\begin{equation}
\left(\frac{Y_{10}}{Y_{126}}\right)_{\textrm{E6}}=\sqrt{\frac{3}{5}}\,,\label{eq:Yukawa-ratio-E6}
\end{equation}
with the understanding that at the GUT scale, the SO(10) contractions ${\bf 16_{F}}\cdot{\bf 16_{F}}\cdot\boldsymbol{10_{H}}$ and ${\bf 16_{F}}\cdot{\bf 16_{F}}\cdot{\bf \overline{126}_{H}}$ normalized in such a way that the two SM doublets (one in $\boldsymbol{10_{H}}$ and the other in ${\bf \overline{126}_{H}}$) contribute to the top quark mass with the same Clebsch-Gordon factor. If we were to write the ${\cal G}_{321}$-invariant Yukawa interactions involving the four Higgs doublets $H_{u/d,10/126}$ contained in ${\bf 10_{H}}$ and ${\bf \overline{126}_{H}}$, they would have the form
\begin{eqnarray}
&& \bar{Q}_{L}\left(Y_{u}^{10}H_{u,10}^{*}+Y_{u}^{126}H_{u,126}^{*}\right)u_{R}+\bar{Q}_{L}\left(Y_{d}^{10}H_{d,10}+Y_{d}^{126}H_{d,126}\right)d_{R} \nonumber \\
&& +\bar{L}_{L}\left(Y_{e}^{10}H_{d,10}+Y_{e}^{126}H_{d,126}\right)e_{R}+{\rm h.c.}\,.
\label{eq:G321-yukawas}
\end{eqnarray}
At the unification scale, the matching relations are as follows:
\begin{equation}
Y_{u}^{10}=Y_{d}^{10}=Y_{e}^{10}=Y_{10};Y_{u}^{126}=Y_{d}^{126}=-\frac{1}{3}Y_{e}^{126}=Y_{126}\,.
\label{eq:G321-yukawas-matching}
\end{equation}
Combining these two expressions, we obtain the fermion mass formulas
\begin{equation}
m_{t}\!=\!{v_{10}^{u}Y_{10}\!+\!v_{126}^{u}Y_{126}},\,m_{b}\!=\!{v_{10}^{d}Y_{10}\!+\!v_{126}^{d}Y_{126}},\,m_{\tau}\!=\!{v_{10}^{d}Y_{10}\!-\!3v_{126}^{d}Y_{126}}\,.\label{eq:fmasses2}
\end{equation}
In addition, we have the Dirac neutrino mass which is given by
\begin{eqnarray}
m_{\nu_D}\! = \! v_{10}^u Y_{10} \!-\!3 v_{126}^u Y_{126}\, .
\end{eqnarray}
Note however that we do not consider a direct breaking of the SO(10) symmetry to the SM group ${\cal G}_{321}$; the purpose of the previous equations is simply to clarify the normalization of the SO(10)-invariant Yukawa couplings that we are considering in eq. (\ref{eq:LY-minimal}). (Furthermore, we consider only two light Higgs doublets, which are necessarily a combination of the four doublets in eq.~(\ref{eq:G321-yukawas}).)

The number indicated in eq.~\eqref{eq:Yukawa-ratio-E6} is quite peculiar since the ratio of Clebsch-Gordon factors is often a rational number (Ref.~\cite{Antusch:2009gu} contains a large list of examples, none of which involves an irrational number). We also cannot avoid commenting on the fact that $\sqrt{3/5}$ is also used to canonically normalize the SM hypercharge; nevertheless, as far as we can tell, this equality is just a coincidence.

The ratio of eq.~\eqref{eq:Yukawa-ratio-E6} was derived with the \texttt{SubgroupCoefficients} function of \texttt{GroupMath} \cite{Fonseca:2020vke} but it can also readily be derived from the available literature. Note in particular that eqs.~(77) and (78) and Table 6 of Ref.~\cite{Babu:2015psa} directly imply that $Y\times{\bf 27_{F}}\cdot\boldsymbol{{\bf 27_{F}}}\cdot\left\langle \boldsymbol{351_{H}^{\prime}}\right\rangle $ contains the terms $\left[1/\left(2\sqrt{10}\right)Yv_{10}^{u}-1/\left(2\sqrt{6}\right)Yv_{126}^{u}\right]tt^{c}$\footnote{The individual values of the two Clebsch-Gordon factors, including their relative sign, are convention-depend and therefore unphysical; the absolute value of their ratio is not.}. 

In the following, we will consider the consequences of the above relation. However, it is beyond the scope of the present work to present a fully realistic ${\rm E_6}$ model for Yukawa unification as this would entail several challenges\footnote{For example, a realistic scalar sector would necessarily have a large number of scalars transforming as ${\bf (2,\pm1/2)}$ or ${\bf (1,0)}$ under the electroweak group, whose vevs affect fermion masses. Providing masses to all the scalars is another challenge. Also, on top of the ${\bf 16_{F}}$ (three generations of it), one would have vector-like fermions transforming as ${\bf 10_{F}}$ and ${\bf 1_{F}}$ which can mix with the spinor representation, complicating the identification of what are the SM fermions. We thank Vasja Susi\v{c} for his helpful comments on this and other ${\rm E_6}$-related topics.}. 

Let us also mention that in our earlier work \cite{Letter}, we have considered some of the implications of requiring the simple relation $Y_{10}=Y_{126}$ at the scale where the gauge couplings unify. While we do not have a mechanism that would prescribe this relation, we will nevertheless consider here in more detail some of its consequences. It is worth keeping in mind that $\sqrt{3/5}\approx0.77$ is not far off from 1, hence the two Yukawa unification conditions, $Y_{10}=CY_{126}$ with $C=\sqrt{3/5}$ or 1, should not lead to dramatically different results.

\subsection{Matching conditions in the 422 breaking chain}

Having introduced the Yukawa unification conditions in our SO(10) model from a top-down perspective, we then seek the relations of the low-energy Yukawa couplings in different breaking chains of SO(10). We first note that the field content needed to enforce the $\mathcal{D}$ parity symmetry yields a unification scale that is unacceptably low, making the proton lifetime too short in  the 422D and the 3221D breaking chains, as can be seen from Table~\ref{tab:scales} and the relevant discussion in the subsection 3.4. We thus ignore these two possibilities in our next discussion. In fact, we will also not discuss the 3221 breaking chain of this particular SO(10) model; some of the elements have been presented in Ref.~\cite{Letter} and others will be postponed to a forthcoming paper. Thus, for illustration, we will discuss in the following only the evolution of the Yukawa couplings in the 422 breaking chain. 

One should first recall that enforcing Yukawa unification can be seen as finding a solution for the system of RGEs of Yukawa couplings satisfying the boundary conditions and the initial conditions obtained from the experimental observables. However, even in a model as simple as a 2HDM, the general RGEs of the Yukawa couplings, which can be read from Appendix B, do not admit an approximate solution like the ones for the gauge couplings discussed in section 3. Therefore, we will leave the discussion of the numerical evaluation of all the Yukawa couplings between different energy scales to the end of the present section, and we first concentrate here on the boundary conditions.

The boundary conditions for Yukawa couplings at the GUT scale $M_U$, where the full SO(10) is restored, relate the couplings in a very specific way which is dictated by appropriate Clebsch–Gordon coefficients originated from the decomposition of the tensor product for the fermion bilinear and the scalar field representation. Below the SO(10) scale, given that the gauge symmetry group is smaller and less constraining, there can be more than two Yukawa couplings, as shown in eqs.~(\ref{eq:Yuk-int}). For the 422 breaking chain, in which the Yukawa couplings can be identified as $Y_{10}^{{422}}$, $Y_{126}^{{422}}$ and $Y_{R}^{{422}}$, the matching conditions can be read from Ref.~\cite{Aulakh:2002zr} which gives:
\begin{eqnarray}
Y_{10}(M_U)=\frac{1} { \sqrt 2 }  Y_{10}^{{422}} (M_U) \, , \, \, Y_{126}(M_U)=\frac{1} { 4\sqrt 2 }  Y_{126}^{{422}} (M_U) =\frac{1} { 4}  Y_{R}^{{422}} (M_U)  \,  .
\label{eq:matching-Y_MU}
\end{eqnarray}
Note that the numerical factors shown here are not intrinsically physical since they depend on how one contracts the ${\rm SU(4)_C \times SU(2)_L \times  SU(2)_R}$ group indices which, incidentally,  are not shown in eqs.~(\ref{eq:Yuk-int}). Obviously, whatever convention is adopted, it must be followed consistently. In the present case, this means that the factors of $\sqrt{2}$ and $4$ shown above must drop out when matching the 321 and 422 Yukawa couplings at $M_I$.

As was mentioned in 4.1, the Yukawa unification in non-SUSY SO(10) can be defined as $Y_{10}=C Y_{126}$, with $C$ the ratio of CG coefficients decomposing the scalar representation of higher symmetry into SO(10) multiplets ${\bf 10_H}$ and ${\bf \overline{126}_H}$\footnote{The special case where $C=1$ has been studied in~\cite{Letter} for a simplified SO(10) model with a real ${\bf 10_H}$ representation without implying any further unification of the scalar representation.}. Motivated by ${\rm E_6}$ in eq.~(\ref{eq:Yukawa-ratio-E6}), we take this factor to be $\sqrt{3/5}$, which, after combining the GUT-scale matching condition in eqs.~(\ref{eq:matching-Y_MU}), implies that the 422-Yukawa couplings at $M_U$ must fell on the line $\ell(M_U)$ in the two-dimensional parameter space ($Y_{10}^{422}(M_U)$, $Y_{126}^{422}(M_U)$) defined by
\begin{eqnarray}
\frac{Y_{10}^{422}(M_U)}{Y_{126}^{422}(M_U)}=\frac14 \sqrt{\frac35} \, .
\label{eq:Yuk-constraints2}
\end{eqnarray}

Comparing eqs.~\eqref{eq:G321-yukawas-matching} and (\ref{eq:matching-Y_MU}), we obtain from eq.~(\ref{eq:fmasses2}) the following fermion masses in the 422-symmetric phases
\begin{eqnarray}
m_t \!=\! \frac{v_{10}^u}{\sqrt 2 } Y_{10}^{422} \!+ \! \frac{v_{126}^u} {4\sqrt 2 } Y_{126}^{422} , \,
m_b \!=\! \frac{v_{10}^d}{\sqrt 2 } Y_{10}^{422} \!+ \! \frac{v_{126}^d} {4\sqrt 2 } Y_{126}^{422}   , \,
m_\tau\! = \! \frac{v_{10}^d}{\sqrt 2 } Y_{10}^{422} \!- \! \frac{3v_{126}^d} {4\sqrt 2 } Y_{126}^{422} \, , 
\label{eq:fmasses3}
\end{eqnarray}
in addition to the Dirac/Majorana neutrino masses written as
\begin{eqnarray}
m_{\nu_D} =  \frac{v_{10}^u}{\sqrt 2 } Y_{10}^{422} - \frac{3v_{126}^u} {4\sqrt 2 } Y_{126}^{422}\, , m_{\nu_R}=\frac14 v_R Y_{R}^{422} \, .
\end{eqnarray}

We can now match the intermediate-scale fermion mass matrices in eq.~(\ref{eq:fmasses3}) to the low-energy ones in eq.~(\ref{eq:fmasses}), as the consistency between both theories implies that the masses predicted from the low-energy effective theory and the high-energy theory should be the same at the symmetry breaking scale. It follows that for the breaking chains 422, the matching conditions of the Yukawa couplings at the intermediate scale read
\begin{eqnarray}
Y_t v_u = v_{10}^u Y_{10}^{422} \!+ \! \frac{1}{4} v_{126}^u Y_{126}^{422} \, , \, Y_b v_d = v_{10}^d Y_{10}^{422} \!+ \! \frac{1}{4} v_{126}^d Y_{126}^{422}\, , \, Y_\tau v_d = v_{10}^d Y_{10}^{422} \!- \! \frac{3}{4} v_{126}^d Y_{126}^{422}\, .
\label{eq:matching}
\end{eqnarray}
The above matching conditions contain six free parameters: the four vevs of the Higgs bi-doublets and two intermediate-scale Yukawa couplings. With the three electroweak-scale Yukawa couplings $Y_{t,b,\tau}(M_Z)$ in the 2HDM obtained from the experimental inputs, we actually have enough degrees of freedom to be able to fix these free parameters by the Yukawa couplings RGEs, as has been done in the literature, see Refs.~\cite{Letter,fermionsrunningSO10}. We will discuss such a numerical fitting procedure in detail in the next subsection. However, by imposing the constraints from the scalar potential, such as forbidding dangerous {flavor changing neutral currents (FCNCs)}, we find that the allowed parameter spaces can be largely reduced as will be discussed shortly.

Finally, we emphasize that $Y_R$ is not a free parameter which contributes to the running of other Yukawa couplings. This is because, for every possible set of ($Y_{10}^{422} (M_I)$, $Y_{126}^{422} (M_I)$), there is a uniquely determined $Y_R^{422}(M_I)$ defined by the GUT-scale matching condition in eqs.~(\ref{eq:matching-Y_MU}) as their values at $M_I$ and at $M_U$ are related by their RGEs
\begin{eqnarray}
    Y_{126}^{422}(M_U)=\sqrt{2} Y_R^{422}(M_U) \, .
    \label{eq:mathchingYR}
\end{eqnarray}
Therefore, in practice, we scan for all the possible values of $Y_R(M_I)$ to satisfy the above relation (within a certain accuracy), together with ($Y_{10}^{422} (M_I)$, $Y_{126}^{422} (M_I)$) to determine the initial conditions at $M_I$ for solving the RGEs from the intermediate scale to the GUT scale. 

\subsection{The evolution of Yukawa couplings}

In this subsection, we give the details of the numerical fitting procedure for the parameter space allowing to address the possibility of Yukawa coupling unification, following Ref.~\cite{Letter}, where two breaking patterns of a non-SUSY SO(10) model with a real ${\bf 10_H}$ representation were discussed. The analysis is restricted to the 422 case, and is based on numerically solving the RGEs and the matching conditions in eqs.~(\ref{eq:matching}) simultaneously. 

In our numerical evaluation of the RGEs, the Yukawa couplings at the electroweak scale chosen to be the $Z$ boson mass $M_Z=91.2$ GeV, have to be fitted with the physical observables which are the top, bottom and tau masses using the relations in eq.~(\ref{eq:fmasses}). The following input values of the $\overline{\text{MS}}$ running fermion masses in the SM \cite{PDG,Bednyakov:2016onn} (we again ignore here the related experimental uncertainties) will be used
\begin{eqnarray}
\label{eq:exp-masses}
[m_t(M_Z), m_b(M_Z), m_\tau (M_Z)] = [168.3, 2.87,  1.73]~{\rm GeV}\, . 
\end{eqnarray}
We convert these inputs into the corresponding masses in the 2HDM by using the appropriate RGEs in the evolution from the scale of the fermion masses to the scale $M_Z$. With the value of $\tan\beta$ and the fermion masses at the electroweak scale $M_Z$, one can obtain the Yukawa couplings $Y_{t,b,\tau}(M_Z)$ in the 2HDM, which are then evaluated from $M_Z$ to $M_I$ by the Mathematica program SARAH \cite{SARAH} similar to what we did in the case of the evolution of the gauge couplings in section 3.4. 

As was discussed in section 2, after complexifying the ${\bf 10_H}$ field by introducing an extra ${\rm U(1)_{PQ}}$ symmetry, we can separate the up- and down-type Higgs component of the bi-doublet field $\Phi_{10}$ in our intermediate-scale left-right symmetric model. As a result, we will have a few more free parameters, which are the vevs $v_{10}^u$ and $v_{10}^d$ instead of a single vev $v_{10}$ in Ref.~\cite{Letter}, and also the relative phases between them, for fitting all the experimental inputs. Counting on the freedom of modifying the scales $M_I$ and $M_U$ by appropriate threshold corrections when enforcing gauge coupling unification, it turns out that within some corners of the huge possible parameter space,  we will always be capable of finding solutions for Yukawa coupling unification, unless there are additional constraints from the scalar potential. One such example is the constraints from FCNCs when matching the intermediate-scale left-right model to the low-energy 2HDM, which will be discussed shortly after this subsection. 

In the 2HDM, when electroweak symmetry breaking is achieved, the ${\rm SU(2)}_L$ gauge bosons $W_L$ will acquire masses from the vevs of both Higgs doublets.  This implies a relation between $\tan\beta$ and the SM vev given by $v_u^2+v_d^2=v_{\rm SM}^2\approx246^2$ at the scale $M_Z$. Similarly, in the intermediate left-right model, the electroweak symmetry was broken by the vevs of bi-doublets which then gives the following relation
\begin{eqnarray}
    \left( v_{10}^u\right)^2 +\left( v_{10}^d\right)^2 +\left( v_{126}^u\right)^2 +\left( v_{126}^d\right)^2 = v_u^2+v_d^2=v_{\rm SM}^2 \, .
\end{eqnarray}
In the absence of knowledge of technical details about the intermediate-scale scalar potential, this is the only constraint that we would have for constraining the parameter space.

With the above equation, we can eliminate one free vev. Furthermore, with the matching conditions and the Yukawa coupling conditions defined in eqs.~(\ref{eq:yukawa-ratio}) and eq.~(\ref{eq:matching-Y_MU}), we can eliminate one free 422-Yukawa coupling. Note that all the other Yukawa couplings and vevs in the 2HDM can be computed from the sole parameter $\tan\beta$ by the masses of the top and bottom quarks and the tau lepton that are experimentally given. As a result, we have $\tan\beta$, one 422-Yukawa coupling and three vevs, in total five free parameters, when solving the three eqs.~(\ref{eq:matching})\footnote{Rigorously speaking, we should take into account the effects from the runnings of vevs for $v_u$ and $v_d$ when solving the matching conditions in eqs.~(\ref{eq:matching}) at the intermediate scale, which makes about 10\% deviation from their electroweak-scale values after running to $M_I$ by their RGEs.}. We can thus numerically scan for some definite values of the two free parameters, $\tan\beta$ and $Y_U$ which is the free Yukawa coupling at the GUT scale, to get the numerical solutions of these matching conditions for obtaining Yukawa unification.

Therefore, differently from the case discussed in Ref.~\cite{Letter} where in addition to the different matching condition at $M_U$?, the parameter space is very constrained because of the fact that the field ${\bf 10_H}$ is real. We conclude that the model with a complexified ${\bf 10_H}$ field is more general and has a much larger parameter space, thus allowing for Yukawa unification that is not restricted to high values of $\tan\beta$ anymore as found in Ref.~\cite{Letter}.

\begin{figure}[!t]
\begin{center}
\includegraphics[width=7.8cm,height=5.9cm,clip]{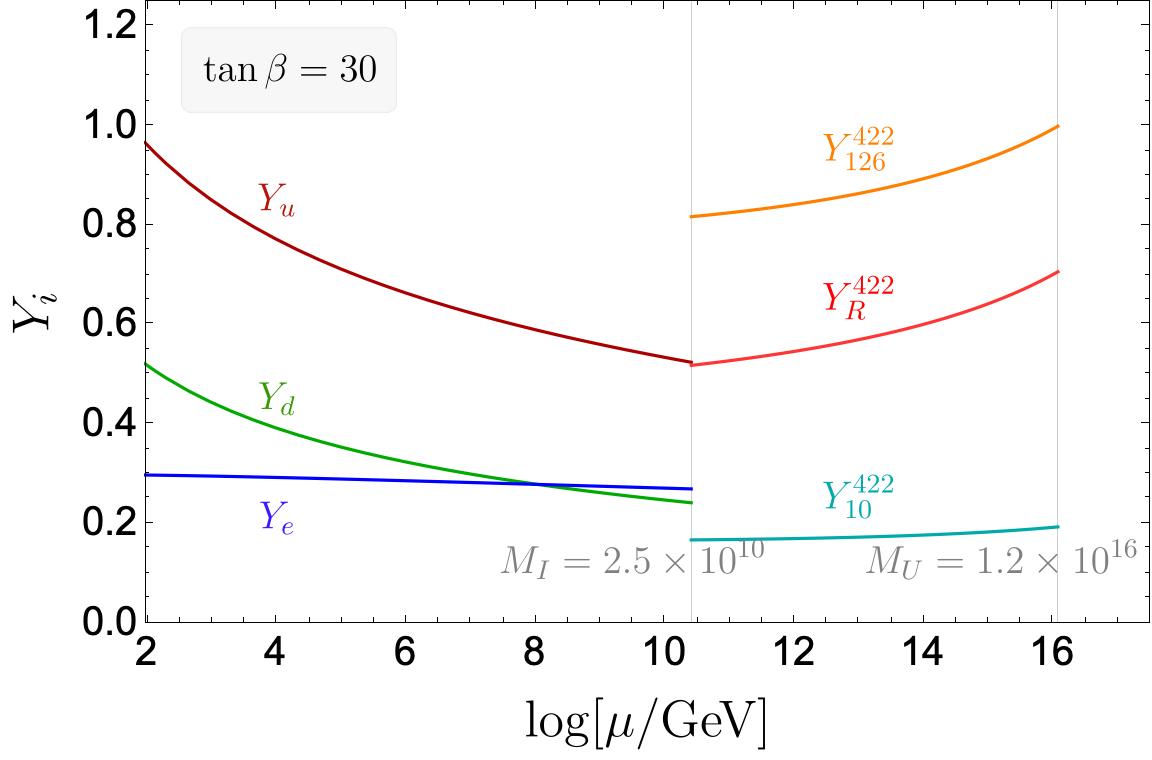}
\includegraphics[width=7.8cm,height=5.9cm,clip]{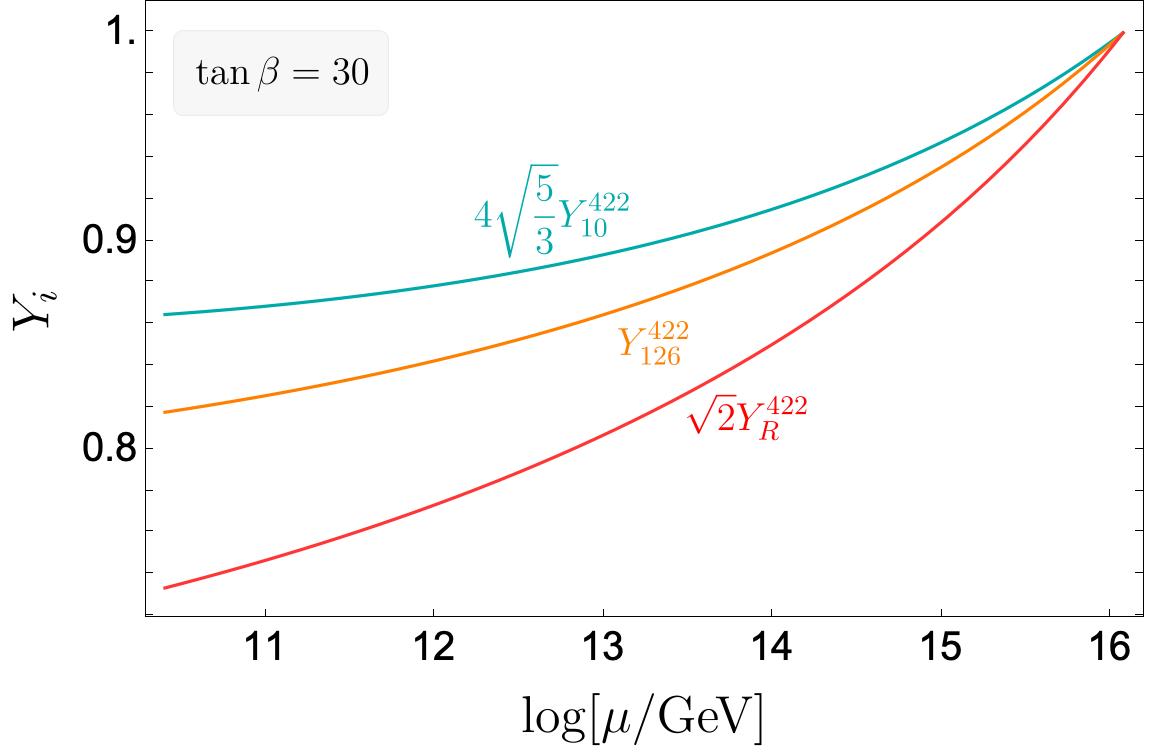} \\[3mm]
\includegraphics[width=7.8cm,height=5.9cm,clip]{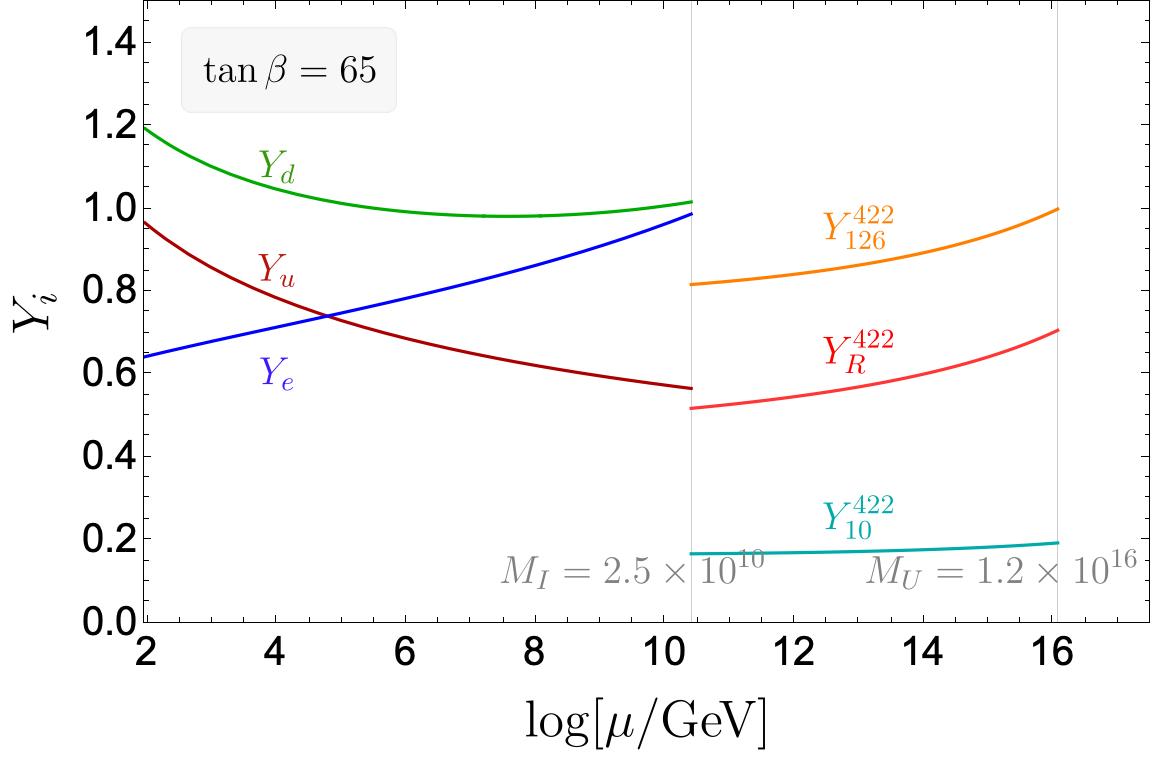}
\includegraphics[width=7.8cm,height=5.9cm,clip]{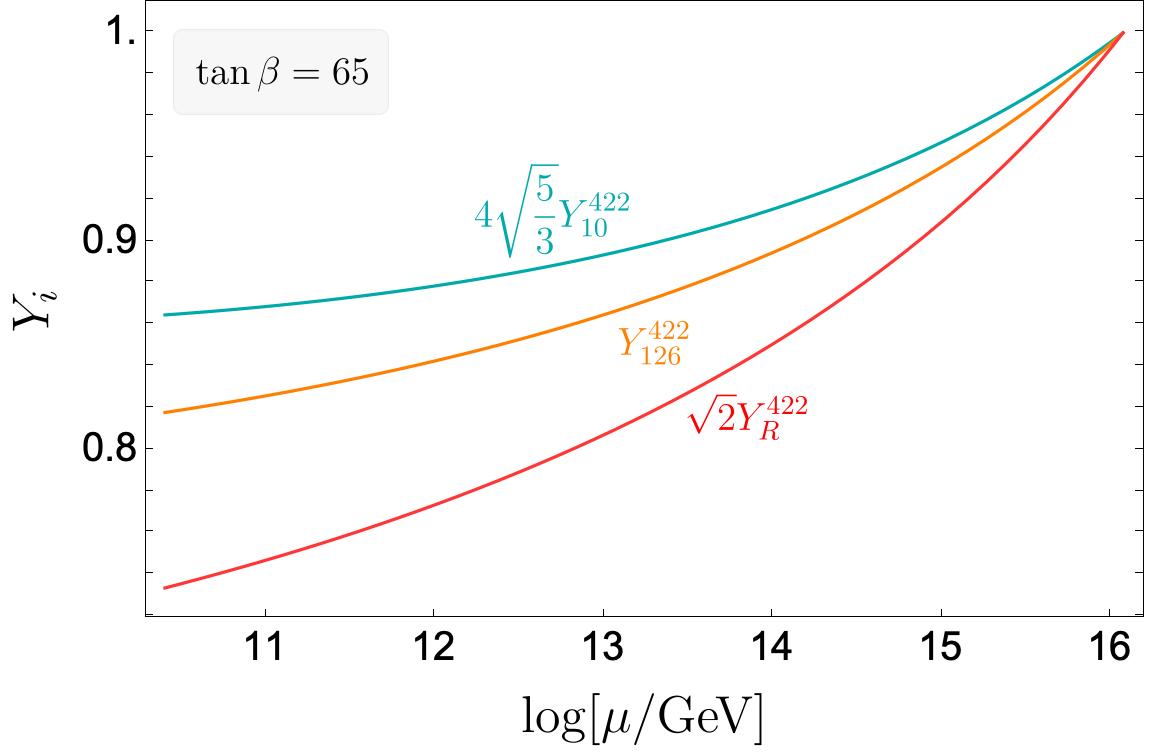} 
\caption[]{The runnings of Yukawa couplings in the 422 breaking chains of our non-SUSY SO(10) model including the threshold corrections of gauge couplings, where the E6 factor in eq.~(\ref{eq:Yukawa-ratio-E6}) has been used to define the Yukawa unification at the GUT scale. Because of the large parameter spaces allowed, we only show here two particular examples when $\tan\beta=30$ (in the top) and $\tan\beta=65$ (in the bottom).}
\label{fig:YU}
\end{center}
\end{figure}

Because of the largely allowed parameter spaces, we show in Fig.~\ref{fig:YU} only two particular examples of the sets of parameters needed to achieve Yukawa coupling unification: one for $\tan\beta=30$ (in the top panels) and the other for $\tan\beta=65$ (in the bottom panels) when $Y_{126}^{422}(M_U)=1$, where the GUT-scale matching conditions motivated from ${\rm E_6}$ in eqs.~(\ref{eq:matching-Y_MU})-(\ref{eq:Yuk-constraints2}) have been applied to numerically solve the RGEs of Yukawa couplings from $M_Z$ to $M_U$. In Table~\ref{tab:Yukawas}, we explicitly list the important free parameters for Yukawa unification.

\begin{table}[!h]
    \centering
    \begin{tabular}{|c|ccc|ccc|cc||cccc|}
    \hline
       scale  & & $M_Z$ & & & \hspace*{-1.2cm}$M_I$ \hspace*{-1.2cm} & &  \hspace*{0.5cm}  $M_U$\hspace*{-0.5cm} & & 
       & &   \hspace*{-.5cm}$M_I$\hspace*{1cm}  & \\ \hline
       $\tan\beta$ & $Y_{t}$ &  $Y_{b}$ & $Y_{\tau}$ & $Y_{10}^{422}$ &  $Y_{126}^{422}$ & $Y_{R}^{422}$ &    $Y_{10}^{422}$ & $Y_{126}^{422}$ & $v_{10}^{u}$ & $v_{10}^d$ & $v_{126}^u$ & $v_{126}^d$ \\
       \hline
       30 & 0.97 & 0.35 & 0.20 & 0.17 & 0.82 & 0.52 & 0.19 & 1.0 & 204.9 & 12.7 & 105.1 & -0.33  \\ \hline
       65 & 0.97 & 1.19 & 0.64 & 0.17 & 0.82 & 0.52 & 0.19 & 1.0 & 194.3 & 16.0 & 118.2 & 0.09   \\ \hline
    \end{tabular}
    \caption{The set of third generation fermion Yukawa couplings at the scales $M_Z$, $M_I$ and $M_U$, and the relevant vevs at the electroweak and intermediate mass scales at the two-loop level that lead to both gauge coupling and Yukawa coupling unification in our non-SUSY 
    SO(10) model with intermediate 422 breaking.}
    \label{tab:Yukawas}
\vspace*{-2mm}
\end{table}

\subsection{Matching conditions with constraints from FCNCs}

At the intermediate scale, the two bi-doublets ($\Phi_{10}$ and $\Sigma_{126}$) first split into four intermediate-scale Higgs doublets (denoted as $H_{u/d,10}$ and $H_{u/d,126}$ in eq.~(\ref{eq:G321-yukawas})), and then two linear combinations of them become light forming the two Higgs doublets ($H_u$ and $H_d$) of the low-energy 2HDM, while the other two linear combinations acquire masses at the intermediate-scale~\cite{Deshpande:1990ip}. Without presenting the technical details about the splitting of bi-doublets in the scalar sector and to simplify our model, we adopt a simple parameterization to forbid the FCNCs for the four intermediate-scale Higgs doublets in eq.~(\ref{eq:G321-yukawas}) in the mass eigenstates assuming no complex phases involved as
\begin{eqnarray}
&
\begin{pmatrix}
H_u \\ H_u^{\rm heavy}
\end{pmatrix} = \begin{pmatrix}
\cos\theta_U & \sin\theta_U \\  -\sin\theta_U  & \cos\theta_U 
\end{pmatrix} \begin{pmatrix}
H_{u,10} \\ H_{u,126}
\end{pmatrix} \ , \nonumber \\
 &
\begin{pmatrix}
H_d \\ H_d^{\rm heavy}
\end{pmatrix} = \begin{pmatrix}
\cos\theta_D & \sin\theta_D \\  -\sin\theta_D  & \cos\theta_D 
\end{pmatrix} \begin{pmatrix}
H_{d,10} \\ H_{d,126}
\end{pmatrix} ,
\end{eqnarray}
where the $H_u$ and $H_d$ are the admixtures of two scalar doublets coupling only to teh isospin up/down fermionic sector~\footnote{More general combinations of the 4 scalars of the type
\begin{eqnarray}
H_u &=& \alpha_1^u H_{u,10}+\alpha_2^u H_{u,126}+\beta_1^u H_{d,10}^*+\beta_2^u H_{d,126}^* \, , \nonumber \\
H_d &=& \alpha_1^d H_{d,10}+\alpha_2^d H_{d,126}+\beta_1^d H_{u,10}^*+\beta_2^d H_{u,126}^* \, , \nonumber
\end{eqnarray}
are highly constrained by FCNC~\cite{Fukuyama:2002vv,Deshpande:1990ip}.}, which will be identified as the two Higgs fields in low-energy 2HDM, and $H_u^{\rm heavy}$ and $H_d^{\rm heavy}$ are the doublets acquiring intermediate-scale masses via the interactions like ${\rm tr}(\Phi^2\Delta_R^2)$. Indeed, this assumption implies that the up/down-type Higgs doublet consists of the up/down components of the two Higgs bidoublets ($\Phi_{10}$ and $\Sigma_{126}$) at $M_I$, which can be seen as the definition of the mixing angle $\theta_{U/D}$ 
\begin{eqnarray}
\cos \theta_U = \frac{v_{10}^u}{v_u} \, , \, 
\sin \theta_U = \frac{v_{126}^u}{v_u} \, , \,
\cos \theta_D = \frac{v_{10}^d}{v_d} \, , \, 
\sin \theta_D = \frac{v_{126}^d}{v_d} \, . \,
\end{eqnarray}

The above parameterization includes the constraints from FCNCs when matching the intermediate-scale left-right model to the low-energy 2HDM, so equivalently, we can express the matching conditions for Yukawa couplings at $M_I$ derived in eqs.~(\ref{eq:matching}) by the mixing angles $\theta_{U/D}$ as
\begin{eqnarray}
\label{eq:topyukawa}
Y_t (M_I)  &=& \cos\theta_U Y_{10}^{422}(M_I)+ \frac14 \sin\theta_U {Y_{126}^{422}}(M_I)  \, ,  \\
 Y_b (M_I) &=& \cos\theta_D Y_{10}^{422}(M_I) +  \frac14 \sin\theta_D {Y_{126}^{422}}(M_I)\, , \\
 Y_\tau (M_I) &=& \cos\theta_D Y_{10}^{422} (M_I) -  \frac34 \sin\theta_D{Y_{126}^{422}}(M_I) \, .
\end{eqnarray}

Because above the intermediate scale the bottom quark will couple exactly the same way to the Higgs bi-doublets as the tau lepton does, we can eliminate one free parameter $\theta_D$ from the last two matching conditions for $Y_b(M_I)$ and $Y_\tau(M_I)$, and get a relation for the Yukawa couplings at $M_I$:
\begin{eqnarray}
 \left( Y_{10}^{422} (M_I)\right)^2 &=& \frac{\left( Y_{126}^{422} (M_I) \right)^2 \left( 3 Y_b(M_I)+Y_\tau(M_I) \right)^2 }{16\left[ \left( Y_{126}^{422}(M_I) \right)^2 - \left(Y_b (M_I)- Y_\tau (M_I)\right)^2 \right] } \, .
 \label{eq:Yuk-MI-constraint1}
\end{eqnarray}
This equation defines a curve $\gamma(M_I)$ in the parameter space ($Y_{10}^{422} (M_I)$, $Y_{10}^{422} (M_I)$) as a function of $\tan\beta$. Note that eq.~(\ref{eq:Yuk-MI-constraint1}) also implies the lower bound for $Y_{10}^{422} (M_I)$ by
\begin{eqnarray}
Y_{10}^{422} (M_I) > {\frac{ 3 Y_b(M_I)+Y_\tau(M_I) }{4
}} \, , 
\label{eq:lowerboundY10}
\end{eqnarray}
while requiring $Y_{10}^{422} (M_I)<\sqrt{4\pi}$ implies the lower bound of $Y_{126}^{422} (M_I)$ by
\begin{eqnarray}
Y_{126}^{422} (M_I) > {\frac{  Y_b(M_I)-Y_\tau(M_I) }{\sqrt{1- \frac{\left(3Y_b(M_I) + Y_\tau(M_I)\right)^2}{64\pi} }} } \, .
\label{eq:lowerboundY126}
\end{eqnarray}

\begin{figure}[!ht]
\begin{center}
\mbox{
\includegraphics[width=14cm,clip]{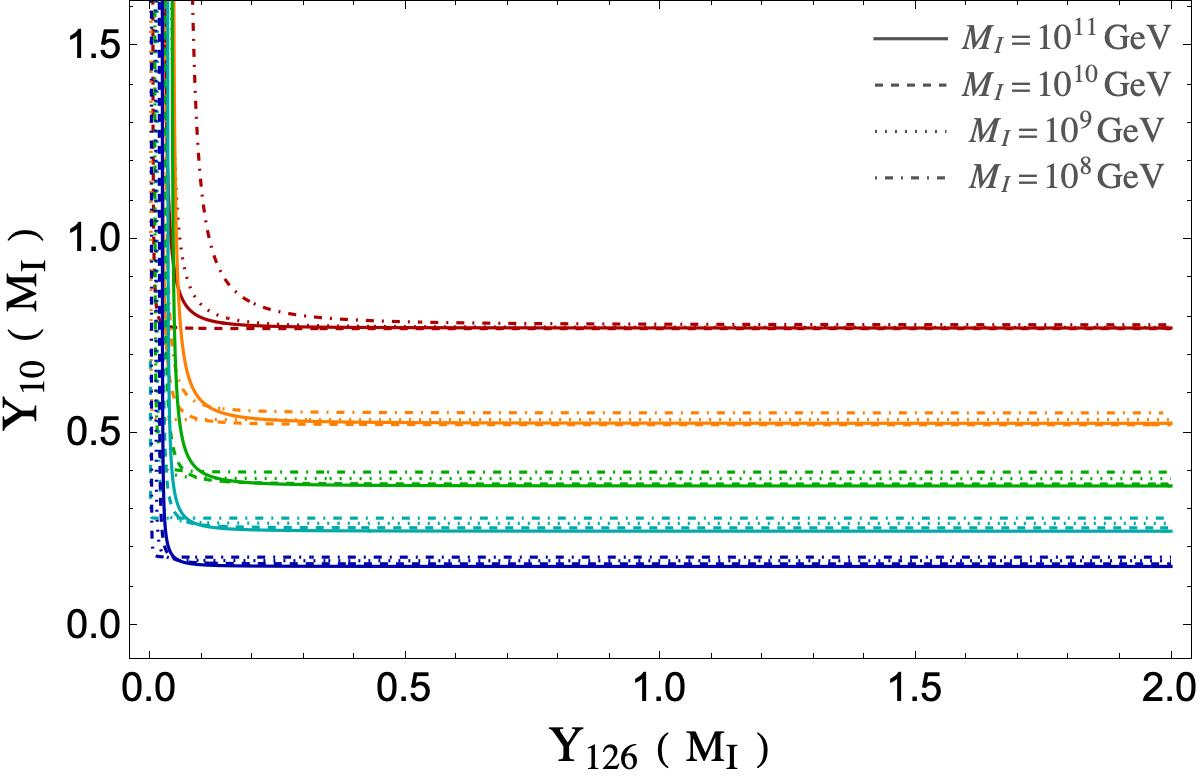} 
}
\end{center}
\vspace*{-0.5cm}
\caption[]{The curves $\gamma(M_I)$ defined in eq.~(\ref{eq:Yuk-MI-constraint1}) when $\tan\beta$ is taken to be 60 (red), 50 (orange), 40 (green), 30 (cyan), 20 (blue), where the parameter $M_I$ is chosen to be $10^{11}$ GeV (solid), $10^{10}$ GeV (dashed), $10^9$ GeV (dotted), $10^8$ GeV (dot-dashed).}
\label{fig:curveMI}
\end{figure}

For a straightforward comparison, we show in Fig.~\ref{fig:curveMI} several curves $\gamma(M_I)$ depicted by eq.~(\ref{eq:Yuk-MI-constraint1}) for certain values of $\tan \beta$ and the intermediate scale $M_I$ taken from $10^8$ to $10^{11}$ GeV, where the minimum of $Y_{10}^{422} (M_I)$ and $Y_{126}^{422} (M_I)$ are given by eqs.~(\ref{eq:lowerboundY10})--(\ref{eq:lowerboundY126}) correspondingly.

If we assume all the vevs are positive, i.e. $0<\theta_D < \pi/2 $, from eqs. (60)--(61) we can separate the intermediate-scale Yukawa couplings $Y_{10}^{422} (M_I)$ and $Y_{10}^{422} (M_I)$ as
\begin{eqnarray}
 Y_{10}^{422} (M_I) \cos \theta_D &=& \frac{1}{4}\left(3Y_b(M_I)+Y_{\tau}(M_I) \right) \, , \nonumber \\ 
 Y_{126}^{422} (M_I) \sin \theta_D &=& Y_b(M_I)-Y_{\tau}(M_I)  \, .
 \label{eq:constraint-YUMI}
\end{eqnarray}
It suggests that if the coupling $Y_{126}^{422} (M_I)$ had the same positive sign as $Y_{10}^{422} (M_I)$ in order to be able to be unified at $M_U$, then at $M_I$ we must have  $Y_b(M_I)-Y_{\tau}(M_I)>0$. Indeed, it is only an artifact by assuming the positivity of vevs, as the latter relation is nothing but 
\begin{eqnarray}
\sqrt{2} (m_b-m_{\tau})=(Y_b-Y_{\tau}) v_d = 4 Y_{126} v_{126}^d  \, ,
\end{eqnarray}
from subtracting the SO(10) mass matrices in eq.~(\ref{eq:fmasses2}) if we are matching the SO(10) directly to the 2HDM. 

\begin{figure}[!ht]
 \begin{center}
 \includegraphics[width=13.5cm,clip]{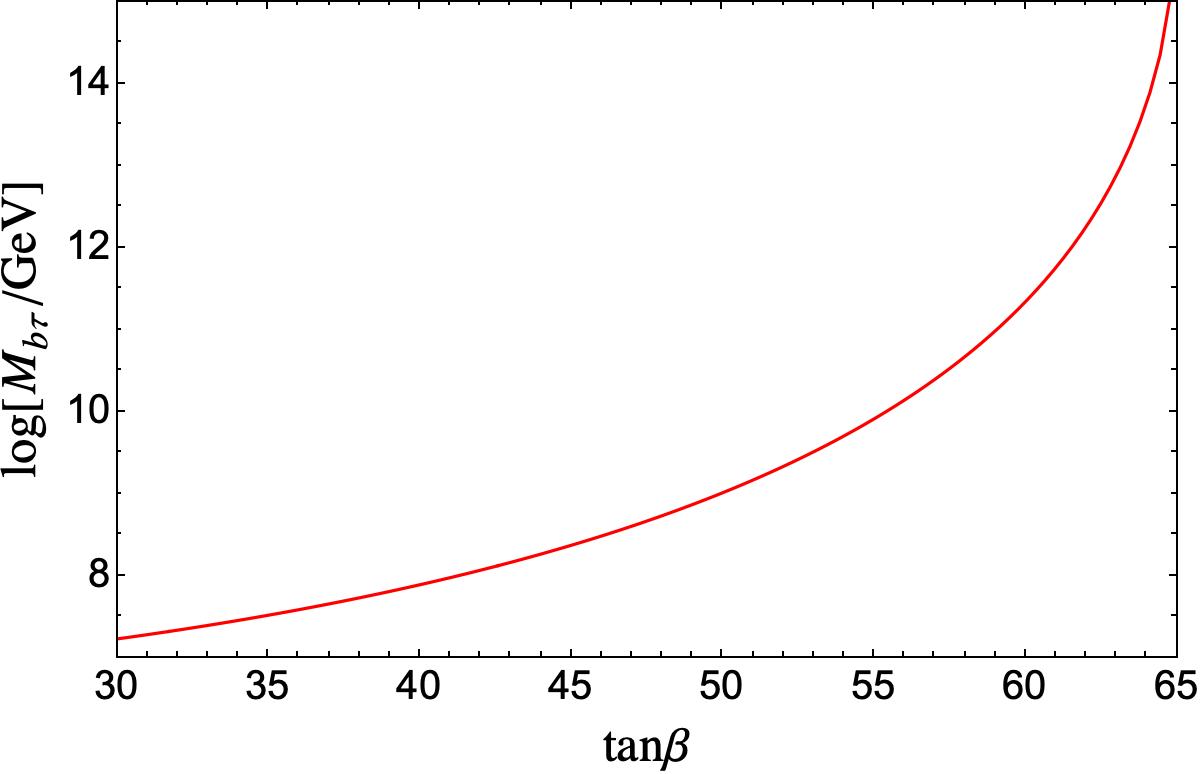}
 \end{center}
 \vspace*{-.5cm}
 \caption[]{The unification scale for the bottom quark and tau lepton ($M_{b\tau}$) in in the 2HDM which is determined by the point at which the curves for their RG running from the weak scale $M_Z$ upwards intersect. }
 \label{fig:Mbtau}
 \end{figure}

As was discussed in section 3, the intermediate scale $M_I$ and the unification scale $M_U$ are totally fixed by the input parameters $\tan\beta$ and the threshold corrections, irrelevant of the Yukawa couplings. However, once the intermediate-scale Yukawa couplings are switched on, what we deduce from eq.~(\ref{eq:constraint-YUMI}) is that an upper bound on $M_I$ exists such that $Y_b(M_I)>Y_{\tau}(M_I)$. As both $Y_b$ and $Y_{\tau}$ are functions of $\tan \beta$ only, we can thus derive a scale $M_{b\tau}$ as a function of $\tan \beta$, determined (within some accuracy) by the point at which the curves for their RG running from the weak scale $M_Z$ upwards intersect so that $Y_b(M_{b\tau})=Y_{\tau}(M_{b\tau})$. Then the assumption of eq.~(\ref{eq:constraint-YUMI}) translate to 
\begin{eqnarray}
    M_I \leq M_{b\tau} \quad \text{\rm (assuming the positivity of vevs)} \, .
    \label{eq:bounds-pos-vevs}
\end{eqnarray}
This is exemplified in Fig.~\ref{fig:Mbtau} where the scale $M_{b\tau}$ leading to the unification of the bottom quark and tau lepton couplings in 2HDM are shown as a function of the input value of $\tan\beta$ at the low scale. This scale increases with increasing values of $\tan\beta$ and, in order to have reasonably high values of $M_I>10^{10}$ GeV, needs rather large $\tan\beta$ values\footnote{Note that we cannot have much higher values of $\tan\beta$, i.e. $\tan\beta<65$ in general, to avoid the bottom quark Yukawa couplings running into a non-perturbative regime at high energy scales.},  $\tan\beta > 55$.  


The parameter $\theta_U$ for determining the top Yukawa coupling in the 2HDM, on the other hand, cannot be eliminated without further assumptions. Thus, in practice, we treat it as a free parameter that should fit the mass of the top quark. Again by assuming the positivity of vevs, from eq.~(\ref{eq:topyukawa}) we can estimate that the top Yukawa coupling in the 2HDM should lie in the region of
\begin{eqnarray}
{\rm Min}\left[Y_{10}^{422}(M_I),\frac{Y_{126}^{422}(M_I)}{4}\right] \leq Y_t (M_I) \leq \sqrt{\left(Y_{10}^{422}(M_I)\right)^2 + \left(\frac{Y_{126}^{422}(M_I)}{4}\right)^2} \, ,
\label{eq:bound-Yt1}
\end{eqnarray}
which, when combined with eq.~(\ref{eq:lowerboundY10})--(\ref{eq:lowerboundY126}), gives
\begin{eqnarray}
Y_t (M_I) \!>\! {\rm Min}\left[\frac{ 3 Y_b (M_I)\!+\!Y_\tau (M_I) }{4} \frac{Y_b (M_I)\!-\!Y_\tau  (M_I)}{4\sqrt{1 \!-\! \frac{\left(3Y_b (M_I) \!+\! Y_\tau (M_I)\right)^2}{64\pi} }} \right] \!=\! \frac{Y_b (M_I)\!-\!Y_\tau  (M_I)}{4\sqrt{1 \!-\! \frac{\left(3Y_b  (M_I)\!+\! Y_\tau (M_I)\right)^2}{64\pi} }} \, .
\label{eq:bound-Yt2}
\end{eqnarray}
This criterion thus helps us check easily whether a parameter $\theta_U$ exists for fitting the mass of the top quark.

\subsection{Numerical results for Yukawa unification}

In principle, with the RGEs obtained for the 422 breaking chain of our SO(10) model, we can run all the 422-Yukawa couplings on the curves $\gamma(M_I)$ from $M_I$ to $M_U$ to get a new curve $\gamma(M_U)$. The intersections of the curve $\gamma(M_U)$ with the line $\ell(M_U)$ defined in eq.~(\ref{eq:Yuk-constraints2}) at the GUT scale $M_U$ thus define the solutions admitting the Yukawa unification in the 422 breaking chain. 

When evaluating the curve $\gamma(M_I)$ to the GUT scale, the exact values of $M_I$ and $M_U$ will be determined by numerical solving the RGEs to ensure the unification of gauge couplings as done in section 3.4. The randomly-taken threshold corrections would thus bring some uncertainties in determining the exact values of the two scales $M_I$ and $M_U$ which eventually affect the curves $\gamma(M_I)$ and $\gamma(M_U)$. However, as can be seen from the analytical results in Fig.~\ref{fig:curveMI}, the curves $\gamma (M_I)$ almost remain intact when varying the scales $M_I$, suggesting that the threshold corrections of gauge couplings only make a tiny difference in determining the curves $\gamma(M_I)$ and similarly to $\gamma (M_U)$, contrary to what happens in gauge coupling unification, i.e. Fig.~1. We can thus safely choose some random-sampling threshold corrections when visualizing the curves $\gamma(M_U)$ as shown in Fig.~\ref{fig:curveMU} below, where for each $\tan\beta$ we explicitly show the uncertainty regions allowed by varying the parameters of threshold corrections from $\eta_i = \ln(M_i/\mu) \in [-1,1]$. 

\begin{figure}[!ht]
\vspace*{2mm}
\begin{center}
\mbox{ \includegraphics[width=13.5cm,clip]{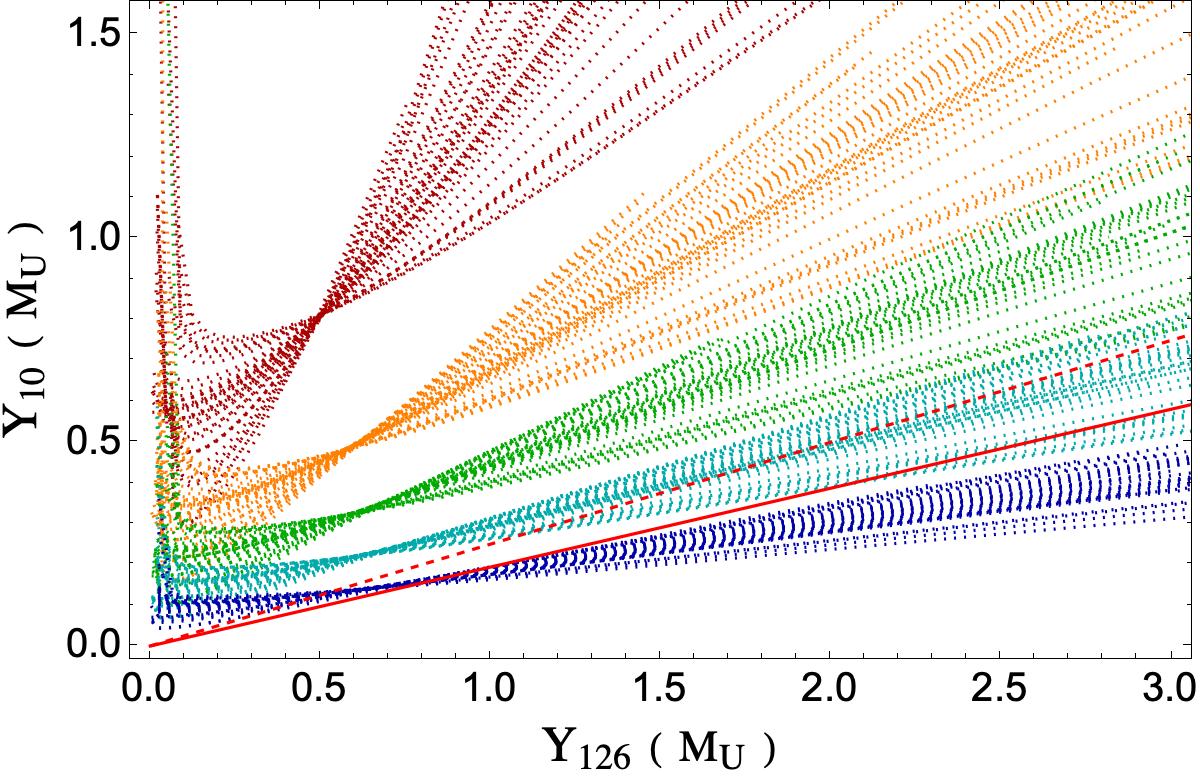}  }
\end{center}
\vspace*{-0.5cm}
\caption[]{
The curves $\gamma(M_U)$ obtained from numerically evaluating the curves $\gamma(M_I)$ from $M_I$ to $M_U$ by the RGEs of 422 breaking chain, where the scales $M_I$ and $M_U$ are determined by enforcing the gauge unification with randomly taking threshold corrections for  $\eta_i = \ln(M_i/\mu) \in [-1,1]$. The uncertainties of random threshold correction thus generate the uncertainties of these curves $\gamma(M_U)$, which are plotted for the parameter $\tan \beta$ corresponding to 60 (dark red), 50 (orange), 40 (green), 30 (cyan), and 20 (blue). The red line $\ell (M_U)$ corresponds to the condition of Yukawa unification motivated by ${\rm E_6}$ unification given in eq.~(\ref{eq:Yuk-constraints2}). The intersections of $\ell (M_U)$ and $\gamma (M_U)$ thus define the solutions for Yukawa unification motivated from ${\rm E_6}$.}
\label{fig:curveMU}
\end{figure}

When considering a different scenario of Yukawa unification for the 422 breaking chains of the non-SUSY SO(10) models, one can merely change the slope of the line $\ell(M_U)$ by defining the ratios of $Y_{10}$ and $Y_{126}$ in eq.~(\ref{eq:yukawa-ratio}), while the curves $\gamma (M_U)$ remain the same. To be compared with the ${\rm E_6}$ case, we also show the condition of Yukawa unification for the ratio $C=c_{10}/c_{126}=1$, which is presented by the dashed red line in Fig.~\ref{fig:curveMU}.

As a consistency check, we must combine all the conditions that we derive to constrain the parameter space, including the proton decay bound of eq.~(\ref{eq:bounds-protondecay}), the GUT-scale matching condition in eq.~(\ref{eq:matching-Y_MU})-(\ref{eq:Yuk-constraints2}), the lower bounds of the Yukawa couplings in eqs.~(\ref{eq:lowerboundY10})-(\ref{eq:lowerboundY126}), and the perturbative bound when requiring that all Yukawa couplings must be smaller than $\sqrt{4\pi}$ at all energy scales.

One then immediately finds that these constraints also influence the intermediate scale $M_I$, especially through eq.~(\ref{eq:bounds-pos-vevs}). Thus, the unification of third generation Yukawa couplings also has a repercussion on gauge coupling unification. This refines the naive statement that we initially made in section 3, namely that the contributions of the Yukawa couplings hardly affect the RGEs of the gauge couplings and, hence, their unification.

\begin{figure}[!ht]
\begin{center}
\mbox{
\includegraphics[width=13cm,clip]{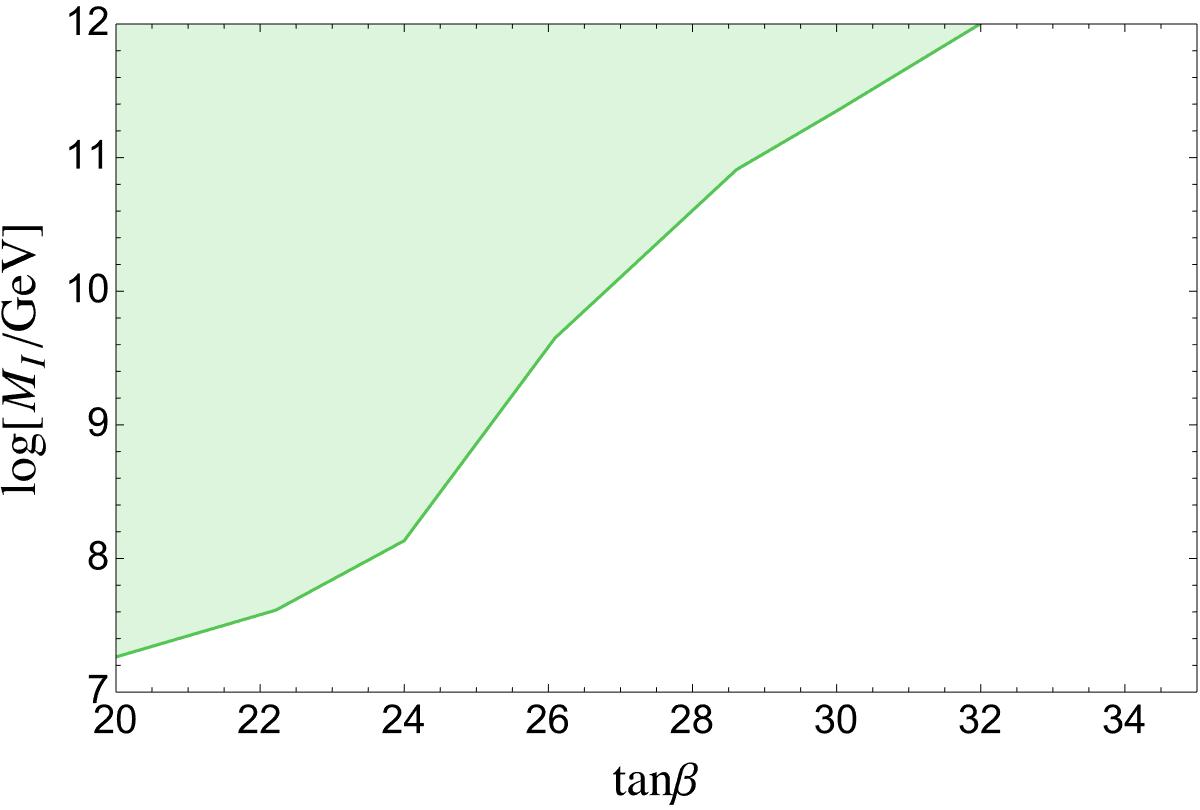}
}
\end{center}
\vspace*{-0.5cm}
\caption[]{Values of the intermediate scale $M_I$ and $\tan \beta$ consistent with gauge and Yukawa unification (the green region), using the ${\rm E_6}$ factor in our non-SUSY SO(10) model.}
\label{fig:TBMI}
\end{figure}

Including all the constraints, and enforcing the unification of the Yukawa couplings with the ${\rm E_6}$ ratio, one can visualize the numerical solutions in the 422 breaking chain of our non-SUSY SO(10) model in Fig.~\ref{fig:TBMI}. It shows, in green, the parameter region in which both gauge and Yukawa coupling unification can be achieved in the plane 
$[\tan\beta , \log (M_I/{\rm GeV})]$ as 
$\tan\beta$ is the most important  parameter in determining unification in the two cases. As can be seen, for each $\tan\beta$ value, there can be multiple solutions depending on the exact threshold corrections resulting in the different intermediate scale $M_I$ and at a later stage, the unification scale $M_U$. Thus,  after considering the constraints of FCNCs, in addition to all the other constraints, the parameter $\tan\beta$ is again very constrained in this 422 intermediate breaking model and only relatively lower values (compared to those discussed in the earlier analysis of Ref.~\cite{Letter}), $\tan\beta \lsim 30$ for $M_I \lsim 10^{12}$ GeV, are favored\footnote{We should note that in this green region in which both gauge and Yukawa unification occur, the vevs of the Higgs bi-doublets, for example, $v_{126}^d$, are complex and are negative according to eq.~\ref{eq:bounds-pos-vevs} and Fig.~\ref{fig:Mbtau}. If we require the vevs to be positive, unification occurs only for lower values of $M_I$ and higher values of $\tan\beta$.  The corresponding region will not intersect with the green region.}

As a preliminary conclusion, the constraints from FCNCs largely reduce the allowed parameter spaces for Yukawa coupling unification motivated by ${\rm E_6}$ symmetry. This only favors lower values of $\tan\beta$ in the 422 breaking chain for instance. Thus, the 422 breaking chain of our non-SUSY minimal SO(10) model is very constrained, with the only parameter which can be varied being the value of the input $\tan\beta$ of the low-energy 2HDM. This renders the model quite predictive. The other nice feature is that Yukawa unification, with a common  coupling at the high scale being naturally of order unity, implies that a condition at the high scale has an impact on the low energy parameters such as $\tan\beta$.

\section{Conclusions}

The unification of fundamental forces plays an extremely important role in particle physics.  A wide range of studies have dealt with the unification of the three gauge couplings of the SM either by sticking to the minimal SU(5) gauge group and extending the SM particle spectrum, as is the case in Supersymmetric theories, or keeping the SM particle content and extending the unifying gauge symmetry group.  In this last option, the SO(10) group has been the most widely studied as it is the simplest one beyond the minimal SU(5) group. It possibly leads to a left-right symmetry group and it has a  fundamental representation of dimension 16 which could contain all SM fermions plus an additional Majorana neutrino. If the mass of the latter particle is high enough, ${\cal O}(10^{12}\!-\! 10^{14})$ GeV, one could explain the pattern of masses and mixing of the SM light neutrino species and address the problem of the baryon asymmetry in the universe by invoking a leptogenesis triggered by this additional heavy neutrino. Unification is achieved by considering that this large mass of the Majorana neutrinos is in fact due to the intermediate scale of the breaking of SO(10) into the SM group via an intermediate step, corresponding, for instance, to the Pati-Salam or the minimal left-right symmetry groups. This is achieved by including the threshold effects of the additional Higgs and gauge bosons at this intermediate scale $M_I$, which then modify the renormalization group evolution of the coupling constants and make them intersect at a single point, the unification scale $M_U$. 

It is very tempting to extend the unification paradigm to the case of the Yukawa couplings of fermions, in particular those of the third generation which are heavy enough to allow for a perturbative treatment at the low energy scale. This has been attempted in an earlier analysis in Ref.~\cite{Letter} {in both the Pati-Salam and the minimal left-right intermediate schemes, which showed that, {ignoring constraints from flavor changing neutral currents},  one can achieve the Yukawa unification in the context of a low energy two-Higgs doublet model in which the ratio of the two vevs is very high, $\tan\beta \approx 60$, and reproduce the hierarchy of the fermion masses of the third generation from the running of Yukawa couplings.}

In this paper, we generalized our previous analysis made in Ref.~\cite{Letter} to the case with a complex $10_H$ field, where a ${\rm U(1)_{PQ}}$ global symmetry was introduced to forbid the Yukawa couplings with the field ${\bf 10_H^*}$, in order to relax the parameter space in the previous over-constrained model which also changes the RGEs of the gauge couplings. We then derived the analytical approximate solutions of the RGEs of gauge couplings enforced by unification at the two-loop level.  The procedure in our chosen non-SUSY SO(10) model with an intermediate scale can also be applied for any breaking patterns of SO(10). The uncertainties of our approximation were also discussed, including the constraints from proton decay experiments. All our approximate analytical results have been compared with the numerical results given in Tables~\ref{tab:sol} and Table~\ref{tab:scales}, and a good agreement was found. 

We have then discussed the possibility of unifying the Yukawa couplings of third generation heavy fermions at the high scale which, in the present context,  implies a relation between the fermion couplings to the scalar representations 10 and 126. Specializing to the Pati-Salam intermediate SO(10) breaking chain,  we have considered {the particular case where the coupling is obtainable in an ${\rm E_6}$ model where the previous two scalars are part of a single multiplet and which leads to the relation $Y_{10}=\sqrt{3/5}\, Y_{126}$.  We concluded that Yukawa unification is a very strong constraint which,  when imposing the absence of flavor changing neutral currents at tree--level induced by the two light Higgs doublet fields, is achieved only for $\tan \beta$ values that are not too large. Our non-SUSY SO(10) model is thus very predictive and can be testified by future electroweak-scale experiments.}

Our present exploratory analysis raises rather interesting questions which require further attention and studies of the subject. In particular, there are still some phenomenological issues to be discussed within this model, such as the problem of the stability of the electroweak vacuum and the origin of neutrino masses. Because our low-energy effective theory is based on a 2HDM scenario, we must constrain our scalar potential to enforce a stable vacuum that is bounded from below as, for instance, discussed in Ref.~\cite{2HDM-vacuum}. 
We expect the discussions held in these references to also apply in our case as we are dealing with the same Type-II 2HDM scenarios. On the other hand, if the neutrinos are to acquire masses from a Type-I/II see-saw mechanism, the scale of the right-handed neutrino mass, which is assumed to be of the order of the intermediate scale $M_I$, cannot be too small as to avoid unnatural fine-tuning in the determination of the light neutrino masses.  These neutrino masses thus contribute to setting another constraint on the intermediate scale.  All these aspects and others need further attention and we plan to address them in future work.
\bigskip 

\noindent {\bf Acknowledgements:} \smallskip

\noindent We thank Vasja Susi\v{c} for discussions related to the ${\rm E_6}$ symmetry group. AD is supported by the Estonian Research Council (ERC) grants MOBTT86 and PRG803, and by the Junta de Andalucia through the Talentia Senior program and the grants A-FQM-211-UGR18, P18-FR-4314 with ERDF and PID2021-128396NB-I00. RF acknowledges the financial support from MCIN/AEI (10.13039/501100011033) through grant number PID2019-106087GB-C22 and from the Junta de Andalucía through grant number P18-FR-4314 (FEDER). RO and MR are supported by European Regional Development Fund through the CoE program grant TK133.

\clearpage

\setcounter{equation}{0}
\renewcommand{\theequation}{A.\arabic{equation}}
\setcounter{table}{0}
\renewcommand{\thetable}{A.\arabic{table}}

\section*{Appendix A: Lists of useful coefficients}

\subsection*{A1: $\beta$ coefficients for different gauge groups and representations}\label{subsecA1}

The one-loop and two-loop $\beta$ coefficients $a_i$ and $b_{ij}$, can be calculated from Ref.~\cite{two-loop-RGEs} in the general case. We list the values of the $\beta$ coefficients for some particular gauge groups ${\cal G}_I$ with the considered scalar representations that are relevant for our discussions and which are given in Table~\ref{tab:beta}. 

\begin{table}[h]
\centering
\renewcommand{\arraystretch}{1.2}
\begin{tabular}{|c|c|c|}
\hline
${\cal G}_I$ & $a_i$ & $b_{ij}$  \\
\hline
\rule{0pt}{2.5\normalbaselineskip}
       \ ${\cal G}_{321}{(\rm SM)}$ \ & $\begin{pmatrix}-7 \vspace{1.2mm} \\ -\frac{19}{6} \vspace{1.2mm} \\ \frac{41}{10} \vspace{1.2mm} \end{pmatrix}$  & 
       $  \begin{pmatrix}  -26 & \frac{9}{2} & \frac{11}{10} \vspace{1.2mm} \\  12 & \frac{35}{6} & \frac{9}{10} \vspace{1.2mm} \\ \frac{44}{5} & \frac{27}{10} & \frac{199}{50} \vspace{1.2mm}   \end{pmatrix}$  
    \\
\hline
\rule{0pt}{2.5\normalbaselineskip}
       \ ${\cal G}_{321}{(\rm 2HDM)}$ \ & $\begin{pmatrix} -7 \vspace{1.2mm} \\ -3 \vspace{1.2mm} \\ \frac{21}{5} \vspace{1.2mm} \end{pmatrix}$  & 
       $ \begin{pmatrix}   -26 & \frac{9}{2} & \frac{11}{10} \vspace{1.2mm} \\  12 & 8 & \frac{6}{5} \vspace{1.2mm} \\ \frac{44}{5} & \frac{18}{5} & \frac{104}{25}  \vspace{1.2mm} \end{pmatrix}$ 
       \\
\hline
\rule{0pt}{2.5\normalbaselineskip}
       ${\cal G}_{422}$ & $\begin{pmatrix} -\frac{7}{3} \vspace{1.2mm} \\ 2 \vspace{1.2mm} \\ \frac{28}{3} \vspace{1.2mm} \end{pmatrix} $  & $\begin{pmatrix}  \frac{2435}{6} & \frac{105}{2} & \frac{249}{2} \vspace{1.2mm} \\  \frac{525}{3} & 73 & 48 \vspace{1.2mm} \\ \frac{1245}{2} & 48 & \frac{835}{3} \vspace{1.2mm}  \end{pmatrix}$ 
       \\
\hline
\rule{0pt}{2.5\normalbaselineskip}
      ${\cal G}_{422}\times {\cal D}$
       & $\begin{pmatrix}\frac{2}{3}\vspace{1.2mm} \\ \frac{28}{3}\vspace{1.2mm} \\ \frac{28}{3}\vspace{1.2mm} \end{pmatrix}$  & $\begin{pmatrix}  \frac{3551}{6} & \frac{249}{2} & \frac{249}{2} \vspace{1.2mm}\\  \frac{1245}{2} & \frac{835}{3} & 48 \vspace{1.2mm}\\ \frac{1245}{2} & 48 & \frac{835}{3} \vspace{1.2mm}  \end{pmatrix}$  \\
\hline
\rule{0pt}{3.\normalbaselineskip}
     ${\cal G}_{3221}$ & $\begin{pmatrix} -7 \vspace{1.2mm} \\ -\frac{8}{3}\vspace{1.2mm} \\ -2\vspace{1.2mm} \\ \frac{11}{2} \vspace{1.2mm} \end{pmatrix}$  & $\begin{pmatrix} -26 & \frac{9}{2} & \frac{9}{2} & \frac{1}{2} \vspace{1.2mm}\\ 12 & \frac{37}{3} & 6 & \frac{3}{2} \vspace{1.2mm}\\ 12 & 6 & 31 & \frac{27}{2} \vspace{1.2mm}\\ 4 & \frac{9}{2} & \frac{81}{2} & \frac{61}{2} \vspace{1.2mm} \end{pmatrix}$ 
   \\
     \hline
     \rule{0pt}{3.\normalbaselineskip}
     ${\cal G}_{3221}\times {\cal D}$ &  $\ \begin{pmatrix} -7 \vspace{1.2mm} \\ -\frac{4}{3}\vspace{1.2mm}\\-\frac{4}{3}\vspace{1.2mm}\\7\vspace{1.2mm} \end{pmatrix} \ $  & $\ \begin{pmatrix} -26 & \frac{9}{2} & \frac{9}{2} & \frac{1}{2} \vspace{1.2mm}\\ 12 & \frac{149}{3} & 6 & \frac{27}{2} \vspace{1.2mm}\\ 12 & 6 & \frac{149}{3} & \frac{27}{2} \vspace{1.2mm}\\ 4 & \frac{81}{2} & \frac{81}{2} & \frac{115}{2} \vspace{1.2mm} \end{pmatrix} \  $ \\
     \hline
 \end{tabular}
  \caption[]{The coefficients $a_i$ and $b_i$ of the $\beta$ functions of the RGEs of the gauge couplings $\alpha_i$ for the different breaking schemes that we are considering.} 
\label{tab:beta}
 \end{table}

\subsection*{A2: The two-loop $\theta_i^{\cal G}$ coefficients in our approximations}\label{subsecA2}

At two loop level, the solutions of the two-loop RGEs of gauge couplings takes the general implicit form of eq.~(\ref{eq:RGE2l}): 
\begin{eqnarray}
\alpha_{i, {\cal G}}^{-1}(\mu) = \alpha_{i,{\cal G}}^{-1}(\mu_0) -\frac{a_i^{\cal G}}{2\pi}\ln \frac{\mu}{\mu_0} + \gamma_i^{\cal G} + \Delta_{i, Y}^{\cal G} \, ,
\end{eqnarray}
where the two-loop contributions are functions of gauge couplings $\alpha^{-1}_{j, {\cal G}}(\mu)$ read
\begin{eqnarray}
\gamma_i^{\cal G}=-\frac{1}{4\pi}\sum_{j}\frac{b_{ij}^{\cal G}}{a_j^{\cal G}}\ln \frac{\alpha_{j,{\cal G}}(\mu)}{\alpha_{j,{\cal G}}(\mu_0)} \, .
\end{eqnarray}
This two-loop factor can be approximated by expanding the variables $\alpha^{-1}_{j, {\cal G}}(\mu)$ using the one-loop RGEs~\cite{Langacker:1992rq}:
\begin{eqnarray}
\gamma_i^{\cal G} \approx -\frac{1}{4\pi}\sum_{j}\frac{b_{ij}^{\cal G}}{a_j^{\cal G}}\ln \frac{\alpha^{-1}_{j,{\cal G}}(\mu_0)}{\alpha^{-1}_{j,{\cal G}}(\mu_0)-{a_j^{\cal G}}t} = -\frac{1}{4\pi}\sum_{j}\frac{b_{ij}^{\cal G}}{a_j^{\cal G}}\ln \left( 1+ a_j^{\cal G} t \, {\alpha_{j,{\cal G}}(\mu)}\right) \, ,
\end{eqnarray}
where we define $t=\frac{1}{2\pi}\ln \frac{\mu}{\mu_0}$. 

In Grand Unified Theories, all the gauge couplings intersect at the unification scale $M_U$ for the value of $\alpha_U$, so we can approximate the gauge couplings at an arbitrary high scale $\mu$ to be the universal gauge couplings $\alpha_U$ at the GUT scale. Now the two-loop factors $\gamma_i^{\cal G}$ become independent of the gauge couplings at the high scale so we can express them by the $\theta_i$ coefficients as in~Refs.~\cite{Langacker:1992rq,Pokorski:2019ete}:
\begin{eqnarray}
\gamma_i^{\cal G} \approx -\frac{\alpha_{U}}{8\pi^2}
 \theta_{i}^{\cal G}
\ln \frac{\mu}{\mu_0} ~~~~ {\rm and}~~~ \theta_{i}^{\cal G} \equiv \sum_j b_{ij}^{\cal G}\frac{\ln (1+a_{j}^{\cal G}\alpha_U t)}{a_{j}^{\cal G}\alpha_U t} \, .
\end{eqnarray}

In summary, the above equation shows the leading-order corrections of the two-loop $\beta$ coefficients $b_{ij}$ to the full two-loop RGEs, which provides the possibility to obtain analytical solutions for the original implicit differential equations. The coefficients $\theta_i$ are a combination of two-loop $\beta$ coefficients $b_{ij}^{\cal G}$ scaling by the one-loop $\beta$ coefficients $a_j^{\cal G}$ times universal coupling $\alpha_U$ and the logarithmic scales $t$. We therefore define the following combination to simplify the common factor appearing in the coefficient $\theta_i^{\cal G}$ between the scales $M_a$ and $M_b$ as:  
\begin{eqnarray}
\Theta_{ab} \equiv  \frac{1}{2\pi}\alpha_U \ln\left(\frac{M_a}{M_b}\right) \, ,
\end{eqnarray}
where $M_a$ is the high scale to be identified as either the GUT scale $M_U$ or the intermediate scale $M_I$ later, while $M_b$ is the reference low scale to be identified as either the intermediate scale $M_I$ or the Electroweak scale $M_Z$. These scaling factors will finally appear in the four constant terms $C_{{\cal G}_{I}}$, $\Delta_{31}^{{\cal G}_{321}}$, $\Delta_{32}^{{\cal G}_{321}}$ and $\Delta_{3_{I}2L_{I}}^{{\cal G}_{I}}$ from definition eq.~(\ref{eq:def-Delta}), and they will be determined from solving eqs.~(\ref{eq:mi})-(\ref{eq:au}) in section 3.2. We summarize the explicit form of the corresponding coefficients $\theta_i^{\cal G}$ for the symmetry groups and representations we considered in Table~\ref{tab:thetai}.

\begin{table}[h!]
\centering
\renewcommand{\arraystretch}{1.8}
\begin{tabular}{|c|c|}
\hline
${\cal G}_I$ &  $\theta_i^{\cal G} $ \\
\hline
\rule{0pt}{3.\normalbaselineskip}
       \ ${\cal G}_{321}{(\rm SM)}$ \ &
       $\begin{pmatrix}
       -\frac{44 \ln \left(1-7 \Theta_{IZ}\right)}{35 \Theta_{IZ}}-\frac{81 \ln \left((6-19 \Theta_{IZ})/{6}\right)}{95 \Theta_{IZ}}+\frac{199 \ln \left((10+41 \Theta_{IZ})/{10}\right)}{205 \Theta_{IZ}} 
       \vspace{1.2mm}
       \\
       -\frac{12 \ln (1-7 \Theta_{IZ})}{7 \Theta_{IZ}}-\frac{35 \ln \left((6-19 \Theta_{IZ})/{6}\right)}{19 \Theta_{IZ}}+\frac{9 \ln \left((10+41 \Theta_{IZ})/{10}\right)}{41 \Theta_{IZ}} 
       \vspace{1.2mm} \\
      \frac{26 \ln \left(1-7 \Theta_{IZ}\right)}{7 \Theta_{IZ}}-\frac{27 \ln \left((6- 19 \Theta_{IZ})/{6}\right)}{19 \Theta_{IZ}}+\frac{11 \ln
   \left((10+41 \Theta_{IZ})/{10}\right)}{41 \Theta_{IZ}}
   \vspace{1.2mm} \end{pmatrix}$ 
    \\
\hline
\rule{0pt}{3.\normalbaselineskip}
       \ ${\cal G}_{321}{(\rm 2HDM)}$ \ &
$\begin{pmatrix}
       -\frac{44 \ln \left(1-7 \Theta_{IZ}\right)}{35 \Theta_{IZ}}-\frac{6 \ln \left(1-3 \Theta_{IZ}\right)}{5 \Theta_{IZ}}+\frac{104 \ln \left((5+21 \Theta_{IZ})/{5}\right)}{105 \Theta_{IZ}} 
       \vspace{1.2mm}
       \\
       -\frac{12 \ln (1-7 \Theta_{IZ})}{7 \Theta_{IZ}}-\frac{8 \ln \left(1-3 \Theta_{IZ}\right)}{3 \Theta_{IZ}}+\frac{2 \ln \left((5+21 \Theta_{IZ})/{5}\right)}{7 \Theta_{IZ}} 
       \vspace{1.2mm} \\
      \frac{26 \ln \left(1-7 \Theta_{IZ}\right)}{7 \Theta_{IZ}}-\frac{3 \ln \left(1- 3 \Theta_{IZ}\right)}{2 \Theta_{IZ}}+\frac{11 \ln
   \left(5+21 \Theta_{IZ})/{5}\right)}{42 \Theta_{IZ}}
   \vspace{1.2mm} \end{pmatrix}$
       \\
\hline
\rule{0pt}{3.\normalbaselineskip}
       ${\cal G}_{422}$ & 
$\begin{pmatrix}
       -\frac{2435 \ln \left(1-7 \Theta_{UI}/3\right)}{14 \Theta_{UI}}+\frac{105 \ln \left(1+2 \Theta_{UI}\right)}{4 \Theta_{UI}}+\frac{747 \ln \left(1+28 \Theta_{UI}/{3}\right)}{56 \Theta_{UI}} 
       \vspace{1.2mm}
       \\
       -\frac{75 \ln (1-7 \Theta_{UI}/3)}{ \Theta_{UI}}+\frac{73 \ln \left(1+2 \Theta_{UI}\right)}{2 \Theta_{UI}}+\frac{36 \ln \left(1+28 \Theta_{UI}/{3}\right)}{7 \Theta_{UI}} 
       \vspace{1.2mm} \\
      -\frac{3735 \ln \left(1-7 \Theta_{UI}/3\right)}{14 \Theta_{UI}}+\frac{24 \ln \left(1+ 2 \Theta_{UI}\right)}{ \Theta_{UI}}+\frac{835 \ln
   \left(1+28 \Theta_{UI}/{3}\right)}{28 \Theta_{UI}}
   \vspace{1.2mm} \end{pmatrix}$
       \\
\hline
\rule{0pt}{3.\normalbaselineskip}
      ${\cal G}_{422}\times {\cal D}$
       & $\begin{pmatrix}
       \frac{3551 \ln \left(1+2 \Theta_{UI}/3\right)}{4 \Theta_{UI}}+\frac{747 \ln \left(1+28 \Theta_{UI}/{3}\right)}{28 \Theta_{UI}} 
       \vspace{1.2mm}
       \\
       \frac{3735 \ln \left(1+2 \Theta_{UI}/3\right)}{4 \Theta_{UI}}+\frac{979 \ln \left(1+28 \Theta_{UI}/{3}\right)}{28 \Theta_{UI}} 
       \vspace{1.2mm} \\
      \frac{3735 \ln \left(1+2 \Theta_{UI}/3\right)}{4 \Theta_{UI}}+\frac{979 \ln \left(1+28 \Theta_{UI}/{3}\right)}{28 \Theta_{UI}}
   \vspace{1.2mm} \end{pmatrix}$  \\
\hline
\rule{0pt}{3.\normalbaselineskip}
     ${\cal G}_{3221}$ &  
     $\begin{pmatrix}-\frac{27\ln\left(1-8\Theta_{UI}/3\right)}{16\Theta_{UI}}-\frac{9\ln(1-2\Theta_{UI})}{4\Theta_{UI}}+\frac{26\ln(1-7\Theta_{UI})}{7\Theta_{UI}}+\frac{\ln\left(1+11\Theta_{UI}/2\right)}{11\Theta_{UI}}\vspace{1.2mm}\\
-\frac{37\ln\left(1-8\Theta_{UI}/3\right)}{8\Theta_{UI}}-\frac{3\ln(1-2\Theta_{UI})}{\Theta_{UI}}-\frac{12\ln(1-7\Theta_{UI})}{7\Theta_{UI}}+\frac{3\ln\left(1+11\Theta_{UI}/2\right)}{11\Theta_{UI}}\vspace{1.2mm}\\
-\frac{9\ln\left(1-8\Theta_{UI}/3\right)}{4\Theta_{UI}}-\frac{31\ln(1-2\Theta_{UI})}{2\Theta_{UI}}-\frac{12\ln(1-7\Theta_{UI})}{7\Theta_{UI}}+\frac{27\ln\left(1+11\Theta_{UI}/2\right)}{11\Theta_{UI}}\vspace{1.2mm}\\
-\frac{27\ln\left(1-8\Theta_{UI}/3\right)}{16\Theta_{UI}}-\frac{81\ln(1-2\Theta_{UI})}{4\Theta_{UI}}-\frac{4\ln(1-7\Theta_{UI})}{7\Theta_{UI}}+\frac{61\ln\left(1+11\Theta_{UI}/2\right)}{11\Theta_{UI}}\vspace{1.2mm}
\end{pmatrix}$
   \\
\hline
\rule{0pt}{3.\normalbaselineskip}
     ${\cal G}_{3221}\times {\cal D}$ &  
     $\begin{pmatrix}-\frac{27\ln\left(1-4\Theta_{UI}/3\right)}{4\Theta_{UI}}+\frac{26\ln(1-7\Theta_{UI})}{7\Theta_{UI}}+\frac{\ln\left(1+7\Theta_{UI}\right)}{14\Theta_{UI}}\vspace{1.2mm}\\
-\frac{167\ln\left(1-4\Theta_{UI}/3\right)}{4\Theta_{UI}}-\frac{12\ln(1-7\Theta_{UI})}{7\Theta_{UI}}+\frac{27\ln\left(1+7\Theta_{UI}\right)}{14\Theta_{UI}}\vspace{1.2mm}\\
-\frac{167\ln\left(1-4\Theta_{UI}/3\right)}{4\Theta_{UI}}-\frac{12\ln(1-7\Theta_{UI})}{7\Theta_{UI}}+\frac{27\ln\left(1+7\Theta_{UI}\right)}{14\Theta_{UI}}\vspace{1.2mm}\\
-\frac{243\ln\left(1-4\Theta_{UI}/3\right)}{4\Theta_{UI}}-\frac{4\ln(1-7\Theta_{UI})}{7\Theta_{UI}}+\frac{115\ln\left(1+7\Theta_{UI}\right)}{14\Theta_{UI}}\vspace{1.2mm}
\end{pmatrix}$ \\
     \hline
 \end{tabular}
   \caption[]{The coefficients $\theta_i$ for the symmetry groups and different breaking schemes that we are considering in our study. 
   } 
\label{tab:thetai}
 \end{table}
 
 \clearpage
 
\subsection*{A3: Some constant coefficients for the SO(10) breaking chains}\label{subsecA3}
We have shown in subsection 3.1 that the two-loop RGEs with the boundary conditions defined as the gauge coupling unification in eq.~(\ref{eq:boundary}) and the matching conditions with an intermediate scale, {e.g. eqs.~(\ref{eq:match-gauge1})-(\ref{eq:match-gauge2})},  will have the solutions in eqs.~(\ref{eq:mi})-(\ref{eq:au}). These solutions are only dependent on the four constant coefficients $C_{{\cal G}_{I}}$, $\Delta_{31}^{{\cal G}_{321}}$, $\Delta_{32}^{{\cal G}_{321}}$ and $\Delta_{3_{I}2L_{I}}^{{\cal G}_{I}}$, where $\Delta_{ij}^{\cal G}$ gives the difference between the $\beta$ coefficients of the gauge coupling $\alpha_{i,{\cal G}}^{-1}$ and those of $\alpha_{j,{\cal G}}^{-1}$:
\begin{eqnarray}
\Delta_{ij}^{\cal G}=\frac{a_i^{\cal G}-a_j^{\cal G}}{2\pi}+\frac{\theta_i^{\cal G}-\theta_j^{\cal G}}{8\pi^2} \alpha_U \, .
\label{eq:def-delta}
\end{eqnarray}

As explained in the main text, for each intermediate symmetry group ${\cal G}_I$ it is enough to consider a particular combination of the $\Delta_{ij}^{\cal G}$, which we call $C_{{\cal G}_{I}}$. They can be proven to have the following forms for the typical breaking chains ${\cal G}_{I}={\cal G}_{422}$ and ${\cal G}_{I}={\cal G}_{3221}$:
\begin{eqnarray}
C_{{\cal G}_{\rm 422}}={3\Delta^{{\cal G}_{\rm 422}}_{42_R}}/{(5\Delta^{{\cal G}_{\rm 422}}_{42_L})} \, , \ \
C_{{\cal G}_{\rm 3221}}={(3\Delta^{{\cal G}_{\rm 3221}}_{32_R}+2\Delta^{{\cal G}_{\rm 3221}}_{3 B\!-\!L})}/({5\Delta^{{\cal G}_{\rm 3221}}_{32_L})} \, ,
\end{eqnarray}
which are basically a combination of {the difference between the $\beta$ coefficients of the gauge couplings of intermediate symmetry group ${\cal G}_I$}. At one-loop level, we can neglect all the two loop coefficients $\theta_i^{\cal G}$ by setting $\alpha_U=0$ in eq.~\ref{eq:def-delta}, so the coefficients $\Delta_{ij}^{\cal G}$ are merely constants. At two-loop order, because the coefficients $\Delta_{ij}^{\cal G}$ are a functions of the set of variables $(\ln\left({M_I}/{M_Z}\right), \ln\left({M_U}/{M_I}\right), \alpha_U)$, eqs.~(\ref{eq:mi})-(\ref{eq:au}) are implicit functions and were solved approximately using eqs. ~(\ref{eq:approx2.1})-(\ref{eq:approx2.2}). For this approximation, we need to calculate the derivatives $\left.\frac{\partial F}{\partial \alpha_U}\right|_{\alpha_U=0}$ and $\left.\frac{\partial G}{\partial \alpha_U}\right|_{\alpha_U=0}$, {which is equivalent to finding} $\left.\frac{\partial \Delta_{ij}^{\cal G}}{\partial \alpha_U}\right|_{\alpha_U=0}$. These derivatives are independent of the scale factor $t=\frac{1}{2\pi}\ln \frac{\mu}{\mu_0}$, so they are also constants when $\alpha_U=0$. We summarize the numerical values of these coefficients for our considered breaking chains in the following Table~\ref{tab:coefficients}.

\begin{table}[h]
\centering
\renewcommand{\arraystretch}{2.0}
\begin{tabular}{|c|c|c|c|c|c|c|c|c|}
\hline
Breaking chains  & $C_{{\cal G}_{I}}$ & $\Delta_{31}^{{\cal G}_{321}}$ & $\Delta_{32}^{{\cal G}_{321}}$ & $\Delta_{3_{I}2L_{I}}^{{\cal G}_{I}}$ & $\frac{\partial C_{{\cal G}_{I}}}{\partial \alpha_U}$ & $\frac{\partial \Delta_{31}^{{\cal G}_{321}}}{\partial \alpha_U}$ & $\frac{\partial\Delta_{32}^{{\cal G}_{321}}}{\partial \alpha_U}$ & $\frac{\partial\Delta_{3_{I}2L_{I}}^{{\cal G}_{I}}}{\partial \alpha_U}$  \\
\hline
         ${\cal G}_{422} \to {\cal G}_{321}{(\rm SM)}$  & $\frac{21}{13}$  & -$\frac{111}{20\pi}$ & -$\frac{23}{12\pi}$ & $-\frac{13}{6\pi}$ & $\frac{266349}{6760\pi}$ & $-\frac{897}{200\pi^2}$ & $-\frac{587}{120\pi^2}$ & $\frac{1721}{48\pi^2}$
        
    \\
\hline
     ${\cal G}_{422} \to {\cal G}_{321}{(\rm 2HDM)}$  &  $\frac{21}{13}$  & -$\frac{28}{5\pi}$ & -$\frac{2}{\pi}$ & $-\frac{13}{6\pi}$ & $\frac{266349}{6760\pi}$ & $-\frac{231}{50\pi^2}$ & $-\frac{26}{5\pi^2}$ & $\frac{1721}{48\pi^2}$
       \\
\hline
         ${\cal G}_{3221} \to {\cal G}_{321}{(\rm SM)}$  &  $\frac{24}{13}$  & -$\frac{111}{20\pi}$ & -$\frac{23}{12\pi}$ & $-\frac{13}{6\pi}$ & $-\frac{669}{3380\pi}$ & $-\frac{897}{200\pi^2}$ & $-\frac{587}{120\pi^2}$  & $-\frac{145}{24\pi^2}$ 
       \\
 \hline
        ${\cal G}_{3221} \to {\cal G}_{321}{(\rm 2HDM)}$  &  $\frac{24}{13}$  & -$\frac{28}{5\pi}$ & -$\frac{2}{\pi}$ & $-\frac{13}{6\pi}$ & $-\frac{669}{3380\pi}$ & $-\frac{231}{50\pi^2}$ & $-\frac{26}{5\pi^2}$ & $-\frac{145}{24\pi^2}$ 
       \\
 \hline

 \end{tabular}
  \caption[]{The four constant coefficients $C_{{\cal G}_{I}}$, $\Delta_{31}^{{\cal G}_{321}}$, $\Delta_{32}^{{\cal G}_{321}}$ and $\Delta_{3_{I}2L_{I}}^{{\cal G}_{I}}$, and their corresponding derivatives appearing in the solutions of the RGEs of SO(10) in eqs.~(\ref{eq:mi})-(\ref{eq:au}) for our considered breaking chains. The numerical results presented here are those when $\alpha_U$ is taking to zero, which is relevant for calculating the two-loop solutions in eqs.~(\ref{eq:approx2.1})-(\ref{eq:approx2.2}).  } 
\label{tab:coefficients}
 \end{table}

\clearpage

\section*{Appendix B: Two-loop RGEs of the low-energy 2HDM}

The RGEs for the three gauge couplings $g_1, g_2, g_3$ are: 
\begin{eqnarray}
16\pi^2\dd{g_1}{t}&=&\frac{21}{5}g_1^3+\frac{g_1^3}{800\pi^2}\left(208g_1^2+180g_2^2+440g_3^2 -85Y_t^2-25Y_b^2 -75 Y_{\tau}^2\right)  \, , \\ 
16\pi^2\dd{g_2}{t}&=&- 3 g_2^3+\frac{g_2^3}{160 \pi ^2} \left(12 g_1^2+80 g_2^2+120 g_3^2-15 Y_t^2-15 Y_b^2-5 Y_{\tau
   }^2\right) \, , \\
16\pi^2\dd{g_3}{t}&=&-7 g_3^3-\frac{g_3^3}{160 \pi ^2} \left(-11 g_1^2-45 g_2^2+260 g_3^2+20
   Y_t^2+20 Y_b^2\right) \, ,
\end{eqnarray}
while those for the third generation Yukawa couplings $Y_t, Y_b, Y_\tau$, are:  
\begin{eqnarray}
&& 16\pi^2\dd{Y_t}{t}=Y_t \left(3 Y_t^2-\frac{17 g_1^2}{20}-\frac{9 g_2^2}{4}-8 g_3^2\right)+\frac{1}{2}Y_t \left(Y_b^2 +3 Y_t^2\right) \, \nonumber \\
&&  +\frac{1}{16 \pi ^2}\left.\Bigg[Y_b^2 Y_t \left(-\frac{9 Y_b^2}{4}-\frac{3 Y_{\tau  }^2}{4}-\frac{41 g_1^2}{240}+\frac{33  g_2^2}{16}+\frac{16 g_3^2}{3}-2 \lambda _3+2 \lambda _4\right) \right. \nonumber \\
&&   + Y_t^3 \left(-\frac{27}{4}Y_t^2+\frac{223}{80} g_1^2+\frac{135}{16} g_2^2+16 g_3^2-12  \lambda _2\right) -\frac{1}{4} \left(Y_b^2 Y_t^3+Y_b^4 Y_t-6 Y_t^5\right) \nonumber  \\
&& +Y_t   \left(-\frac{27 Y_t^4}{4}-\frac{9}{4} Y_b^2 Y_t^2+\frac{1}{8} \left(17 g_1^2+45 g_2^2+160 g_3^2\right) Y_t^2+\frac{1267 g_1^4}{600}-\frac{9}{20} g_2^2 g_1^2+\frac{19}{15} g_3^2   g_1^2 \right. \nonumber  \\
&& -\frac{21 g_2^4}{4}-108 g_3^4+9 g_2^2 g_3^2  +6 \lambda _2^2+\lambda _3^2+\lambda_4^2+6 \lambda _5^2+\lambda _3 \lambda _4\bigg)  \Bigg] \, , \\
&&16\pi^2\dd{Y_b}{t}= Y_b \left(3 Y_b^2+ Y_{\tau }^2-\frac{1}{4}g_1^2-\frac{9}{4}  g_2^2-8 g_3^2\right)+\frac{1}{2}Y_b \left( Y_t^2+3 Y_b^2\right) \, \nonumber \\
&&  +\frac{1}{16 \pi ^2}\left.\Bigg[Y_b Y_t^2 \left(-\frac{9 Y_t^2}{4}-\frac{53 g_1^2}{240}+\frac{33 g_2^2}{16}+\frac{16 g_3^2}{3}-2 \lambda  _3+2 \lambda _4\right) \right. \nonumber  \\
&&   +Y_b^3 \left(-\frac{27 Y_b^2}{4}-\frac{9 Y_{\tau }^2}{4}+\frac{187 g_1^2}{80}+\frac{135 g_2^2}{16}+16 g_3^2-12   \lambda _1\right)-\frac{1}{4}\left( Y_b^3 Y_t^2+Y_b   Y_t^4-6 Y_b^5 \right) \nonumber  \\
&& + Y_b \left(-\frac{27 Y_b^4}{4}-\frac{9 Y_{\tau }^4}{4}-\frac{9}{4} Y_b^2   Y_t^2+\frac{5}{8} \left(g_1^2+9 g_2^2+32 g_3^2\right) Y_b^2+\frac{15}{8} \left(g_1^2+g_2^2\right) Y_{\tau }^2 -\frac{113 g_1^4}{600} \right. \nonumber \\
&& \left. -\frac{27}{20} g_2^2 g_1^2+\frac{31}{15} g_3^2 g_1^2-\frac{21 g_2^4}{4}-108   g_3^4+9 g_2^2 g_3^2+6 \lambda _1^2+\lambda _3^2+\lambda _4^2+6 \lambda _5^2+\lambda _3   \lambda _4\right) \Bigg] \, , \\
&&16\pi^2\dd{Y_{\tau}}{t}=Y_{\tau } \left(3 Y_b^2+ Y_{\tau }^2-\frac{9}{4} g_1^2-\frac{9}{4}g_2^2\right)+\frac{3}{2} Y_{\tau }^3 \, \nonumber \\
&& +\frac{1}{16 \pi ^2}\Bigg[ Y_{\tau }^3 \left(-\frac{27}{4} Y_b^2-\frac{9}{4}Y_{\tau
   }^2 +\frac{387}{80} g_1^2+\frac{135}{16} g_2^2-12 \lambda _1\right)+\frac{3 Y_{\tau }^5}{2} \, \nonumber \\
&& + Y_{\tau } \left(-\frac{9 Y_{\tau }^4}{4}-\frac{27 Y_b^4}{4}-\frac{9}{4} Y_b^2  Y_t^2+\frac{5}{8} \left(g_1^2+9 g_2^2+32 g_3^2\right) Y_b^2+\frac{15}{8} \left(g_1^2+g_2^2\right) Y_{\tau }^2 \right. \, \nonumber \\
&& \left. +\frac{1449  g_1^4}{200}+\frac{27}{20} g_2^2 g_1^2-\frac{21 g_2^4}{4}+6 \lambda _1^2+\lambda_3^2+\lambda _4^2+6 \lambda _5^2+\lambda _3 \lambda _4\right) \Bigg] \, .
\end{eqnarray}



\clearpage



\end{document}